\documentclass[letterpaper,twocolumn,10pt]{article}
\usepackage{usenix-2020-09}

\microtypecontext{spacing=nonfrench}

\usepackage[T1]{fontenc} % Use 8-bit encoding that has 256 glyphs
\usepackage{microtype} % Slightly tweak font spacing for aesthetics
\usepackage[english]{babel} % Language hyphenation and typographical rules
\usepackage[scaled=0.9]{noto-mono}
\usepackage[normalem]{ulem} % Provides strike-through text `\sout'
\usepackage{graphicx} % Required for including pictures
\usepackage{wrapfig} % Allows in-line images such as the example fish picture
\usepackage{subcaption}
\usepackage{hyperref}
\usepackage{xurl} % Break long URL strings in references
\usepackage{mathtools}
\usepackage{amsmath,amsfonts,amsthm}
\usepackage{amssymb} % Required only for NSDI (non-ACM) templates.
\usepackage{mathpartir} % Inference rules, semantics inference
\usepackage{tikz} % lattice, diagram
\usepackage[ruled,linesnumbered,vlined]{algorithm2e}
\SetKwComment{Comment}{// }{}
\SetKwRepeat{Do}{do}{while}
\SetKwProg{Fn}{Function}{:}{end}
\SetCommentSty{}
\usetikzlibrary{positioning}
\usepackage{cleveref}
\usepackage{lastpage} % command: \pageref{LastPage}
\usepackage{zref-totpages} % command: \ztotpages{}
\usepackage{bookmark}
\bookmarksetup{numbered}
\usepackage{enumitem} % Customized lists
\setlist[itemize]{noitemsep} % Make itemize lists more compact
\usepackage{csquotes}
\usepackage{booktabs} % Horizontal rules in tables
\usepackage{threeparttable}
\usepackage{multirow} % Create tabular cells spanning multiple rows
\usepackage{makecell}
\usepackage{xspace} % Add a space unless the macro is followed by punctuations
\usepackage{flushend}
\usepackage{orcidlink} % for ORCID
\graphicspath{{figures/}{figures/experiments/}} % Specifies figure paths

\usepackage{listings}
\newcommand{\name}      {Scylla\xspace}
\newcommand{\para}[1]   {\paragraph{\bf #1 }}

\begin{document}

% don't want date printed
\date{}

% \title{Scaling Data Plane Verification with Intent-based Slicing}
\title{\Large \bf Fine-grained Distributed Data Plane Verification with Intent-based Slicing}

% \author{\rm Paper \#166 - 12 pages body - \ztotpages{} pages total}
% \author{} % placeholder to suppress LaTeX warning
\author{
    {\rm Kuan-Yen Chou}\thanks{Contribution in part while at VMware} \\
    UIUC
    % \email{kychou2@illinois.edu}
    % \orcid{0009-0006-9365-8638}
    \and
    {\rm Santhosh Prabhu}\thanks{Contribution while at VMware} \\
    Microsoft
    % \email{sanprabhu@microsoft.com}
    \and
    {\rm Giri Subramanian} \\
    Broadcom
    % \email{giri-prashanth.subramanian@broadcom.com}
    \and
    {\rm Wenxuan Zhou} \\
    Broadcom
    % \email{wenxuan.zhou@broadcom.com}
    \and
    {\rm Aanand Nayyar}\thanks{Contribution while at VMware} \\
    Lacework
    % \email{aanand.nayyar@lacework.net}
    \and
    {\rm Brighten Godfrey} \\
    Broadcom and UIUC
    % \email{brighten.godfrey@broadcom.com}
    % \orcid{0009-0003-2930-1982}
    \and
    {\rm Matthew Caesar} \\
    UIUC
    % \email{caesar@illinois.edu}
    % \orcid{0000-0001-5955-9229}
} % end author

%     <ccs2012>
%     <concept>
%     <concept_id>10003033.10003083.10003095.10010752</concept_id>
%     <concept_desc>Networks~Error detection and error correction</concept_desc>
%     <concept_significance>500</concept_significance>
%     </concept>
%     <concept>
%     <concept_id>10010520.10010521.10010537.10010541</concept_id>
%     <concept_desc>Computer systems organization~Grid computing</concept_desc>
%     <concept_significance>500</concept_significance>
%     </concept>
%     <concept>
%     <concept_id>10011007.10011074.10011099</concept_id>
%     <concept_desc>Software and its engineering~Software verification and validation</concept_desc>
%     <concept_significance>300</concept_significance>
%     </concept>
%     </ccs2012>

% \ccsdesc[500]{Networks~Error detection and error correction}
% \ccsdesc[500]{Computer systems organization~Grid computing}
% \ccsdesc[300]{Software and its engineering~Software verification and validation}

% \keywords{network verification, intent-based partitioning, distributed processing}


\maketitle
\pagestyle{plain} % Remove the running header (conference & short authors list)

\begin{abstract}
Data plane verification has grown into a powerful tool to ensure network correctness. However, existing methods with monolithic models have memory requirements tied to network sizes, and the existing method of scaling out is too limited in expressiveness to capture practical network features. In this paper, we describe \name, a general data plane verifier that provides fine-grained scale-out without the need for a monolithic network model. \name creates models for what we call intent-based slices, each of which is constructed at the rule-level granularity with only enough to verify a given set of intents. The sliced models are retained and incrementally updated in memory across a distributed compute cluster in response to network updates. Our experiments show that \name makes the scaling problem more granular --- tied to the size of the intent-based slices rather than that of the overall network. This enables \name to verify large, complex networks in minimum units of work that are significantly smaller (in both memory and time) than past techniques, enabling fast scale-out verification with minimal resource requirement.
\end{abstract}

\section{Introduction}
\label{sec:introduction}

% List of related work about data plane verification.
%
% 2005-XieEtAl (Xie et al.)
% 2011-MaiEtAl (Anteater)
% 2012-KazemianEtAl (HSA)
% 2013-KazemianEtAl (HSA/NetPlumber)
% 2013-KhurshidEtAl (VeriFlow)
% 2013-YangLam (AP verifier)
% 2014-ZengEtAl (Libra)
% 2016-BjoernerEtAl (ddNF)
% 2017-HornEtAl (Delta-net)
% 2017-PandaEtAl (VMN, stateful DP)
% 2019-JayaramanEtAl (Azure DCN local contracts, SecGuru/RCDC)
% 2020-ZhangEtAl (APKeep)
% 2020-YuanEtAl (NetSMC, stateful DP)
% 2020-YousefiEtAl (Liveness verification)
% 2023-GuoEtAl (TOBDD)
% 2024-WenEtAl (Medusa)

% Introduce data plane verification and explain the term ``intent''.
Recent years have seen rapid advances in data plane verification (DPV) research~\cite{2005-XieEtAl, 2011-MaiEtAl, 2012-KazemianEtAl, 2013-KazemianEtAl, 2013-KhurshidEtAl, 2013-YangLam, 2014-ZengEtAl, 2016-BjoernerEtAl, 2017-HornEtAl, 2017-PandaEtAl, 2019-JayaramanEtAl, 2020-ZhangEtAl, 2020-YuanEtAl, 2020-YousefiEtAl, 2023-GuoEtAl, 2024-WenEtAl}, alongside several commercial solutions~\cite{vmware-aria, azure-network-watcher, aws-network-verification, google-nic, forward} available for network operators to verify correctness intents.\footnote{We use the term ``intent'' to refer to expected properties of forwarding behavior, similar to ``policy'' or ``invariant'' in some related work.}
% Typically, the intents that the operators wish to check pertain to traffic forwarding behavior, such as reachability of hosts, segmentation of applications, routing waypoints for security enforcement, flow consistency across devices, and absence of forwarding loops. These intents often need to be checked over a variety of data planes, including physical, virtual, and hybrid networks.
%
% Background on the perspective of a VaaS provider.
Early commercial adoption of network verification was tailored for hyperscale cloud providers~\cite{2015-FogelEtAl, 2016-PlotkinEtAl, 2019-JayaramanEtAl}, where ample compute and memory resources can be dedicated to the verifier. However, there is growing interest among non-hyperscale enterprises to adopt this technology for their networks without having to allocate such resources~\cite{2024-WenEtAl, vmware-aria, forward}. In this paper, we address the problem of providing verification solutions for general, non-hyperscale enterprise networks, which presents new restrictions and challenges.

% To provide a solution for different environments, one challenge is scalability under varying resource and cost constraints.
Although existing DPV techniques can function on well-provisioned machines, some enterprise environments have only modest memory, with DPV being one of many network management services. Since ease of deployment is critical, it is thus advantageous to design DPV as a lightweight, on-demand service that can handle limited memory environments, rather than a heavyweight appliance.

% Point out that monolithic modeling is a problem with scalability.
Most resource optimizations in prior work focused on optimizing models with more efficient data structures~\cite{2012-KazemianEtAl, 2013-KhurshidEtAl, 2013-YangLam, 2016-BjoernerEtAl, 2017-HornEtAl, 2023-GuoEtAl} or leveraging network commonalities to reduce redundancy~\cite{2014-MajumdarEtAl, 2016-PlotkinEtAl}. However, these techniques do not address a fundamental issue --- the monolithic nature of symbolic models, where resource demands scale with network size and hinder workload distribution. To address this, it is crucial to decompose a network model into fine-grained slices\footnote{A ``slice'' refers to a model partition representing a partial network in both header and topological spaces, formally defined as a set of hypercubes in the network space defined in \cite{2012-KazemianEtAl}.}, enabling distributed verification. While slicing is common in large-scale data processing (e.g., MPI~\cite{1994-GroppEtAl}, MapReduce~\cite{2004-DeanGhemawat}), it is challenging to apply in DPV due to the interconnected nature of networks --- the computation of DPV may require any part of the network, which is not easy to be neatly divided. As a result, existing scale-out DPV proposals either lack support for common network features~\cite{2014-ZengEtAl, 2024-WenEtAl}, or work only with specific network architectures~\cite{2019-JayaramanEtAl}.

\vspace{-1em}
\para{Method}
%
% Describe the observation that most intents involve only a narrow slice.
%
To efficiently scale out without limiting the network functionality or architecture, we observe that, even in large networks, most intents can be verified against small network slices and do not require a full network model. For example, a reachability intent like \textit{``VM1 in the App tier can reach VM2 in the DB tier''} requires modeling only the overlays and underlays between these two VMs, and only the forwarding rules that handle packets between them. Similarly, checking for consistency between two firewalls does not require other devices to be modeled. This also applies to other intents like segmentation, waypoint, black hole freedom, etc.

% Describe \name's design.

Building on this observation, we propose \name, the first data plane verifier that slices the network model at the intent granularity. Each slice contains only what is necessary to verify its intent, and these slices are distributed and stored in memory across a cluster for verification in parallel. Since each slice is only a tiny fraction of the network, \name enables verification of large networks with minimal resources.

%
% Summarize technical contributions.
%

%\bgnew{Note that while past work has also created slices (particularly, \cite{2014-ZengEtAl} sliced by IP prefix), intent-based slicing is a new kind of slicing with very different scaling and functionality properties that allows us to solve the ``interconnectedness'' problem mentioned above and handle general networks.}

While past work has also partitioned network models, \name's intent-based slicing is different from all prior designs and comes with opportunities and challenges that did not exist in prior work that partitions only the packet header space.
% network updates
For example, Libra~\cite{2014-ZengEtAl} partitions the network by IP prefix, and all intents need reevaluation when network state changes within a partition, while in \name, any network state changes (e.g., rule or topology changes) are incrementally processed based on intents, achieving minimal overhead for continuous verification.
% intent colocation to reduce duplicates
In addition, while \name drastically reduces the individual partition sizes compared to prior work, it is possible for intent slices to overlap. To improve model sharing across intents and to reduce rule duplication across slices, \name supports intent colocation to cluster similar intents together, which further reduces the overall verification time and memory usage.
% intent decomposition and distributed loop detection
Finally, the size of a slice depends on its corresponding intent, and while most are tiny, some may be large, for which we introduce intent decomposition to break down such intents into smaller sub-intents across compute nodes. An extreme case is loop freedom, essentially the only intent in practice that uses the entire model. We adapt a distributed cycle detection algorithm~\cite{2015-RochaThatte} for loop detection through communication between slices.

\vspace{-1em}
\para{Results}
%
% Summarize the evaluation results.
%
% We performed detailed microbenchmarks of \name on a standalone machine for repeatable comparisons, and evaluated the end-to-end performance in a cluster connected to a live production network.
%
In microbenchmarks, \name is 149--1362$\times$ faster while requiring 21--22$\times$ less memory compared to the monolithic designs when verifying a single intent. This demonstrates \name's ability to distribute workloads efficiently and to allow for on-demand use cases (verification upon interactive user request, \S\ref{sec:motivation}).
For batch-processing multiple intents in the largest tested network, \name is 51$\times$ faster and uses 19.6$\times$ less memory than the monolithic design, and 23$\times$ faster with 5$\times$ less memory than Libra's design.
In a live cluster connected to a production network, we show that it takes less than 3.5 minutes to verify 3,000 intents with 5 VMs (each with a single thread), and that \name can scale out easily by adding more VMs to the cluster. To our knowledge, \name is the first fine-grained, scale-out verifier to demonstrate this kind of scalability in terms of the number of intents.
% The intent colocation in \name also consistently improves resource utilization compared to a naïve distribution.

\vspace{-1em}
\para{Contributions}
% Small conclusion for the introduction, emphasizing the contributions.
In summary, \name
% The design of \name results in a scalable verifier that
(1) enables efficient DPV within resource constraints, (2) allows fast, on-demand verification without having to maintain the entire network model, (3) can easily scale out by expanding the compute cluster for higher numbers of intents, and (4) can model full functionality of modern networks, including non-IP rules and header rewrites. Our method fundamentally changes the scaling of the verification problem, from scaling with the network size to scaling with the size of intent slices, facilitating easy scale-out and on-demand resource allocation.

% \para{Ethical Issues}
% This work does not raise any ethical issues.

%%%%%%%%%%%%%%%%%%%%%%%%%%%%%%%%%%%%%%%%%%%%%%%%%%%%%%%%%%%%%%%%%%%%%%%%%%%%%%%%

\section{Motivation}
\label{sec:motivation}

% Point out that monolithic modeling is a problem for scaling out.
Most existing data plane verifiers are \emph{monolithic}, where a single machine hosts and operates on a symbolic model of the entire network~\cite{2005-XieEtAl, 2011-MaiEtAl, 2012-KazemianEtAl, 2013-KhurshidEtAl, 2013-YangLam, 2017-HornEtAl, 2017-PandaEtAl, 2020-ZhangEtAl, 2020-YuanEtAl, 2020-YousefiEtAl}. One may think this monolithic design is good enough, as it has been shown to be tractable in hyperscalers running periodic verification for data center networks, such as Microsoft~\cite{2014-JayaramanEtAl, 2015-LopesEtAl, 2019-JayaramanEtAl}. However, to provide a scalable verifier for general networks,\footnote{By general networks, we refer to networks of arbitrary size and architecture, typically operated by non-hyperscale enterprises or organizations.} we identify the following four design goals where monolithic approaches fall short.

% Explain why the monolithic designs are not good enough.

\vspace{-1em}
% \para{Resource Requirement Barrier with On-premises Deployments}
\para{Low Memory Requirement Even For Large Networks}
Enterprises that deploy network management software on-premises prefer general-purpose cluster machines over special high-memory machines. Moreover, DPV may run as just one component among many management services (e.g., flow telemetry, resource orchestration). For smooth DPV deployment and operation, memory requirements should remain low even for large networks with tens of thousands of physical (e.g, routers, firewalls) and virtual devices (e.g., SDN-based overlays). However, with a monolithic design, memory requirement scales linearly with the network size, sometimes reaching hundreds of GB~\cite{2024-WenEtAl}. While it may be possible for an enterprise to provision such machines, it is a hindrance especially for non-specialist administrators (causing what is known as ``deployment friction''), which can lead to DPV not being deployed or hitting memory limits that cause it to fail.

\vspace{-1em}
% \para{Increased Cost for Cloud Deployments}
\para{Low Ongoing Cost for Cloud Deployments}
Even with DPV deployments on public cloud services\footnote{This refers to verification infrastructures running on a public cloud, while the network being verified might be on-premises.} where resources are plentiful, it is essential to optimize for resource usage to avoid excessive cost. The monolithic design incurs higher cost from cloud services and is thus undesirable. Additionally, even from the perspective of cloud providers who provide verification services for their customers~\cite{azure-network-watcher, aws-network-verification, google-nic}, a monolithic verifier is also suboptimal for both economical and sustainability reasons. As monolithic verifiers require building the entire model and keeping it in memory during verification, this leads to increased operational costs and inefficient energy usage, making the monolithic design unfavorable for the cloud providers' sustainability goals~\cite{microsoft-sustainability, aws-sustainability, google-sustainability, google-datacenter-pue}.

\vspace{-1em}
% \para{Delayed Response When Building/Loading from Disk}
\para{Fast On-demand Verification From Scratch}
One common use case of DPV is \emph{interactive} or \emph{on-demand} verification, where users (e.g., operators, engineers) issue requests to verify the network or candidate network changes while debugging or developing, which demands fast response even with cold start. In this case, a monolithic verifier would need to either build the entire model from scratch, or swap in a built model from disk. As shown in \cite{2014-ZengEtAl} and our experiments (\S\ref{subsec:microbenchmark}), this leads to delayed response due to disk I/O with large networks. Of course, it is possible to keep the monolithic model in memory persistently even without active workload, which then further increases deployment friction.

\vspace{-1em}
% \para{Limited Cores on a Single Machine}
\para{Easy Scale-out with Inexpensive Machines}
Depending on the scale of applications and networks, the number of intents can range from a dozen to several thousands (e.g., to check connectivity for all deployed applications). However, the monolithic design is ultimately limited by the cores available on a single machine, which delays the verification time when higher numbers of intents queue up for the limited cores. While one could distribute the workload by duplicating the monolithic model across machines, it is challenging and costly to keep all copies in sync with network changes.

\vspace{-1em}
\para{To summarize,}
while a monolithic verifier may currently suffice for periodic verification, all the above provide strong practical incentives for a more lightweight and scalable DPV to broaden the scope of use cases. Also, fundamentally changing the scaling behavior of DPV could be valuable independent of specific use cases.

\vspace{-1em}
\para{Limitations of Existing Scale-out DPV}
To achieve the design goals, it is desirable to scale \emph{out} DPV by dividing up the network model across machines.
% Previous proposals of scaling out verification.
Such scale-out verification has received relatively little attention in academia. One related work is Libra~\cite{2014-ZengEtAl}, which proposed partitioning the network model at the granularity of IP subnets and distributing the workload via MapReduce~\cite{2004-DeanGhemawat}. However, this approach can be inefficient when modeling non-IP rules (e.g., Ethernet, VLAN, MPLS): since non-IP rules are orthogonal to the IP space, they would be duplicated across all partitions. Even more problematic are header rewrites (e.g., NAT, tunneling) which cause network flows to jump between partitions. Without communication between partitions, this renders Libra's approach non-functional. Though Libra was successful for verifying IP forwarding~\cite{2014-ZengEtAl}, the approach is not suitable for general networks, where non-IP rules and header rewrites are prevalent. To the best of our knowledge, none of the prior scale-out proposals are general-purpose verifiers that work well for networks of arbitrary size and architecture (see \S\ref{sec:related-work} for a more detailed discussion of related work).

%%%%%%%%%%%%%%%%%%%%%%%%%%%%%%%%%%%%%%%%%%%%%%%%%%%%%%%%%%%%%%%%%%%%%%%%%%%%%%%%

\section{Network Model}
\label{sec:model}

To understand \name, we begin with our data plane model, including the data plane abstraction, packet set encoding, and the symbolic traversal of models. Though these components are not core technical contributions of this work, we describe them here for completeness and to provide context.

\vspace{-0.5em}
\subsection{Data Plane Abstraction (DPA)}
\label{subsec:dpa}

\begin{figure}
    \centering
    \includegraphics[width=0.6\linewidth]{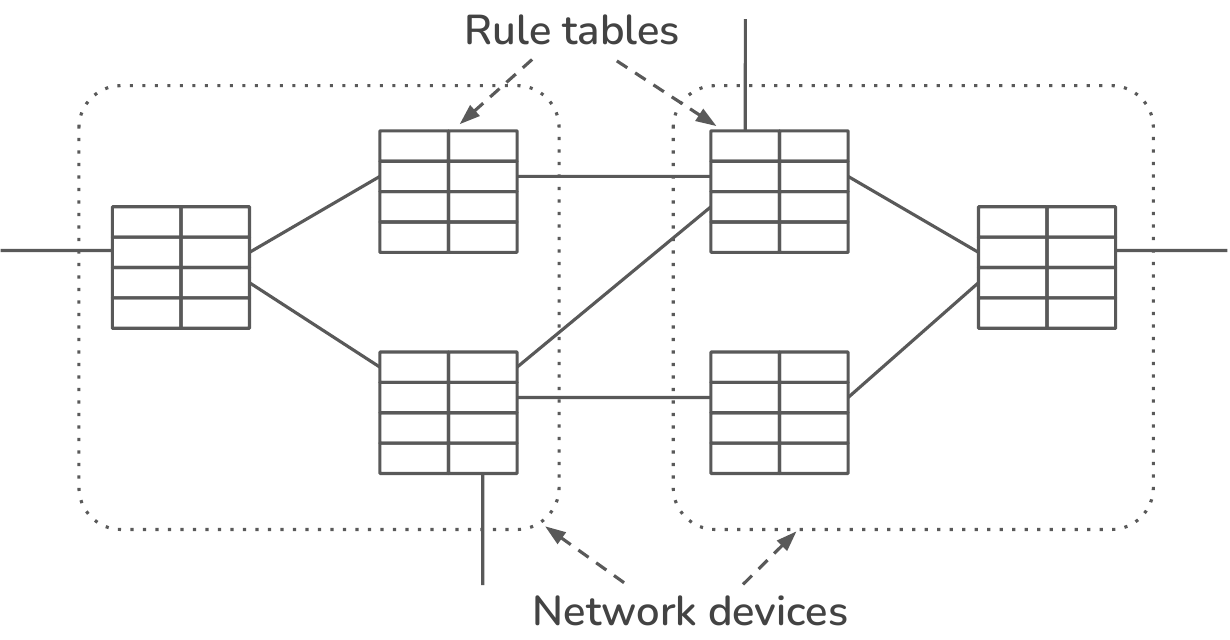}
    \vspace{-0.5em}
    \caption{Rule tables in a network model.}
    \label{fig:rule-table-layout}
    \vspace{-0.5em}
\end{figure}

Each device in the network is modeled by a vendor-agnostic representation called Data Plane Abstraction (DPA). Each DPA contains a JSON object that encodes the entire data plane state of a device, including interfaces, forwarding tables, and other attributes. We provide an example DPA in \S\ref{app:subsec:dpa-examples}.

All DPAs together constitute \name's network model that views the network as a collection of \emph{rule tables}. A rule table contains a prioritized list of \emph{rules}, each of which matches a specific set of packets and executes one or more actions. Possible actions include modifying packet headers, duplicating the packet, dropping it, and forwarding it to other rule tables or external endpoints. In essence, a rule table is analogous to a match-action table in a P4 program~\cite{2014-BosshartEtAl, 2017-BudiuDodd}, which is used to model any lookup table, such as forwarding tables, ACL tables, or VRF tables. As Figure~\ref{fig:rule-table-layout} shows, entities like switches, firewalls, etc., are simply aggregations of close-knit rule tables and hold no semantic significance in \name.

\vspace{-0.5em}
\subsection{Flow Nodes}
\label{subsec:flow-nodes}

\begin{figure}
    \centering
    \includegraphics[width=0.7\linewidth]{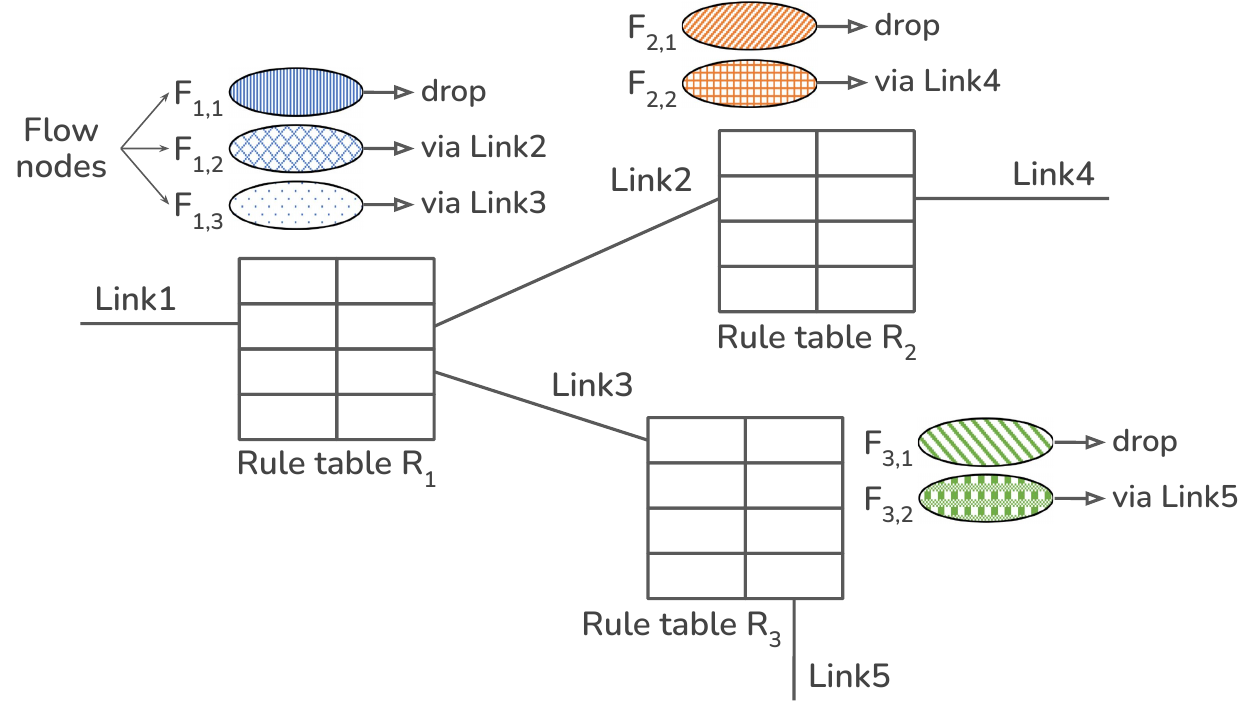}
    \vspace{-0.5em}
    \caption{Example flow nodes of rule tables. Color/pattern indicates different packet sets.}
    \label{fig:flownodes}
    \vspace{-1em}
\end{figure}

To compactly represent the behavior for every packet, \name computes \emph{flow nodes} for each rule table. Like a rule, a flow node consists of a packet set and a sequence of actions. However, the packet set of a flow node represents an equivalence class (EC) of packets at the rule table (i.e., a maximal set of packets to which the same sequence of actions are applied).\footnote{Note an EC here refers to a packet set whose elements exhibit an identical behavior at a single rule table, rather than across the entire network.} In addition, the packet sets of a rule table's flow nodes are mutually disjoint, so no priority is needed. For example, an ACL table typically has two flow nodes: one for the \emph{allow} action and one for the \emph{deny} action. A forwarding table may have multiple flow nodes, each depicting a distinct EC and their corresponding forwarding behavior.
% These flow nodes are used for symbolic traversal to verify intents.
Figure~\ref{fig:flownodes} illustrates the flow nodes for some rule tables.

\vspace{-0.5em}
\subsection{Packet Set Representation}
\label{subsec:packet set}

Over the past decade, there has been great progress in packet set representations.\footnote{Packet sets are sometimes referred to as flows, predicates, or packet filters in related literature.} A packet set may be encoded as a ternary bit-vector~\cite{2012-KazemianEtAl,2013-KazemianEtAl}, a multidimensional trie~\cite{2013-KhurshidEtAl}, a binary decision diagram (BDD)~\cite{2013-YangLam, 2020-ZhangEtAl, 2023-GuoEtAl}, a set of mutually disjoint intervals~\cite{2017-HornEtAl}, or a customized data structure~\cite{2016-BjoernerEtAl}.

In this work, we employ a variant of BDD~\cite{2021-KheradmandPrabhu-blinded} to represent packet sets in the network model. However, it is important to note that designing a data structure for packet sets is orthogonal to the focus of this paper. Regardless of which data structure one uses to encode packet sets, the design benefits of \name still apply. To factor this out of our experiments, we use the same packet set encoding for all evaluated methods.

\vspace{-0.5em}
\subsection{Traversing the Model}
\label{subsec:traversal}

\begin{figure}
    \centering
    \includegraphics[width=0.9\linewidth]{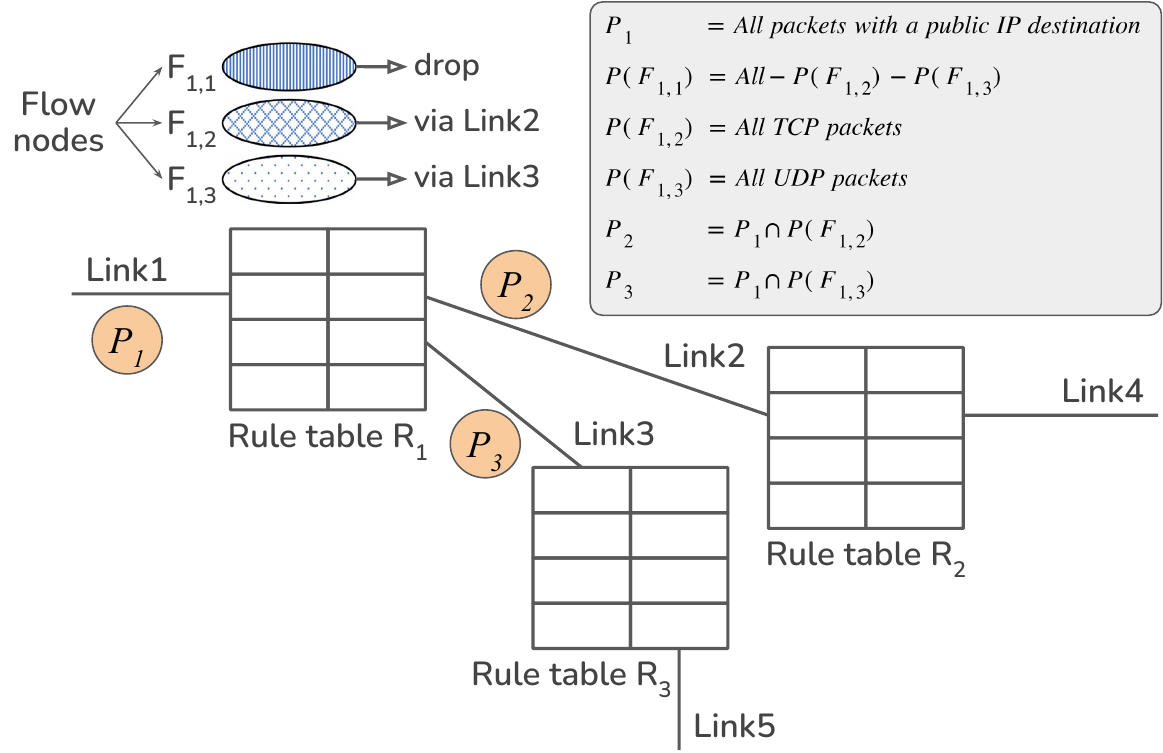}
    \vspace{-0.5em}
    \caption{An example of path traversal at step $(R_1,P_1)$.}
    \label{fig:path-traversal}
    \vspace{-1em}
\end{figure}

\name employs the symbolic model to determine the forwarding behavior of the network. This involves a depth-first traversal of the model's state space.
% The choice of $R_0$ and $P_0$ depends on the specific intent being verified.
Starting from a rule table, $R_0$, and a packet set, $P_0$, this $(R_i,P_i)$ pair acts as a vertex during the traversal. At each step, an unprocessed $(R_i,P_i)$ pair is selected. Then, to obtain the successor pairs, \name computes the intersection of $P_i$ and each flow node of $R_i$ before applying the associated actions to the intersected packet set. If the intersected packet set is empty, it means no packet in $P_i$ would exhibit such behavior represented by the flow node. Otherwise, the actions may transform and eventually drop or forward the packet set to the next rule table, producing a successor pair. See Figure~\ref{fig:path-traversal} as an example.

We note two pertinent model characteristics for \name. First, the model is an aggregation of rule tables. This allows modeling and checking any fragment of a device independently, disregarding the absence of other fragments. Second, it is possible to start the depth-first traversal from a single rule table and, on the fly, determine the rule tables and the rules that are required based on the current traversing states. Consequently, the model can expand in memory as needed.

%%%%%%%%%%%%%%%%%%%%%%%%%%%%%%%%%%%%%%%%%%%%%%%%%%%%%%%%%%%%%%%%%%%%%%%%%%%%%%%%

\section{System Design}
\label{sec:design}

We start with an overview of the pipeline architecture (\S\ref{subsec:architecture}), followed by the details of intent-based slicing (\S\ref{subsec:intent-based-slicing}), how the intents are verified and what types of intents \name can support (\S\ref{subsec:verifying-intents}), and finally the intent colocation methods for further improving model sharing across slices (\S\ref{subsec:intent-colocation}).

\subsection{Architecture Overview}
\label{subsec:architecture}

\begin{figure}
    \centering
    \includegraphics[width=0.82\linewidth]{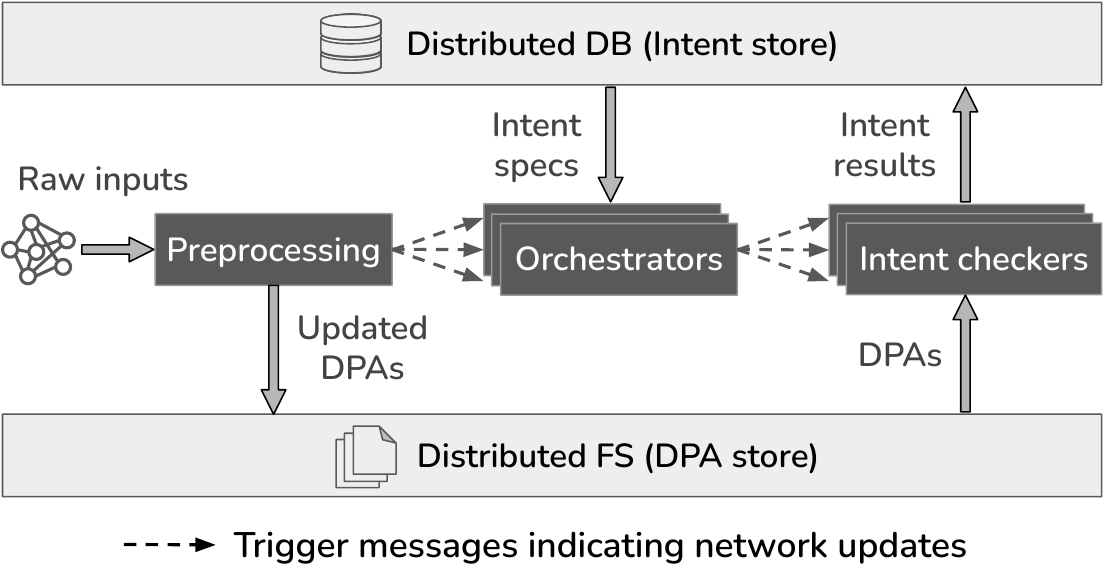}
    \vspace{-0.5em}
    \caption{\name processing pipeline.}
    \label{fig:architecture}
    \vspace{-1em}
\end{figure}

% Introduce the overall pipeline architecture.

As outlined in Figure~\ref{fig:architecture}, \name is designed as a system of event-driven, message-passing functions running in a distributed compute cluster. When a function consumes a message, it executes the corresponding task and generates zero or more output messages that trigger subsequent functions if needed. Additionally, all functions have access to a timeline-aware distributed database for intent specifications and a distributed file system for data storage. \name's major functional units are as follows.

% Describe the Preprocessing function.

At first, the \emph{Preprocessing} function collects raw data from the data plane through SSH, SNMP, and vendor APIs, etc.\footnote{The source of this network data may be a live network for monitoring live events, or a test network to verify network changes before deployment.} This data is parsed, cleaned, and stored as DPAs in the distributed file system, capturing the data plane state as rule tables (\S\ref{sec:model}). The \emph{Preprocessing} function then sends out a message for the devices that have undergone semantic changes (e.g., rule changes). Each message contains only the updated device IDs rather than the actual changes, indicating a network update. As we collect data from the network much more frequently than the data plane changes occur, this method reduces redundant processing significantly.

% Describe the Orchestrator function.

The \emph{Orchestrator} function is a demultiplexer for network updates and intents. When messages from the \emph{Preprocessing} function arrive, the \emph{Orchestrator} fetches the list of configured intents from the database and sends out one message for each intent with the updated device IDs, indicating the intent that \emph{may} need to be checked due to these device updates. To avoid redundant messages, for a new intent or a new network, the \emph{Orchestrator} sends only a single message containing \emph{all} device IDs in the network. In addition, the \emph{Orchestrator} is also responsible for distributing intents to improve model sharing and resource utilization, discussed in \S\ref{subsec:intent-colocation}.

% Describe the Intent checker function.

Each message from the \emph{Orchestrator} is processed by the \emph{Intent checker} function, where it (1) creates or updates the slice based on the updated DPAs as needed by the intent (\S\ref{subsec:intent-based-slicing}), (2) (re-)verifies the intent when necessary (\S\ref{subsec:verifying-intents}), and (3) updates the verification result in the database.

\subsection{Intent-based Slicing}
\label{subsec:intent-based-slicing}

For ease of presentation, we begin by assuming that each slice is created and maintained for one intent. This assumption will be relaxed later, as \name can handle multiple intents with one slice, or multiple slices for one intent.

\vspace{-0.5em}
\subsubsection{Creating an Intent Slice}
\label{subsubsec:create-slice}

\begin{figure}
    \centering
    \includegraphics[width=0.8\linewidth]{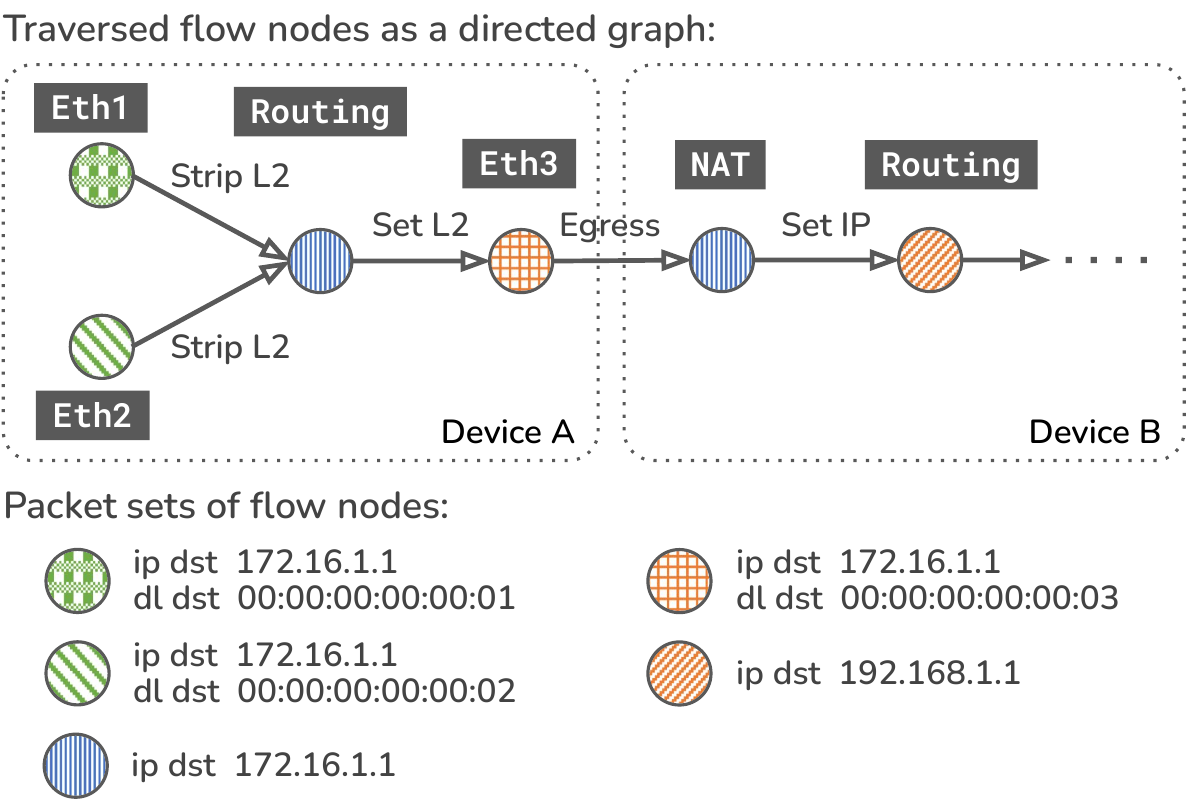}
    \vspace{-0.5em}
    \caption{Intent slice as flow nodes along path traversal.}
    \label{fig:intent-slice}
    \vspace{-1em}
\end{figure}

% High-level idea of on-demand model building.

The key design choices of \name are that a slice is only built during intent verification, and that the intent slice comprises only the part of the network traversed by the intent, in both the header space and the topological space.

% Detailed process of building an intent slice.

Specifically, upon receiving a message from the \emph{Orchestrator}, the \emph{Intent checker} begins to traverse the model based on the starting points specified by the intent, where each starting point is an $(R_0,P_0)$ pair, as defined in \S\ref{subsec:traversal}. At each step of traversing an $(R_i,P_i)$ pair with a newly encountered rule table $R_i$, instead of building the flow nodes of $R_i$ for the entire header space as described in \S\ref{subsec:flow-nodes}, \name builds the flow nodes only for $P_i$, from only the rules of $R_i$ that overlap with $P_i$. In other words, the union of all flow nodes' packet sets becomes the same as $P_i$. This union packet set is called the \emph{universe} of $R_i$, denoted as $U_i$. In real networks, however, it is possible to encounter the same rule table more than once during traversal but with a different packet set, such as the case of VLAN or tunneling. Assuming $R_i = R_{i+n}$, the universe of $R_{i+n}$ \emph{expands} as $U_{i+n} \gets U_i \cup P_{i+n}$. This cycle of expanding universes and building flow nodes repeats until the traversal ends or until the intent is violated. At this point, the slice is built, and the verification result is updated to the database. Concretely, an intent slice refers to the set of all flow nodes traversed by the intent. Figure~\ref{fig:intent-slice} shows an example slice of a reachability intent from Device A to 172.16.1.1.

The slice created is retained in memory, which represents the minimal information required to verify this particular intent since omitting any rule or flow node in the slice would have prevented the traversal. More importantly, even when a device contains many rule tables, or when a rule table has a huge amount of rules, only the specific rules within the rule tables traversed by the intent are built into the slice, which is small in most cases.

\vspace{-0.5em}
\subsubsection{Updating an Existing Slice}
\label{subsubsec:update-slices}

Once an intent slice has been created, \name keeps the slice updated with the subsequent network updates. This step offers an opportunity for optimization. For instance, consider a large firewall with thousands of ACL rules. The \emph{Preprocessing} function suppresses the updates with no actual rule changes (\S\ref{subsec:architecture}). However, even when a rule does change, it oftentimes would affect only a few intents or none at all. To minimize model updates, we develop an efficient method to filter out irrelevant updates based on the existing slice.
% For example, if a rule change applies exclusively to 10.0.0.0/8, it has no impact on an intent whose slice does not involve that subnet, assuming no other devices are updated simultaneously.

When a message is sent by the \emph{Orchestrator} to the \emph{Intent checker} for each network update, containing the updated device IDs, if the existing slice does not include the updated device, the message is soundly ignored. Otherwise, if the slice contains any part of the updated device, the checker fetches the new DPA and compares the rule differences between the old and the new DPAs based on the rule table universes. If any relevant rule\footnote{A rule is ``relevant'' if its packet set overlaps with the rule table's universe.} changes, by definition, the slice could change, and so the intent needs to be re-evaluated with the new slice. To further optimize this rule comparison, instead of iterating through all the rule tables and comparing the rules for each update, the checker maintains the hash value of all relevant rules during traversal and simply compares the hash value with that of the relevant rules in the new DPA to determine if the slice requires updating. Updating a slice can be achieved by simply deleting the old flow nodes of the updating rule tables, since the traversal would automatically build the slice as needed (\S\ref{subsubsec:create-slice}). During this process, some rules or rule tables may become irrelevant altogether, and others may become relevant. Therefore, the latest DPAs can always be fetched from the distributed file system and incorporated into the slice as needed. Any part of the slice that is no longer required by the intent is discarded from memory.

\subsection{Verifying Intents}
\label{subsec:verifying-intents}

\subsubsection{Intent Verification with One Slice}
\label{subsubsec:one-slice-verification}

% Details of intent check implementation.

Upon receiving a network update, the intent checker verifies if the slice requires updating (\S\ref{subsubsec:update-slices}), sets up the starting points, and initiates a symbolic traversal (\S\ref{subsec:traversal}). At each step, the checker invokes a predicate function specific to the intent type, assessing whether the current traversal state (the DFS history, stack, and path) constitutes a violation. \name provides the predicate functions for the intent types in \S\ref{subsubsec:intent-types}. To implement a predicate function for a new intent type, one only needs to define when the traversal state forms a violation, when it will never lead to a violation, and when it remains inconclusive (i.e., the traversal should continue). Example algorithms are provided in \S\ref{app:subsec:verify-intents} as reference.

\vspace{-0.5em}
\subsubsection{Supported Intent Types}
\label{subsubsec:intent-types}

% The class of supported intents.
Moving from a monolithic design to \name's distributed design does not change the scope of intents that are \emph{semantically possible} to be verified. Similar to past approaches~\cite{2013-KhurshidEtAl}, intents in \name are implemented as arbitrary computation that explores the model. In practice, \name factors out routines that are common to many intents, like graph traversal on top of the flow node abstraction. Many intents that are prevalent in past literature and in practice can be naturally supported in this framework; we specifically implemented reachability, segmentation, waypoint, black hole freedom, loop freedom, and flow consistency.
% Intent specification and encoding
Each intent is specified as a JSON object with type-specific parameters, which commonly include things like source devices, target packet sets, or target devices, etc. We show some example intent specifications in \S\ref{app:subsec:intent-examples}.

% Define wide and narrow intents.

Although \name's intent-based slicing \emph{semantically} supports various intents, it works best with intents that involve a small slice, which we describe as \emph{narrow} intents, as opposed to \emph{wide} intents that traverse more of the network. % \footnote{Note here that whether an intent is narrow or wide is relative compared to other intents and the network size.}
Checking wide intents may reduce \name's improvement compared to the monolithic approach due to larger intent slices.\footnote{The worst case reduces \name to exactly the monolithic approach, where the intent slice becomes the entire network.}
However, we observe that most intents from the past literature are narrow~\cite{2005-XieEtAl, 2011-MaiEtAl, 2012-KazemianEtAl, 2013-KazemianEtAl, 2013-KhurshidEtAl, 2013-YangLam, 2014-ZengEtAl, 2017-HornEtAl, 2017-PandaEtAl, 2019-JayaramanEtAl, 2020-ZhangEtAl, 2020-YuanEtAl}, and the same can also be observed in practice. From an industry DPV deployment with 5260 devices, we see that the longest path traversed by the 11 segmentation intents contains only 6 devices.
Nevertheless, for wide intents that are composed of multiple traversals, we provide a generic way to break them down into narrower sub-intents (\S\ref{subsubsec:composite-intent}), which can be verified distributively. For loop freedom, arguably the most challenging intent since it requires analyzing the entire network, we propose a distributed loop detection algorithm (\S\ref{subsubsec:distributed-loop}).

\vspace{-0.5em}
\subsubsection{Composite Intent}
\label{subsubsec:composite-intent}

% Introduce the consistency intent as an example of a composite intent.

While we initially assumed a one-to-one relationship between intents and slices, \name allows one intent to utilize multiple slices distributed across checkers. This approach naturally aligns with the microservice-based architecture. For instance, our prototype supports flow consistency intents, which verifies consistency across flows from one or several starting points. This is an example of a \emph{composite} intent, which involves multiple traversals and can be divided into independent sub-intents, each with their own smaller slice, without splitting the underlying traversal. The results from these sub-intents are then combined to work out the original intent.

\begin{figure}
    \centering
    \includegraphics[width=0.65\linewidth]{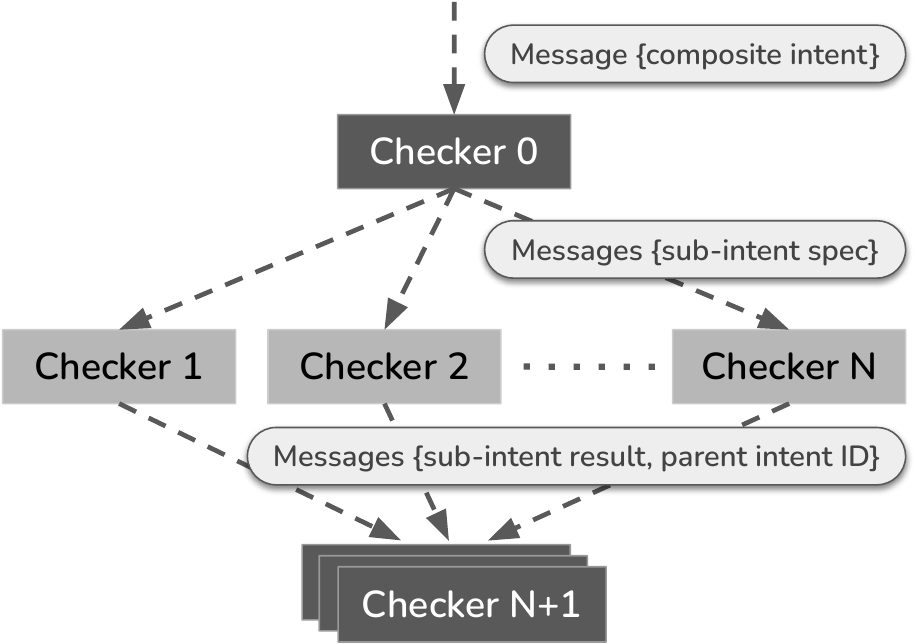}
    \vspace{-0.5em}
    \caption{Example workflow of a composite intent.}
    \label{fig:composite-intent}
    \vspace{-1em}
\end{figure}

Figure~\ref{fig:composite-intent} shows a typical workflow for composite intents, where the intent checkers operate in three different modes: disaggregation, traversal, and aggregation. The disaggregation function (Checker 0) divides the computation into $N$ sub-intents based on the intent specification. In our prototype, we divide consistency intents into sub-intents for each starting pair of rule tables and packet sets, distributed across the cluster (\S\ref{subsec:intent-colocation}). The traversal function (Checker 1 to $N$) verifies these sub-intents and maintains their slices (\S\ref{subsec:intent-based-slicing}). Finally, the aggregation function (Checker $N+1$) combines the sub-intent results to determine the outcome of the original composite intent. While flow consistency is one example, other intents can also be expressed as composite intents. For example, reachability from set $A$ to set $B$ can be expressed as reachability sub-intents from subsets of $A$ to subsets of $B$.

\vspace{-0.5em}
\subsubsection{Distributed Loop Detection}
\label{subsubsec:distributed-loop}

Some intents may not be easily divided into sub-intents with small slices. To the best of our knowledge, loop freedom is the only example in this category. Libra~\cite{2014-ZengEtAl} attempts to divide the intent based on IP prefixes, which is ineffective in practice due to tunneling and non-IP rules that are prevalent in real networks. Instead, we adapt a distributed cycle detection algorithm~\cite{2015-RochaThatte} and extend it with topological slicing. While loop detection does not neatly divide into \emph{independent sub-intents} as in \S\ref{subsubsec:composite-intent}, we can still divide it into \emph{partially independent sub-intents} that require some communication between checkers over several phases. Here is how it works.

\vspace{-1em}
\para{Bounded Traversal and Path Hand-off}
Loop freedom is handled differently in \name. The number of checkers used for loop detection is configurable by users and set before verification. \name then divides the network into segments, each managed by a checker. During loop detection, each checker traverses its own network segment from all possible starting points (\S\ref{subsec:traversal}). However, when a path reaches the segment's boundary, i.e., when the next $(R_i,P_i)$ pair resides in another checker's segment, the path is marked as \emph{external} and sent to the checker responsible for the next segment, where the traversal continues from $(R_i,P_i)$. This process repeats until no external paths are emitted. While paths are shared between checkers, the actual slice built within each checker remains a subset of the assigned segment.

\vspace{-1em}
\para{Dividing Schemes}
How a network is segmented is critical to the performance of distributed loop detection. However, there is no universally optimal dividing scheme. The performance depends on various factors like bandwidth, latency between checkers, disk speed, and the desired optimization objective (e.g., minimizing memory or total run time).
Typically, the loop detection algorithm performs better with balanced segments due to even workload distribution, but it can also benefit from fewer inter-segment links, which reduces communication overhead.
Currently, \name supports two dividing heuristics: \emph{random partitioning}, which randomly assigns devices to segments, and \emph{sparsest cut}, which minimizes the ratio of inter-segment links to the product of all segment sizes.
By minimizing this ratio, the sparsest cut scheme attempts to reduce the communication between checkers, while maintaining relatively balanced segments. Since the sparsest cut problem is a well-known NP-hard problem~\cite{2004-AroraEtAl, 2006-ChawlaEtAl, 2009-AroraEtAl, 2016-Chawla, 2021-JavadiAshkboos}, we adapt a greedy algorithm for a local approximation instead of seeking a global optimum.
% Note: See GraphPartition.cpp for more details of sparsest cut implementation.

For distributed loop detection, we divide the network into segments at the device level. The reason is two-fold. (1) Since the slice of the loop intent covers most of the network, dividing at the rule table level would have the same DPA loaded by multiple checkers, incurring unnecessary I/O overhead. (2) Since a device is a natural cluster of highly related rule tables, dividing at the rule table level can also lead to more frequent communication between segments.

% It is worth noting that the distributed algorithm described above is not limited to loop checks, but general to other types of intents. However, we observed that, in practice, all other types of intents are either composite, or narrow enough where the benefit of parallelism does not outweigh the message passing overhead.

\subsection{Intent Colocation}
\label{subsec:intent-colocation}

\begin{table}
    \centering
    \scalebox{0.9}{
        \begin{tabular}{*2l}
            \toprule
            Schemes   & Description                               \\
            \midrule
            Hashing   & Hashing based on intent specifications.   \\
            K-medoids & K-medoids clustering with distance matrix \\
                      & defined on intent specifications.         \\
            Dynamic   & K-medoids clustering with distance matrix \\
                      & defined on previous intent slices.        \\
            \bottomrule
        \end{tabular}
    }
    \vspace{-0.5em}
    \caption{Intent colocation schemes.}
    \label{tab:dist-schemes}
    \vspace{-1.5em}
\end{table}

% Distributing intents with message routing based on hash digests.

As described in \S\ref{subsec:architecture}, \name is a message-passing compute cluster where message routing is determined by consistent hashing~\cite{1997-KargerEtAl, 2017-NoghabiEtAl}. While the hash value of a message is typically unpredictable, we sometimes need specific messages delivered to and processed on certain nodes, such as for aggregating sub-intent results (\S\ref{subsubsec:composite-intent}) or colocating similar intents for better resource utilization. In these cases, we override the hash function to ensure a deterministic outcome.

% Motivation of colocating intents.
In fact, with thousands of intents, it is inevitable that intent slices overlap with each other. For example, we may have several intents checking reachability from different VMs to a common subnet, or have intents configured for the virtualized networks sharing an underlay.
To avoid duplicating models, it is ideal to use the same slice for similar intents.
% Relaxed definitions for "slices" and "universes".
Previously in \S\ref{subsec:intent-based-slicing}, we defined a slice as the set of flow nodes traversed by an intent. Here we relax the definition to be the set of flow nodes traversed by all intents located in the same node.\footnote{The term ``node'' here refers to a compute node in the cluster.} Similarly, the universe $U_i$ of a rule table $R_i$ is now the union of all universes at that rule table for all intents within the same node. The model of each $(R_i,U_i)$ pair is then kept in memory for as long as at least one intent requires it.

% How to colocate intents with the colocation schemes.

Therefore, \name aims to colocate intents with similar slices on the same node to maximize sharing and minimize model duplication across nodes. To this end, the \emph{Orchestrator} offers several colocation schemes (Table~\ref{tab:dist-schemes}) and computes the hash values for outgoing messages accordingly. Static schemes like hashing and k-medoids create a fixed assignment of intents when the intents remain unchanged, while the dynamic scheme adjusts the intent placement based on current slices. Though current prototype does not yet consider loop freedom for intent colocation, as it uses a different distribution method, it is possible to distribute loop checks along with other intents, which we leave as future work.

\vspace{-0.5em}
\subsubsection{Static Distribution}
\label{subsubsec:static-distribution}

Static schemes are useful for the initial distribution when \name has only the information of intent specifications. Hashing computes digests based on the specification parameters, distributing most intents randomly while assigning the intents with identical parameters to the same checker. This scheme reduces model duplication only for the intents with identical parameters. K-medoids uses the PAM (Partitioning Around Medoids) algorithm~\cite{1990-KaufmanRousseeuw} to cluster similar intents to the same checker by measuring distances between every pair of intents based on the proximity of the endpoints and packet sets in their specifications. This distance suggests how different the two intent slices are if placed in the same checker. Nonetheless, the static schemes are heuristic and may not optimally capture slice dissimilarity from specifications alone.

\vspace{-0.5em}
\subsubsection{Dynamic Distribution}
\label{subsubsec:dynamic-distribution}

\begin{table*}
    \centering
    \scalebox{0.8}{
        \begin{threeparttable}
            \begin{tabular}{*2l *3c p{0.4\linewidth} p{0.23\linewidth}}
                \toprule
                \textbf{Name}                                     & \textbf{Model slicing method} & \textbf{HS? $^1$} & \textbf{TS? $^1$} & \textbf{Distributed?} &
                \textbf{Rebuild the model and recheck intents...} &
                \textbf{Example prior work}                                                                                                                         \\
                \midrule
                \midrule
                \name                                             & Intent-based slicing          & \checkmark        & \checkmark        & \checkmark            &
                Only for the intents whose slice is modified      &
                \name (this work)                                                                                                                                   \\
                \midrule
                Libra                                             & Subnet-based slicing          & \checkmark        & $\times$          & \checkmark            &
                For all updates, but only on the affected subnets &
                \cite{2014-ZengEtAl, 2013-KazemianEtAl}                                                                                                             \\
                \midrule
                Mono+                                             & EC-based  slicing             & \checkmark        & $\times$          & $\times$              &
                For all updates, but only on the affected ECs     &
                \cite{2013-KhurshidEtAl}                                                                                                                            \\
                \midrule
                Mono                                              & No slicing                    & $\times$          & $\times$          & $\times$              &
                For all updates                                   &
                \cite{2005-XieEtAl, 2011-MaiEtAl, 2012-KazemianEtAl, 2013-YangLam, 2017-HornEtAl, 2017-PandaEtAl, 2020-ZhangEtAl, 2020-YuanEtAl}                    \\
                % Xie et al., Anteater, HSA, AP verifier, Delta-net, APKeep, VMN, NetSMC
                \bottomrule
            \end{tabular}
            \begin{tablenotes}
                \item[1] HS: Whether the model slicing happens in the header space.
                TS: Whether the model slicing happens in the topological space.
            \end{tablenotes}
        \end{threeparttable}
    }
    \vspace{-0.5em}
    \caption{Different model slicing methods implemented for evaluation.}
    \label{tab:taxonomy}
    \vspace{-1em}
\end{table*}

After the initial check, the dynamic scheme can be applied. It also uses the PAM algorithm, but now measures distance between intents by the number of different relevant rules along their traversal paths, representing the actual difference between slices. However, intents might traverse the same devices without sharing rules when they utilize different rules or rule tables in the device. To minimize overhead, we first filter out \emph{rule solitary intents} --- those sharing no rules with other intents --- and then cluster the remaining intents using rule-level PAM as described above, before merging the rule solitary intents back into the clusters based on the distances defined as the number of different devices in their slices.

%%%%%%%%%%%%%%%%%%%%%%%%%%%%%%%%%%%%%%%%%%%%%%%%%%%%%%%%%%%%%%%%%%%%%%%%%%%%%%%%

\section{Implementation}
\label{sec:implementation}

We built a \name prototype in C++ and Java, alongside three other data plane verification methods for comparison (Table~\ref{tab:taxonomy}). All methods use Apache Samza~\cite{2017-NoghabiEtAl} for task distribution, FoundationDB~\cite{2021-ZhouEtAl} for intent storage, and HDFS~\cite{2010-ShvachkoEtAl} for DPA storage. These methods differ in (1) how the model is partitioned, (2) whether the model is distributed, and (3) how the model and intents are maintained and updated. We purposely implemented these methods in the same codebase to observe the impact caused only by the three aspects, avoiding performance skew from other factors, such as choices of data structures, modeling granularity, libraries, or distribution frameworks. Additionally, there is no public release of Libra, and the public releases of other works do not support the vendor devices in our test networks (\S\ref{subsec:setup}).

Specifically, the \emph{Libra} method~\cite{2013-KazemianEtAl, 2014-ZengEtAl} builds and distributes slices based on IP subnets before verification begins, and hence could not support header rewrites, as the part of the model needed after header modification may reside in a different slice~\cite{2014-ZengEtAl}. Therefore, we limit the comparison with this method to only the IP-based underlay networks.

\emph{Mono+} builds slices based on ECs but still keeps the entire model in the memory of a single machine. When a network update occurs, it updates the EC slices and rechecks the affected intents. To our knowledge, the only prior work in this category is VeriFlow~\cite{2013-KhurshidEtAl}, which supports non-IP rules but still cannot handle header rewrites. Our \emph{Mono+} implementation simulates VeriFlow's slicing and modeling, but avoids precomputing ECs in order to support header rewrites.

\emph{Mono} uses a monolithic model without slicing. A full network model is built and maintained in memory and all intents are checked against the model upon receiving a network update. Most prior works fall into this category, including AP Verifier~\cite{2013-YangLam}, Delta-net~\cite{2017-HornEtAl}, and APKeep~\cite{2020-ZhangEtAl}.

%%%%%%%%%%%%%%%%%%%%%%%%%%%%%%%%%%%%%%%%%%%%%%%%%%%%%%%%%%%%%%%%%%%%%%%%%%%%%%%%

\section{Evaluation}
\label{sec:evaluation}

\subsection{Experimental Setup}
\label{subsec:setup}

\para{Compute Environment}
We evaluate the performance in two different compute environments (outside the network being verified). The first is a virtual compute cluster, where each node is an Ubuntu VM with 16 vCPUs and 32 GB RAM, used to study \name's end-to-end performance with live network data. The second is a standalone server with Intel i7-9750H and 32 GB RAM, used for repeatable performance tests with pre-collected network data to precisely compare different model slicing methods.

\begin{table}
    \centering
    \scalebox{0.78}{
        \begin{threeparttable}
            \begin{tabular}{l *3r}
                \toprule
                Network       & Devices & Subsequent updates & Total rules $^1$ \\
                \midrule
                N1            & 75      & 2                  & 1,905            \\ % nsxt
                N2            & 354     & 347                & 63,412           \\ % wwtatc
                N3            & 743     & 4,897              & 100,053          \\ % field-demo
                N4            & 3,820   & 12,716             & 5,503,291        \\ % scale
                N5            & 7,767   & 6,188              & 3,527,730        \\ % vmware-it
                N1 (underlay) & 14      & 2                  & 108              \\ % nsxt-underlay
                N2 (underlay) & 103     & 114                & 11,142           \\ % wwtatc-underlay
                N3 (underlay) & 173     & 338                & 14,561           \\ % field-demo-underlay
                N4 (underlay) & 2,360   & 4,594              & 5,452,362        \\ % scale-underlay
                N5 (underlay) & 3,324   & 1,214              & 460,786          \\ % vmware-it-underlay
                \bottomrule
            \end{tabular}
            \begin{tablenotes}
                \item[1] Total number of rules of the initial data plane snapshot.
            \end{tablenotes}
        \end{threeparttable}
    }
    \vspace{-0.5em}
    \caption{Network datasets for evaluation.}
    \label{tab:dataset}
    \vspace{-1em}
\end{table}

\vspace{-1em}
\para{Dataset}
Our dataset comprises 5 different networks (Table~\ref{tab:dataset}). N1 is a small network with virtualized devices for testing a commercial network management suite~\cite{vmware-aria-blinded}. N2 is a lab network with real and emulated devices for developing the same suite's modeling features. N3 is a real network for demonstrating a commercial network verifier. N4 is a large synthetic network designed to test the scalability of the verifier. N5 is a production IT network of a large enterprise with 38K employees. Each network features physical and virtual devices, including distributed firewalls that span across physical devices. We collected network snapshots and subsequent network updates over 1--2 weeks, replaying them as input for evaluation. For Libra experiments, we removed overlay devices and use only the underlay due to its inability to handle header rewrites. While there exist larger networks in practice, testing all of which can be impractical, our dataset effectively captures general performance trends.

\vspace{-1em}
\para{Intents}
For end-to-end experiments, we randomly sampled 3,000 intents from all source-destination pairs for reachability, segmentation, and flow consistency. For microbenchmarks and comparison experiments, we sampled 500 intents per network similarly for reachability, segmentation, black hold freedom, and waypoint for the multi-intent cases, and one for each intent type from the 500 intents for the single-intent cases. As a result, in most of our experiments with multiple intents (\S\ref{subsec:e2e-cluster}-\ref{subsec:libra-compare}), we mixed different intent types to reflect real-world scenarios where various intent types are employed together in practice. In \S\ref{subsec:wide-intents}-\S\ref{subsec:loop-eval}, we focused on evaluating the intent types that involve larger slices.

\subsection{End-to-end Performance on Clusters}
\label{subsec:e2e-cluster}

\begin{figure}
    \centering
    \includegraphics[width=0.55\linewidth]{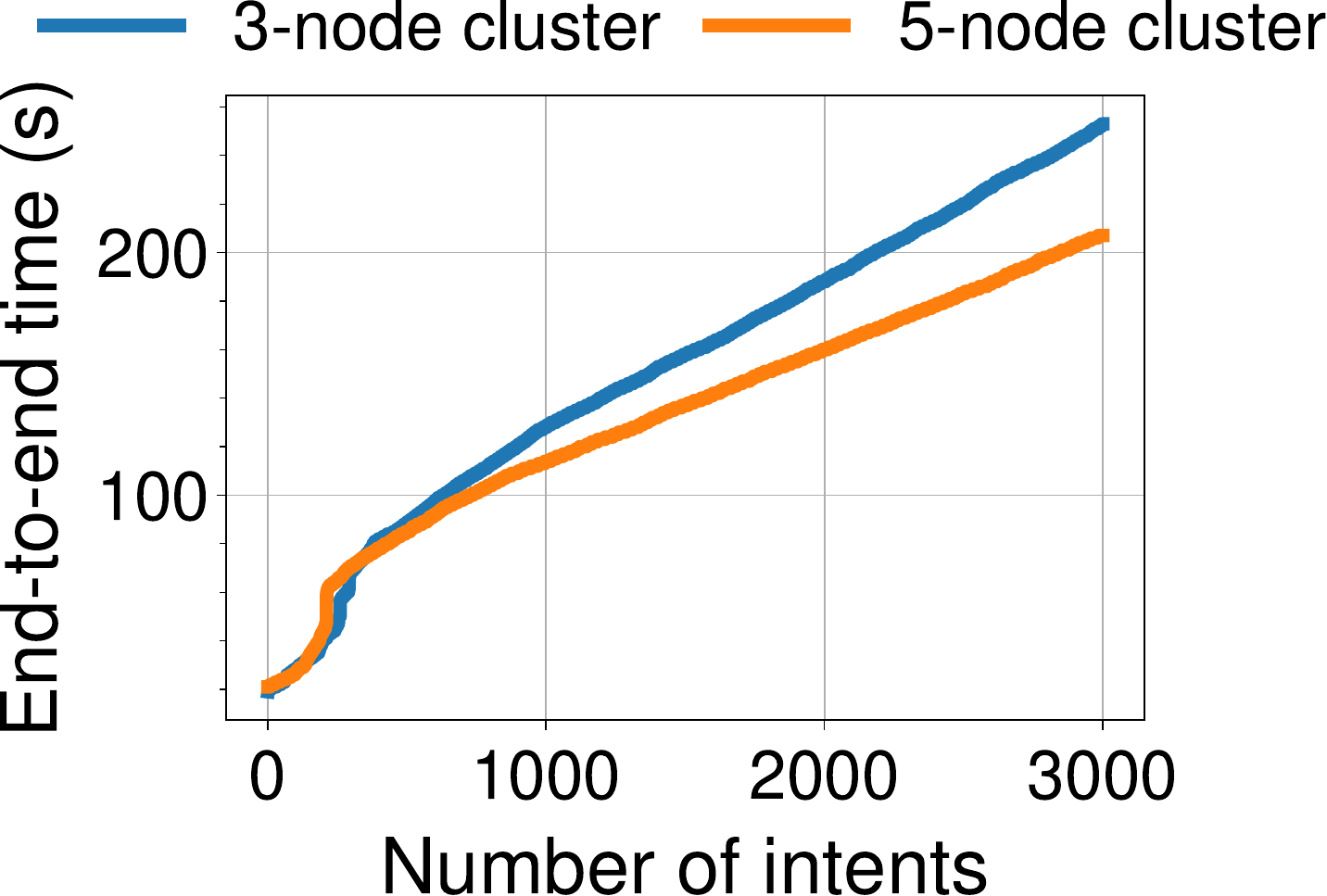}
    \vspace{-0.5em}
    \caption{End-to-end processing time on live clusters.}
    \label{fig:e2e-time}
    \vspace{-1.5em}
\end{figure}

To evaluate the end-to-end performance, we deployed \name on both 3-node and 5-node clusters connected to the live N2 network. Figure~\ref{fig:e2e-time} shows the time from when the \emph{Preprocessing} function finishes to when all intents are verified. The 3-node cluster processed 3,000 intents in 4 minutes and 13 seconds (84 ms/intent), while the 5-node cluster took 3 minutes and 27 seconds (69 ms/intent), both including the network latency overheads. In addition, the total memory usage increased by only 2.5 GiB for 3,000 intents (0.85 MiB/intent) in both clusters, demonstrating that \name can maintain tens of thousands of intents across inexpensive machines efficiently, since intents within the same node usually share much of their intent slices. We note that \name is the first to demonstrate such scale in terms of the number of verified intents.

\subsection{Microbenchmarks}
\label{subsec:microbenchmark}

\subsubsection{Single Intent}
\label{subsubsec:single-intent}

\begin{figure*}
    \centering
    \begin{minipage}[b]{0.49\linewidth}
        \centering
        \begin{subfigure}[t]{0.48\linewidth}
            \includegraphics[width=\linewidth]{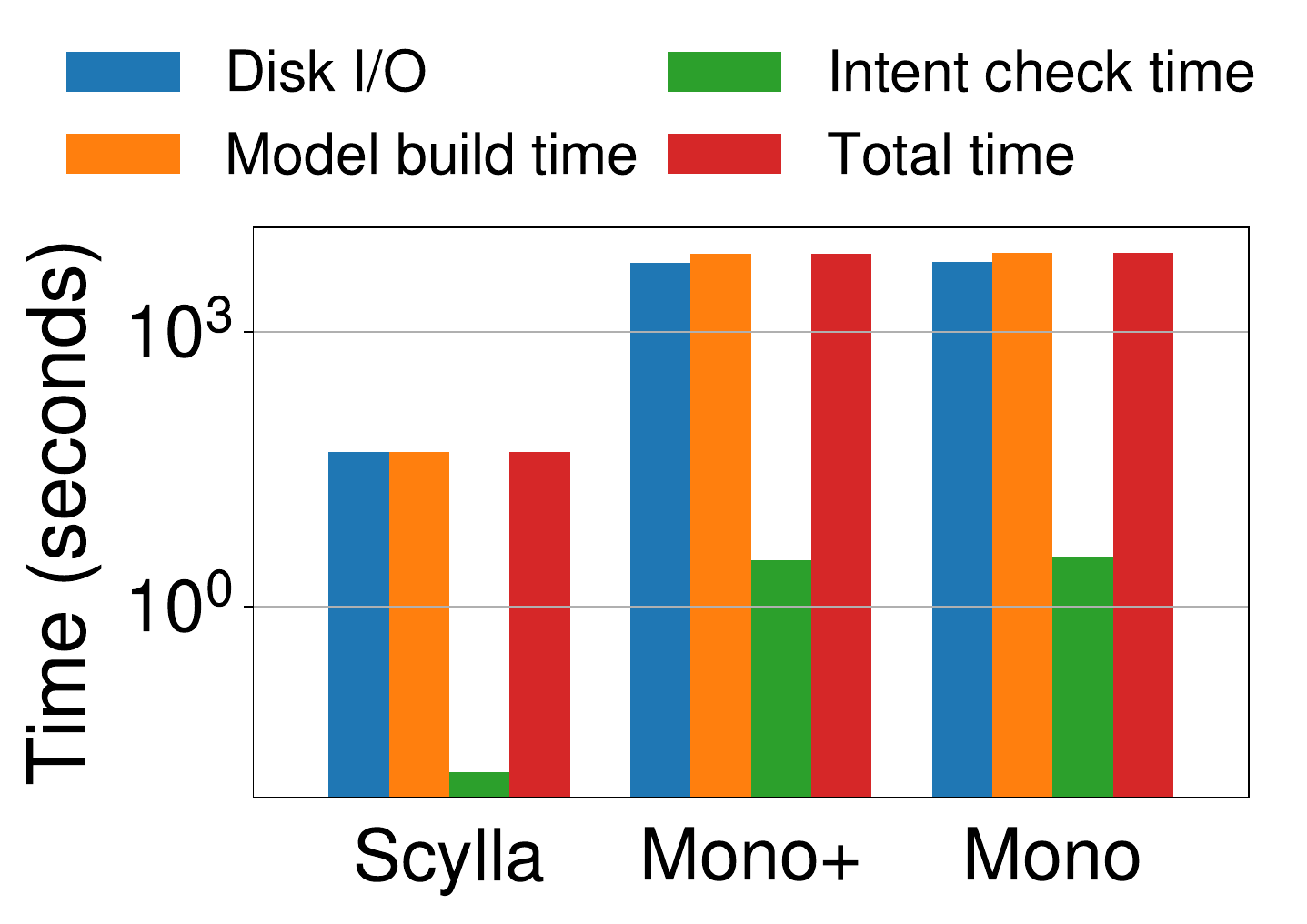}
            \caption{New intent from scratch (N5).}
            \label{fig:single-intent.time.first-run.N5}
        \end{subfigure}
        \hfill
        \begin{subfigure}[t]{0.48\linewidth}
            \includegraphics[width=\linewidth]{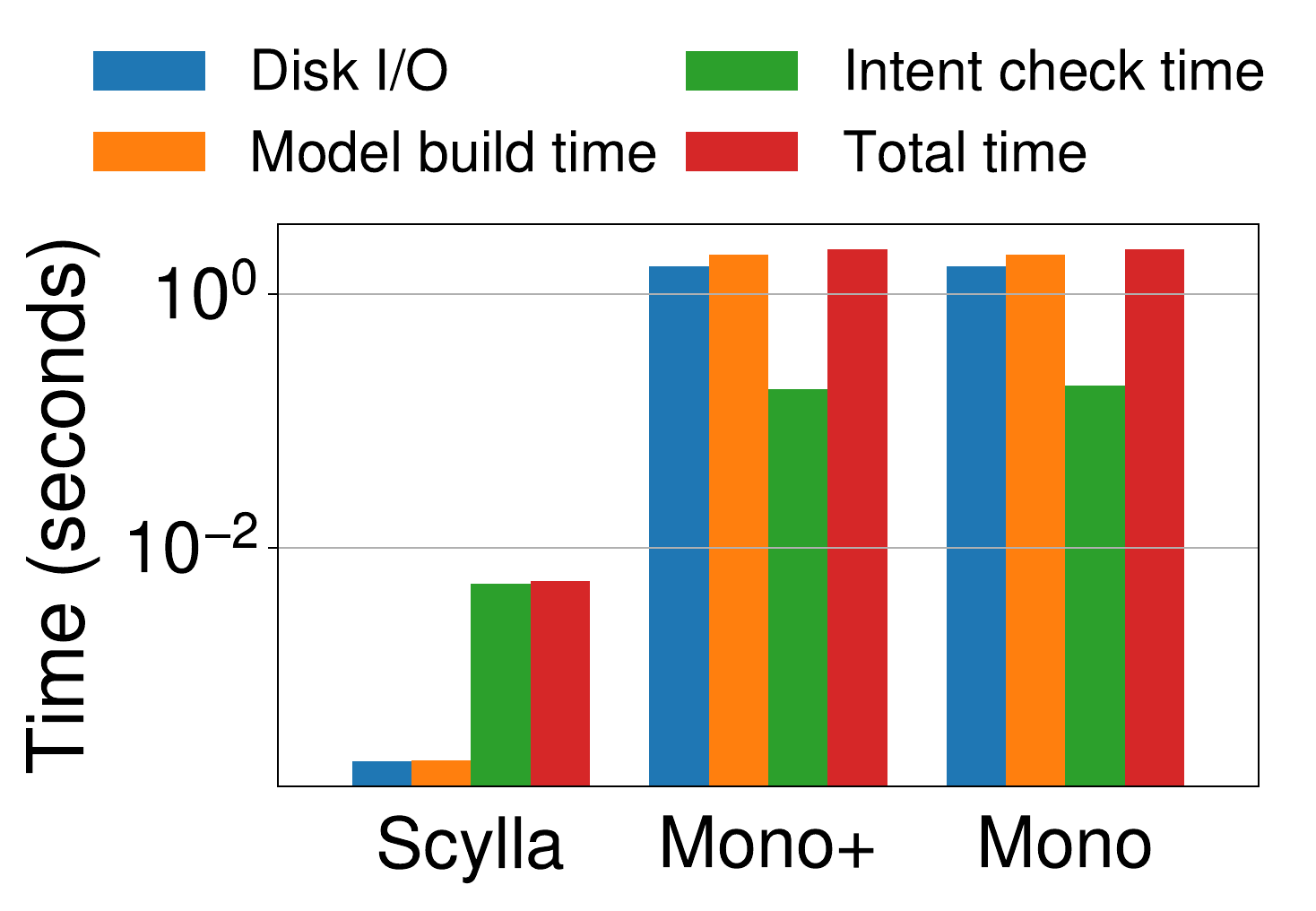}
            \caption{Maintain existing intent (N5).}
            \label{fig:single-intent.time.update.N5}
        \end{subfigure}
        \vspace{-0.5em}
        \caption{Time to verify a single intent (log scale).}
        \label{fig:single-node.time}
    \end{minipage}
    \hfill
    \begin{minipage}[b]{0.49\linewidth}
        \begin{subfigure}[t]{0.48\linewidth}
            \includegraphics[width=\linewidth]{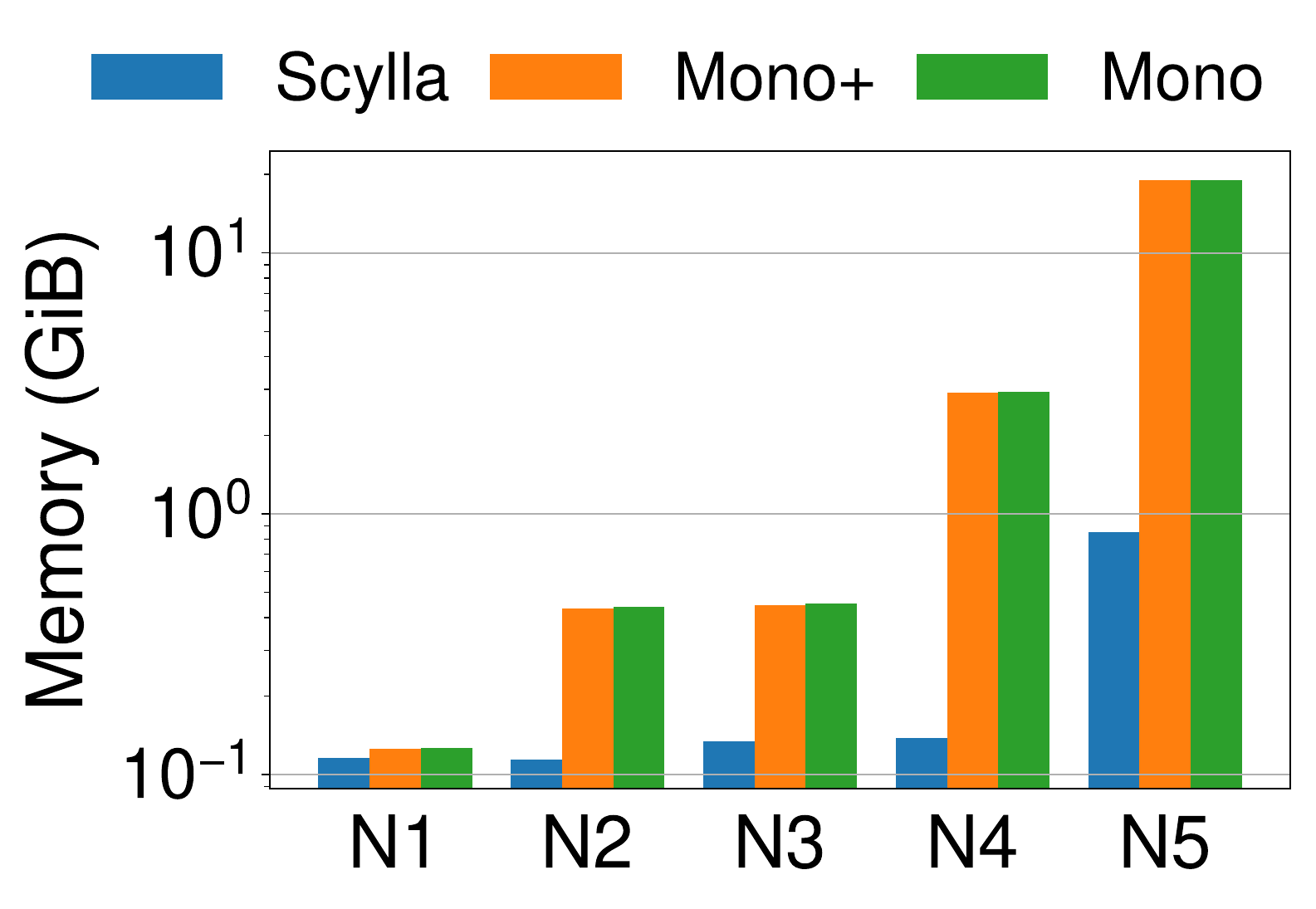}
            \caption{New intent from scratch.}
            \label{fig:single-intent.memory.first-run}
        \end{subfigure}
        \hfill
        \begin{subfigure}[t]{0.48\linewidth}
            \includegraphics[width=\linewidth]{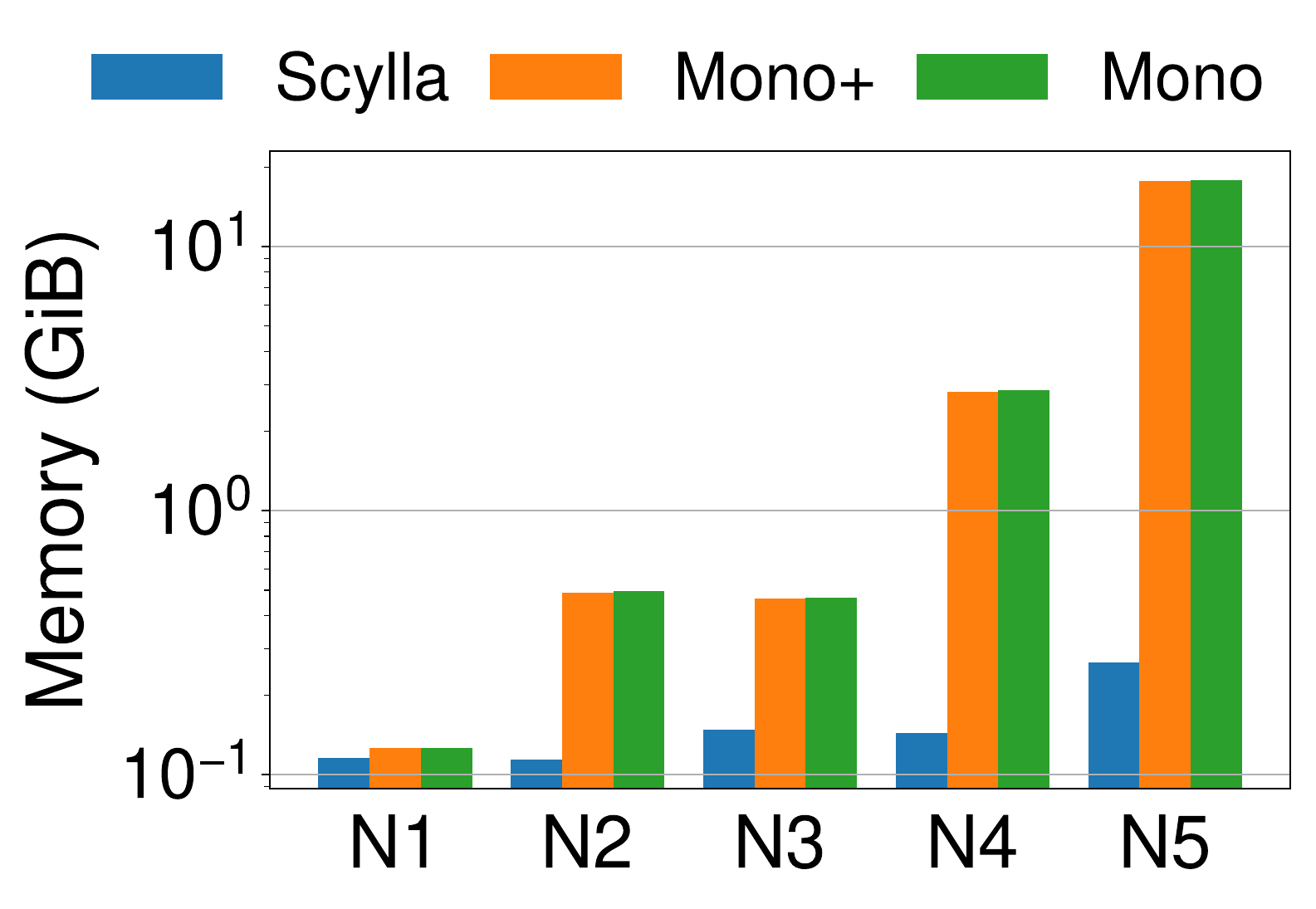}
            \caption{Maintain existing intent.}
            \label{fig:single-intent.memory.update}
        \end{subfigure}
        \vspace{-0.5em}
        \caption{Memory usage for a single intent (log scale).}
        \label{fig:single-node.memory}
    \end{minipage}%
\end{figure*}

To obtain detailed microbenchmarks for verifying a single intent, we replay the network updates on the standalone server and measure the following metrics: (1) \emph{Total time}: time between receiving a network update and having verified the intent, (2) \emph{Intent check time}: $\emph{Total time} - \emph{Model build time}$, i.e., the actual verification time, (3) \emph{Model build time}: time to load DPAs and build model slices, which includes \emph{Disk I/O}, and (4) \emph{Disk I/O}: time spent on file system access.

Figure~\ref{fig:single-node.time}, \ref{fig:single-node.memory} show the time and memory required to verify one intent on a single machine. Due to space constraints, here we show only the results of the largest network in our dataset. Other networks exhibit similar trends (\S\ref{app:subsec:single-intent}). Figure~\ref{fig:single-node.time} shows that model build time, dominated by disk I/O, is the primary bottleneck, and that the actual model traversal for intent checking is comparatively fast, which is consistent with the results from previous work~\cite{2014-ZengEtAl}. This further emphasizes the need to minimize model building and disk I/O for verification, which \name achieves through fine-grained, intent-based slicing.

For larger networks, \name speeds up single-intent verification by 2--3 orders of magnitude ($1362 \times$ in N4 and $149 \times$ in N5), while using only 4.69\% (N4) and 4.49\% (N5) of the memory compared to monolithic methods. The performance gain becomes greater as the network size increases. In addition, since \name updates only the affected slices --- a small part of the model --- most updates were soundly ignored, reducing model construction and I/O overhead, making \name much faster at verifying existing intents.

% For verifying a new intent, \emph{Mono+} performs similarly to \emph{Mono}, as expected, since they both need to build a full network model before checking intents. However, \emph{Mono+} performs slightly better at the intent check time for existing intents, because it only rechecks the intent for the updated ECs. However, for large networks, the overhead from model build time and disk I/O still masks the improvement of intent check time.

These tests verified a \emph{single intent}, useful for microbenchmarks and some use cases (e.g., interactive verification mentioned in \S\ref{sec:motivation}). They also highlight a key contribution of \name: \emph{The minimum resource required to verify an intent is tiny and largely decoupled with the network size}. While larger networks may result in bigger intent slices, the slices are constrained by the path length of intent traversals. With \name, the resource consumption grows with network size at a rate much flatter than previous methods. This fine-grained slicing is key to scaling out and workload distribution (\S\ref{subsec:eval-colocation}).

\vspace{-0.5em}
\subsubsection{Multiple Intents}
\label{subsubsec:multi-intent}

\begin{figure*}
    \centering
    \begin{minipage}[b]{0.49\linewidth}
        \centering
        \begin{subfigure}[t]{0.48\linewidth}
            \includegraphics[width=\linewidth]{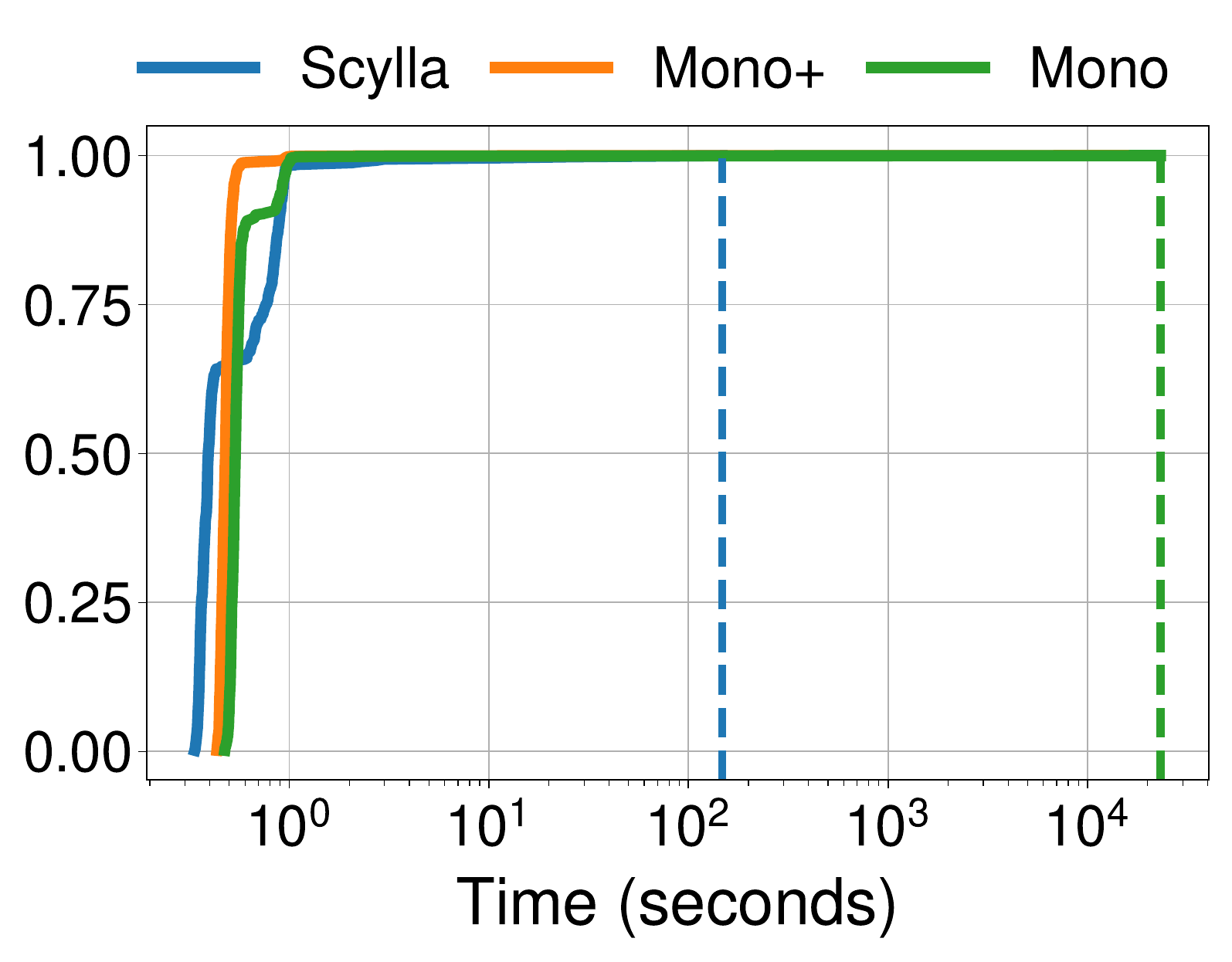}
            \caption{New intents from scratch (N5).}
            \label{fig:multi-intent.time.first-run.N5}
        \end{subfigure}
        \hfill
        \begin{subfigure}[t]{0.48\linewidth}
            \includegraphics[width=\linewidth]{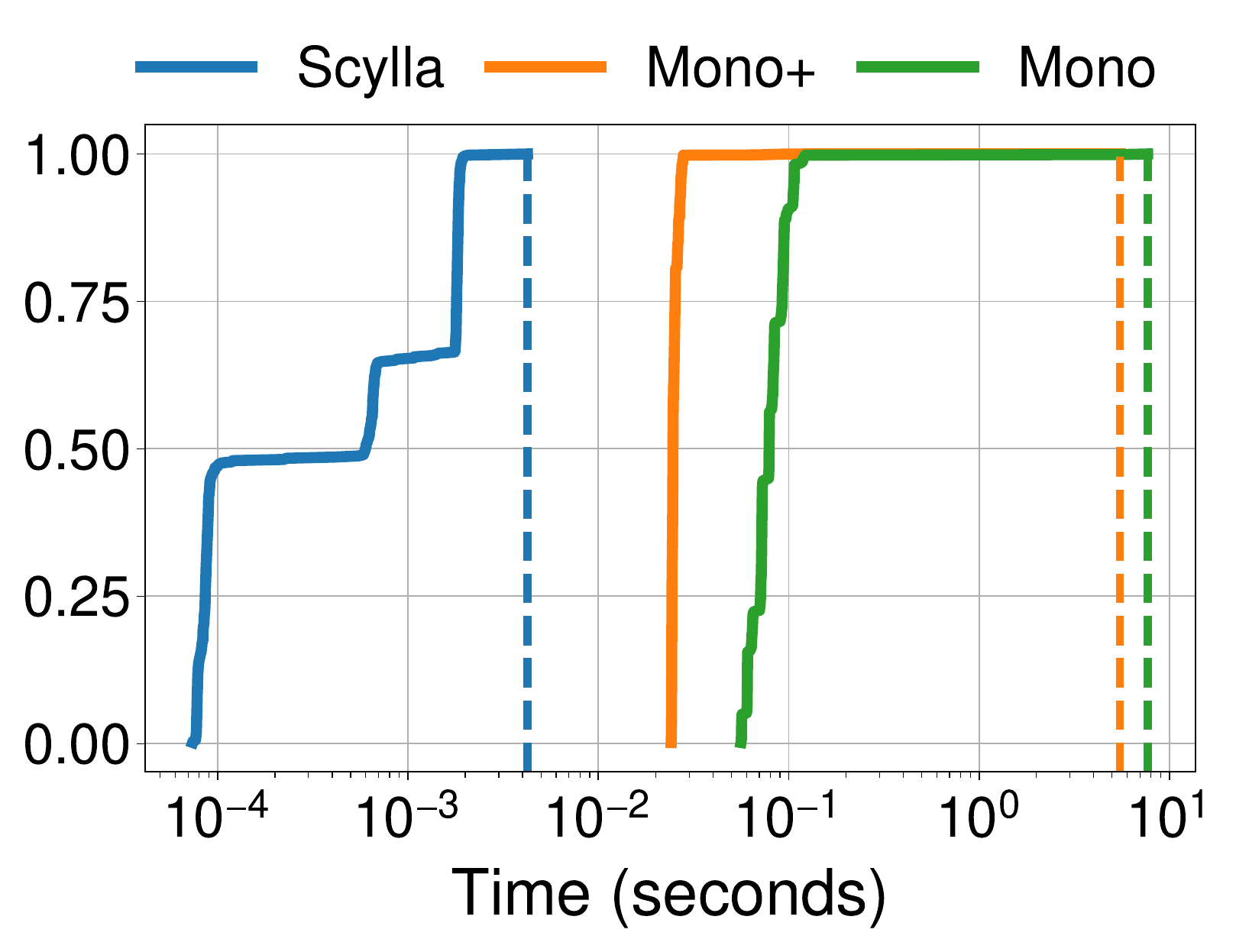}
            \caption{Maintain existing intents (N5).}
            \label{fig:multi-intent.time.update.N5}
        \end{subfigure}
        \vspace{-0.5em}
        \caption{Time to verify 500 intents (CDF, log scale).}
        \label{fig:multi-intent.time}
    \end{minipage}
    \hfill
    \begin{minipage}[b]{0.49\linewidth}
        \begin{subfigure}[t]{0.48\linewidth}
            \includegraphics[width=\linewidth]{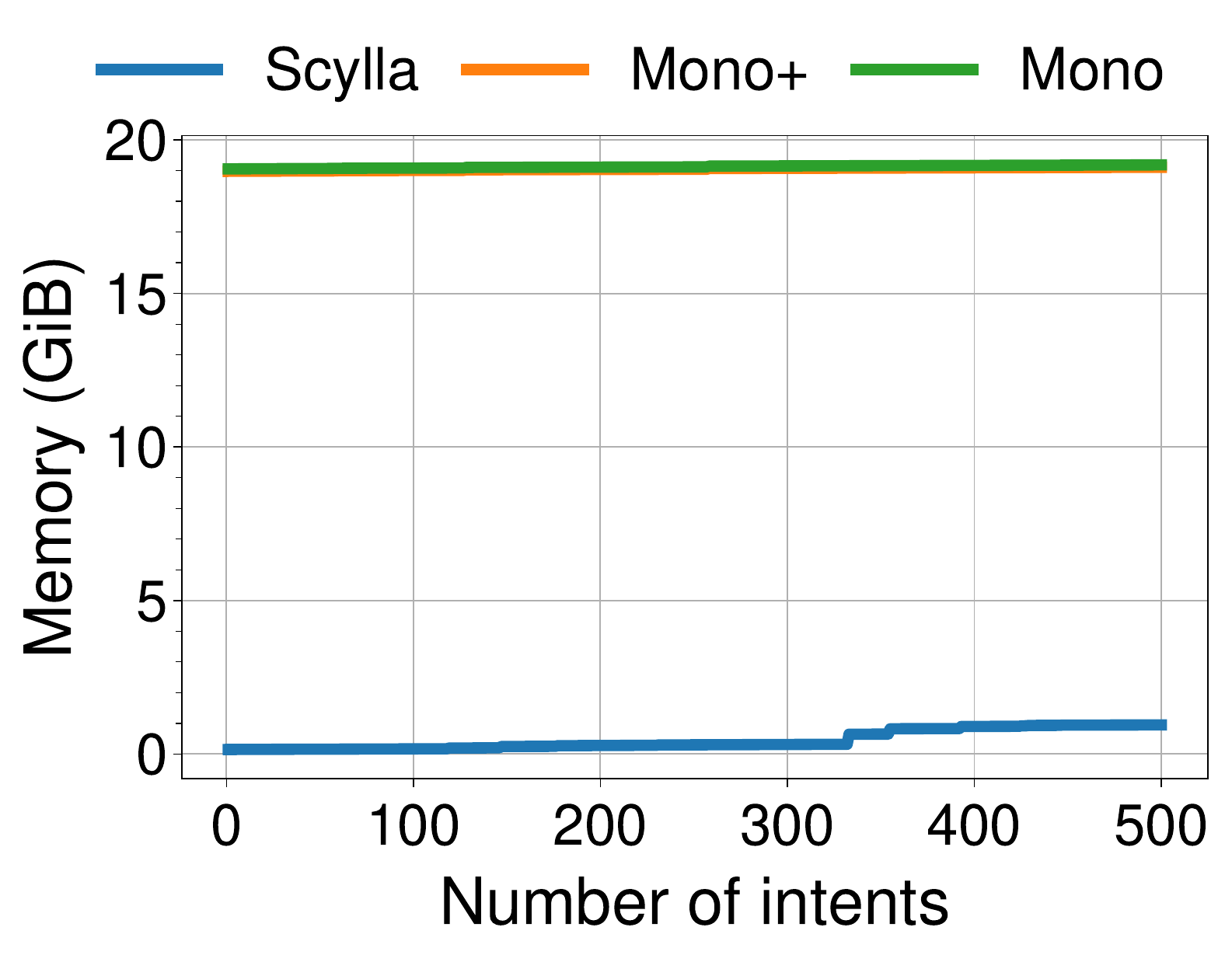}
            \caption{New intents from scratch (N5).}
            \label{fig:multi-intent.memory.first-run.N5}
        \end{subfigure}
        \hfill
        \begin{subfigure}[t]{0.48\linewidth}
            \includegraphics[width=\linewidth]{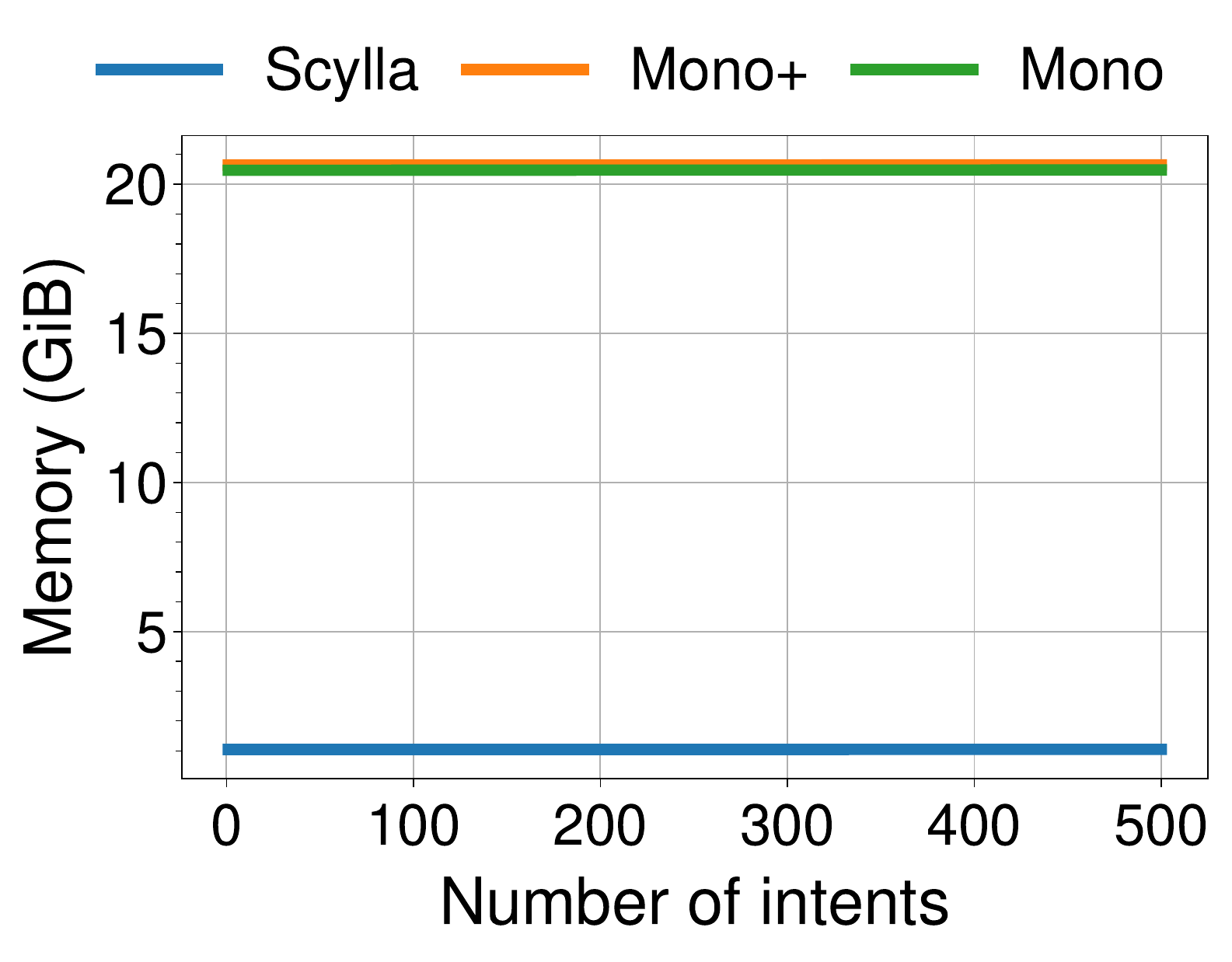}
            \caption{Maintain existing intents (N5).}
            \label{fig:multi-intent.memory.update.N5}
        \end{subfigure}
        \vspace{-0.5em}
        \caption{Memory usage for 500 intents.}
        \label{fig:multi-intent.memory}
    \end{minipage}%
    \vspace{-1em}
\end{figure*}

Verifying multiple intents aligns closer with the intended use case of previous work, as it allows amortizing the cost of building the full model. Here we demonstrate the performance given the amortization with multiple intents. This is similar to what happens within a single node of a live cluster. Figure~\ref{fig:multi-intent.time}, \ref{fig:multi-intent.memory} show the time and memory needed to create and maintain 500 intents across network updates for the largest network in our dataset. Other networks exhibit similar trends (\S\ref{app:subsec:multi-intent}).

\begin{table}
    \centering
    \scalebox{0.73}{
        \begin{threeparttable}
            \begin{tabular}{l *4r}
                \toprule
                Network       & Rules/slice in Libra & Rules in \name      & Total rules & $\varphi$ \\
                \midrule
                N1 (underlay) & 92.000               & 0 $^1$              & 108         & 6.82      \\ % nsxt-underlay
                N2 (underlay) & 6,493.375            & 98                  & 11,142      & 4.66      \\ % wwtatc-underlay
                N3 (underlay) & 10,403.125           & 9                   & 14,561      & 5.72      \\ % field-demo-underlay
                N4 (underlay) & 827,048.125          & 19                  & 5,452,362   & 1.21      \\ % scale-underlay
                N5 (underlay) & 152,134.500          & 34                  & 460,786     & 2.64      \\ % vmware-it-underlay
                \bottomrule
            \end{tabular}
            \begin{tablenotes}
                \item[1] \name utilized zero rules in N1 (underlay) because
                    the initial packet set was dropped at the source hop due
                    to the overlay being stripped away.
            \end{tablenotes}
        \end{threeparttable}
    }
    \caption{Number of rules modeled for an intent slice.}
    \label{tab:rule counts}
    \vspace{-1em}
\end{table}

In Figure~\ref{fig:multi-intent.time.first-run.N5}, after adding the first intent, which requires \emph{Mono+} and \emph{Mono} to build the full network model (indicated by the vertical dashed lines), the verification time for the remaining intents are similar across all three methods. This indicates that in \name, most of the necessary slices are built after the first few intents, with many intents sharing slices with each other. The benefit of \name's design are more evident with large networks (\S\ref{app:subsec:multi-intent}). For the largest network in our dataset with 500 intents, \name is $51\times$ faster for new intents and $120\times$ faster for maintaining existing ones (Figure~\ref{fig:multi-intent.time}), while using only 5.1\% of the memory compared to monolithic methods (Figure~\ref{fig:multi-intent.memory}).

\subsection{Comparison with Libra's Slicing Method}
\label{subsec:libra-compare}

\begin{figure}[ht]
    \centering
    \begin{subfigure}[t]{0.48\linewidth}
        \includegraphics[width=\linewidth]{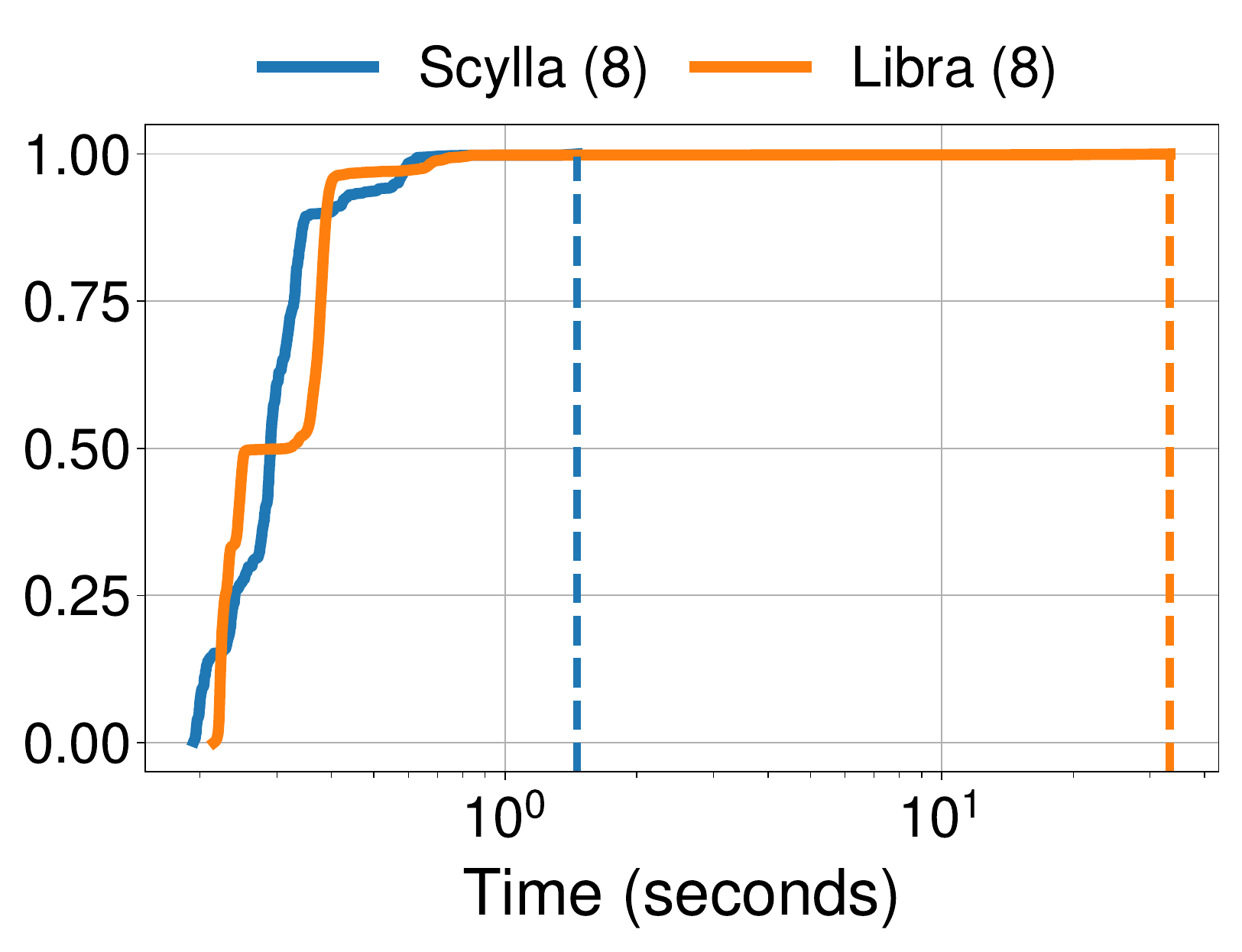}
        \caption{Time CDF (N5 underlay).}
        \label{fig:multi-intent.time.first-run.N5-underlay}
    \end{subfigure}
    \hfill
    \begin{subfigure}[t]{0.48\linewidth}
        \includegraphics[width=\linewidth]{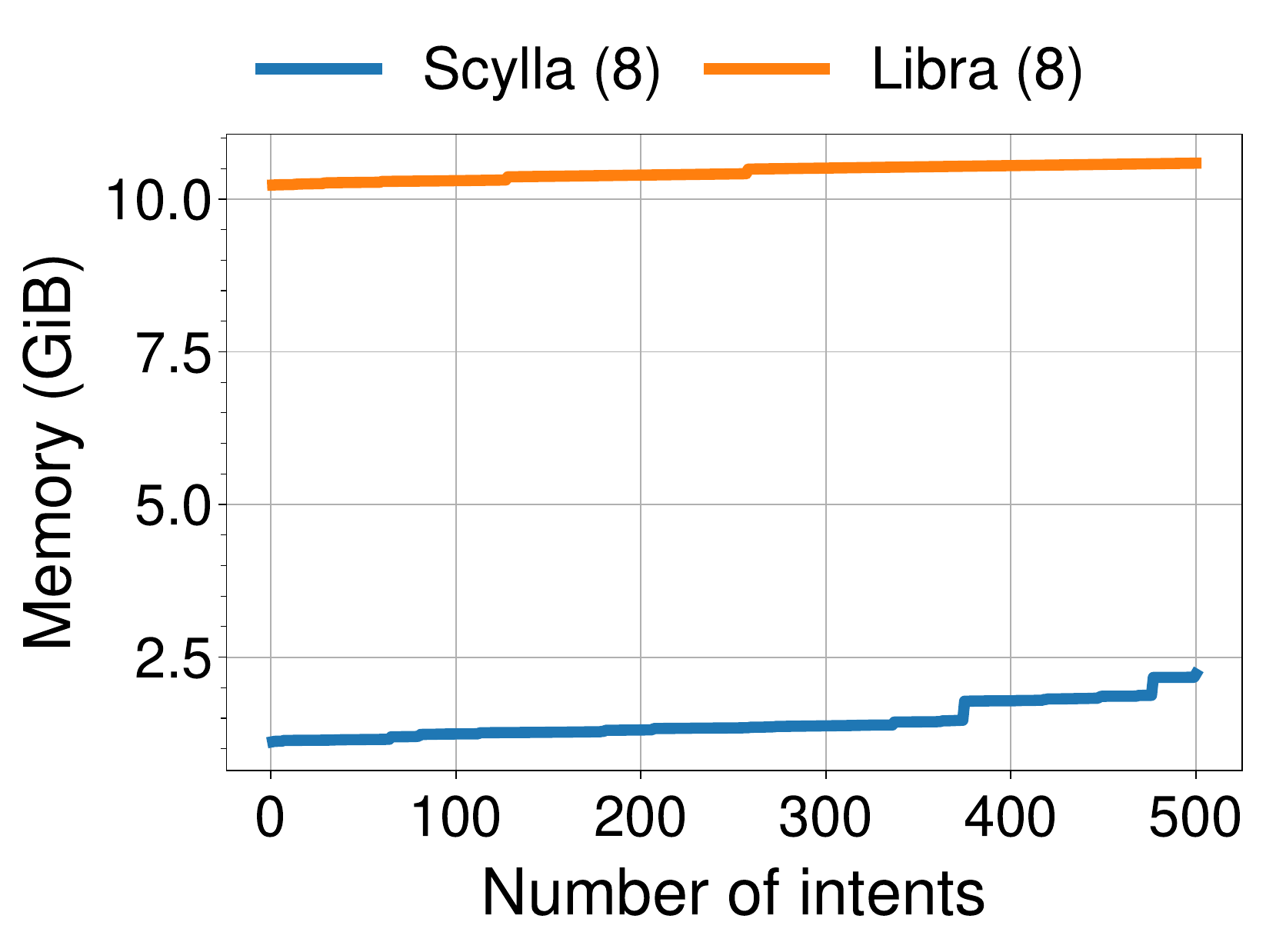}
        \caption{Memory (N5 underlay).}
        \label{fig:multi-intent.memory.first-run.N5-underlay}
    \end{subfigure}
    \vspace{-0.5em}
    \caption{Comparison with Libra (verify 500 new intents).}
    \label{fig:multi-intent.libra.first-run}
    \vspace{-1em}
\end{figure}

% Eliminate the overhead caused by MapReduce.
To factor out the overhead from Libra's use of MapReduce \cite{2004-DeanGhemawat}, which increases verification time as the number of checkers grows~\cite{2014-ZengEtAl}, we run our Libra implementation sequentially for each slice on a single machine and calculate the time and memory that would result from parallel verification without messaging overhead. Thus, the results represent an idealized version of Libra's slicing design without the overhead of distribution framework.
% How we statically partition the header space to simulate Libra.
Specifically, we split the header space into 8 equal-sized blocks based on destination IP subnets. While this may not be the most load-balanced slicing, all performance metrics for Libra (time and memory) are either averaged or summed across slices, so we do not expect significant changes with different slicing.

To quantify slice complexity and size, we calculate the number of rules modeled in memory, shown in Table~\ref{tab:rule counts}. The intent slice in \name (the ``Rules in \name'' column) is significantly smaller than that of Libra, demonstrating the rule duplication issue of Libra described in \S\ref{sec:introduction}, where the same rule may be duplicated in multiple partitions. We define the rule duplication factor $\varphi$ as
$$
\varphi = \frac{(\text{Average \# rules per slice in Libra}) \times \text{\# slices}} {\text{Total \# rules in the network}},
$$
which indicates how often a rule is duplicated across slices. (Less is better; $\varphi = 1$ means no duplication.) Table~\ref{tab:rule counts} shows that Libra's slicing incurs an additional 21\% to 582\% overhead in terms of modeled rules due to rule duplication. In addition, Figure~\ref{fig:multi-intent.libra.first-run} shows that, in the largest network (N5), \name is $23\times$ faster and uses only $1/5$ of the memory compared to Libra. Other networks exhibit similar trends (\S\ref{app:subsec:libra-compare}).

\subsection{Intent Colocation}
\label{subsec:eval-colocation}

There is an inherent conflict between distributing and colocating intents. Distribution aims to balance the load to reduce the verification time but inevitably increases memory usage due to potentially overlapping slices across checkers. Colocation minimizes memory overhead by grouping similar intents but can prolong the verification time due to an imbalanced workload. An extreme case of this is verifying all intents on one checker. As a result, there is no one-size-fits-all solution for how to distribute/colocate intents. A good configuration must balance the time and memory overhead based on the environment and operational priorities.

\begin{table}
    \centering
    \scalebox{0.7}{
        \begin{tabular}{l *6r}
            \toprule
            Network             & Scheme         & Peak time & Peak  & Avg time   & Avg mem. & Rule   \\
                                & (\# nodes)     & (seconds) & mem.  & per intent & per node & reuse  \\
                                &                &           & (GiB) & (seconds)  & (GiB)    & $\psi$ \\
            \midrule
            \multirow{3}{*}{N1} & Hashing (9)    & 1.37      & 0.13  & 0.02       & 0.12     & 3.72   \\
                                & K-medoids (9)  & 2.18      & 0.14  & 0.01       & 0.12     & 4.65   \\
                                & Dynamic (9)    & 3.56      & 0.13  & 0.01       & 0.12     & 5.52   \\
                                % \cmidrule{2-7}
                                % & (1-node \name) & 5.61      & 0.14  & 0.01       & 0.14     & 26.03  \\
            \midrule
            \multirow{3}{*}{N2} & Hashing (9)    & 147.41    & 0.73  & 1.31       & 0.53     & 1.64   \\
                                & K-medoids (9)  & 350.41    & 0.95  & 1.12       & 0.40     & 2.76   \\
                                & Dynamic (9)    & 116.88    & 0.49  & 0.74       & 0.34     & 4.77   \\
                                % \cmidrule{2-7}
                                % & (1-node \name) & 896.11    & 1.03  & 1.79       & 1.03     & 7.48   \\
            \midrule
            \multirow{3}{*}{N3} & Hashing (9)    & 75.07     & 0.57  & 0.58       & 0.45     & 3.04   \\
                                & K-medoids (9)  & 171.72    & 0.79  & 0.48       & 0.32     & 4.81   \\
                                & Dynamic (9)    & 48.99     & 0.52  & 0.24       & 0.32     & 5.40   \\
                                % \cmidrule{2-7}
                                % & (1-node \name) & 397.94    & 0.78  & 0.80       & 0.78     & 16.57  \\
            \midrule
            \multirow{3}{*}{N4} & Hashing (9)    & 11.07     & 0.43  & 0.17       & 0.42     & 1.11   \\
                                & K-medoids (9)  & 33.88     & 0.52  & 0.12       & 0.27     & 1.47   \\
                                & Dynamic (9)    & 40.88     & 0.57  & 0.10       & 0.25     & 1.83   \\
                                % \cmidrule{2-7}
                                % & (1-node \name) & 88.60     & 0.57  & 0.18       & 0.57     & 2.01   \\
            \midrule
            \multirow{3}{*}{N5} & Hashing (9)    & 3,068.27  & 2.36  & 20.87      & 1.47     & 1.01   \\
                                & K-medoids (9)  & 4,225.43  & 2.99  & 21.49      & 1.00     & 1.15   \\
                                & Dynamic (9)    & 2,829.27  & 1.84  & 18.90      & 1.22     & 1.20   \\
                                % \cmidrule{2-7}
                                % & (1-node \name) & 4748.37   & 3.06  & 9.50       & 3.06     & 2.63   \\
            \bottomrule
        \end{tabular}
    }
    \vspace{-0.5em}
    \caption{Performance with intent colocation schemes.}
    \label{tab:dist-scheme-performance}
    % \vspace{-1em}
\end{table}

To test the effect of different colocation schemes (\S\ref{subsec:intent-colocation}), we assigned 500 intents across 9 checkers, each verifying the assigned intents from scratch using a single core.
% In addition, we also verify the 500 intents in one checker as a baseline for extreme colocation.
The results are shown in Table~\ref{tab:dist-scheme-performance}. We define the rule reuse rate $\psi$ as
$$
\psi = \frac{\sum_i (\text{\# times a rule is traversed in } Checker_i)}
            {\sum_i (\text{\# rules in } Checker_i)},
$$
which shows the average number of times a modeled rule is used by the symbolic traversal of any intent. A higher $\psi$ typically indicates better model sharing among intents, reduced rule duplication, and less memory overhead. Here are some key observations:
(1) For resource utilization (last 3 columns), \emph{Dynamic} outperformed \emph{K-medoids}, which outperformed \emph{Hashing}. This trend is general but not always true. For example, higher $\psi$ does not always mean lower average memory because some rules occupy more memory than others due to more complex and elaborate packet sets that need to be encoded.
(2) \emph{Hashing} and \emph{Dynamic} generally have lower peak time and memory, which are linked to how evenly the intents are distributed. A more even distribution usually leads to lower peak time and memory.
% Hashing expectedly produced fair evenness. Rule-based dynamic clustering is often better than static spec-level k-medoids because of the better ``resolution'' in intent dissimilarities derived from rule traces, though it still depends on the intent composition and may not always be the case.
%
% (3) For N5, the 1-node baseline had the best average time without distribution. The reason is N5 has large distributed firewalls, each of which is modeled as a single DPA file, so if two intents traversing the same distributed firewall get assigned to different checkers, both checkers need to fetch the DPA file even though they may only build a tiny part of it into the memory. Therefore, the I/O operations took the majority of the time, and the duplication of I/O time led to the result. Nonetheless, when there are more intents, this would eventually be amortized.

% Summarize the benefits of the colocation schemes.

Consequently, with the colocation schemes that enhances utilization and intent-based slicing, the system can efficiently fit intents within clusters of varying resource constraints. \name can leverage inexpensive hardware to accommodate as many intents as possible within available resources. This is ideal for heterogeneous deployments with no prior knowledge about resource availability, or for non-hyperscale organizations to allocate resources on demand as networks grow.

\subsection{Performance with Wide Intents}
\label{subsec:wide-intents}

\begin{figure}
    \centering
    \begin{subfigure}[t]{0.48\linewidth}
        \includegraphics[width=\linewidth]{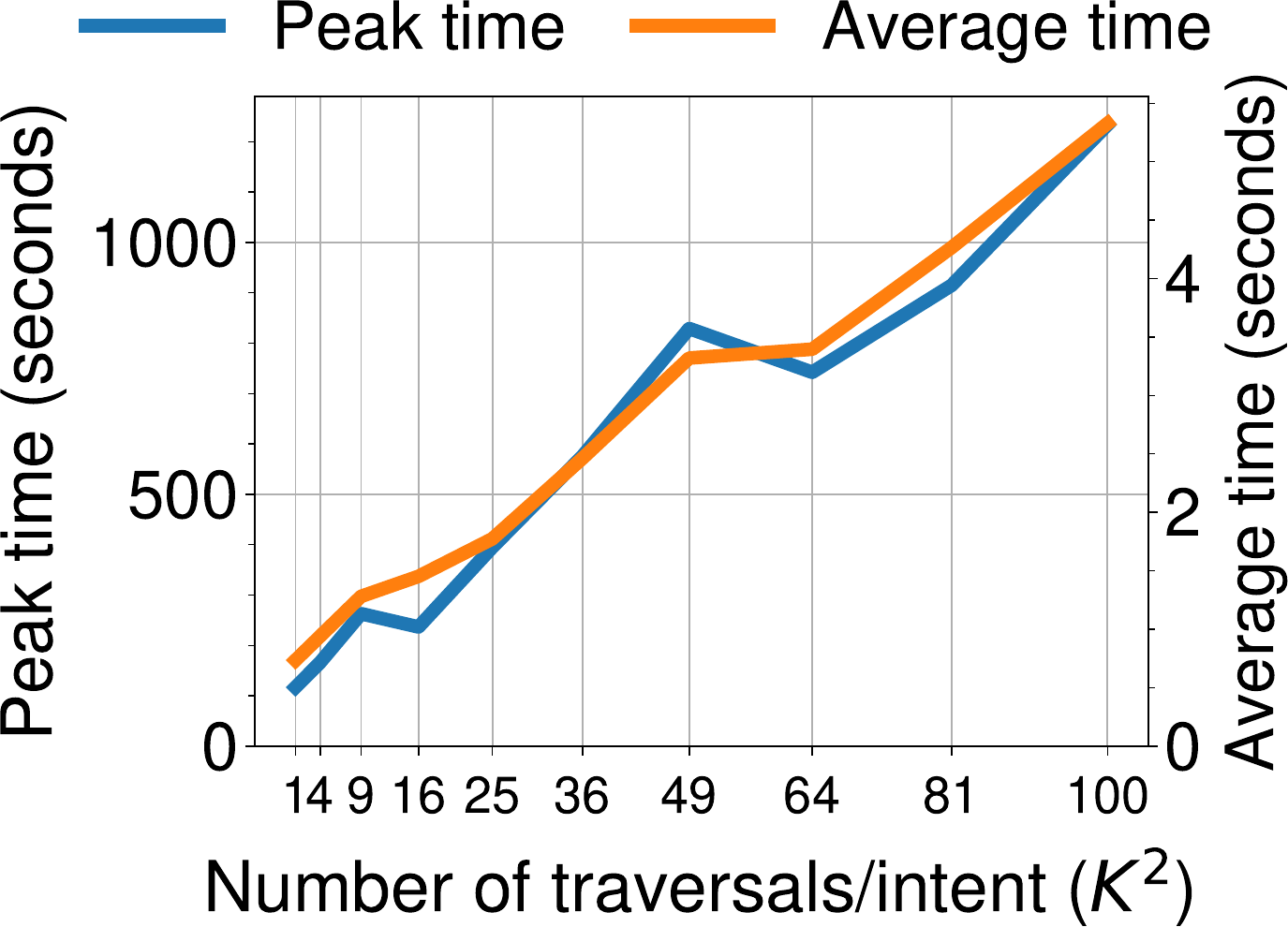}
        \caption{Peak time \& Average time.}
        \label{fig:widening.time.N2}
    \end{subfigure}
    \hfill
    \begin{subfigure}[t]{0.48\linewidth}
        \includegraphics[width=\linewidth]{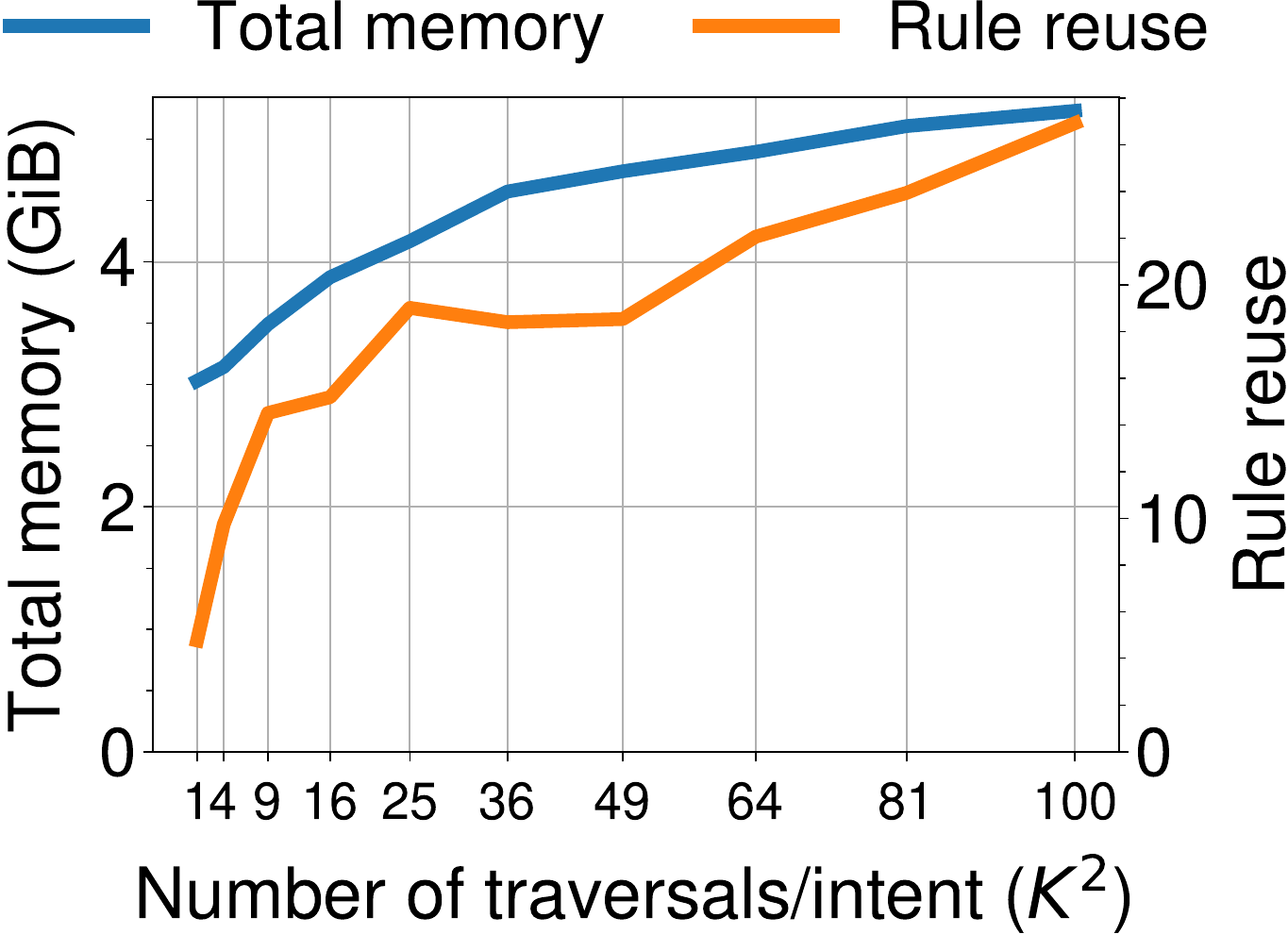}
        \caption{Total memory \& Rule reuse.}
        \label{fig:widening.memory.N2}
    \end{subfigure}
    \vspace{-0.5em}
    \caption{Performance of verifying 500 intents with gradually increasing width (2 y-axes).}
    \label{fig:widening}
    \vspace{-1em}
\end{figure}

Our earlier experiments used a mix of intent types and demonstrated that intents are typically narrow. Now, to test \name's limits, we create increasingly wider intents by generating 500 intents as described in \S\ref{subsec:setup}, but with $K$ sources and $K$ destinations, which we call the intent \emph{width}. Normally, \name would decompose these intents into narrower sub-intents (\S\ref{subsubsec:composite-intent}), but here we disable this feature to force the intents to be wide. In general, verifying an intent with width $K$ requires $K^2$ traversals without aggregation. With $K$ from 1 to 10, we verify the 500 intents with 9 checkers using the \emph{Dynamic} colocation scheme.
Figure~\ref{fig:widening} shows that peak and average times increase linearly with the number of traversals, while total memory grows sublinearly due to the increased slice sharing between wide intents (indicated by the increasing rule reuse rate $\psi$). Therefore, although wide intents should ideally be decomposed (\S\ref{subsubsec:composite-intent}, \S\ref{subsubsec:distributed-loop}), \name's performance degrades gracefully if decomposition is not possible.

\subsection{Distributed Loop Detection}
\label{subsec:loop-eval}

\begin{figure*}
    \centering
    \begin{subfigure}[t]{0.23\linewidth}
        \includegraphics[width=\linewidth]{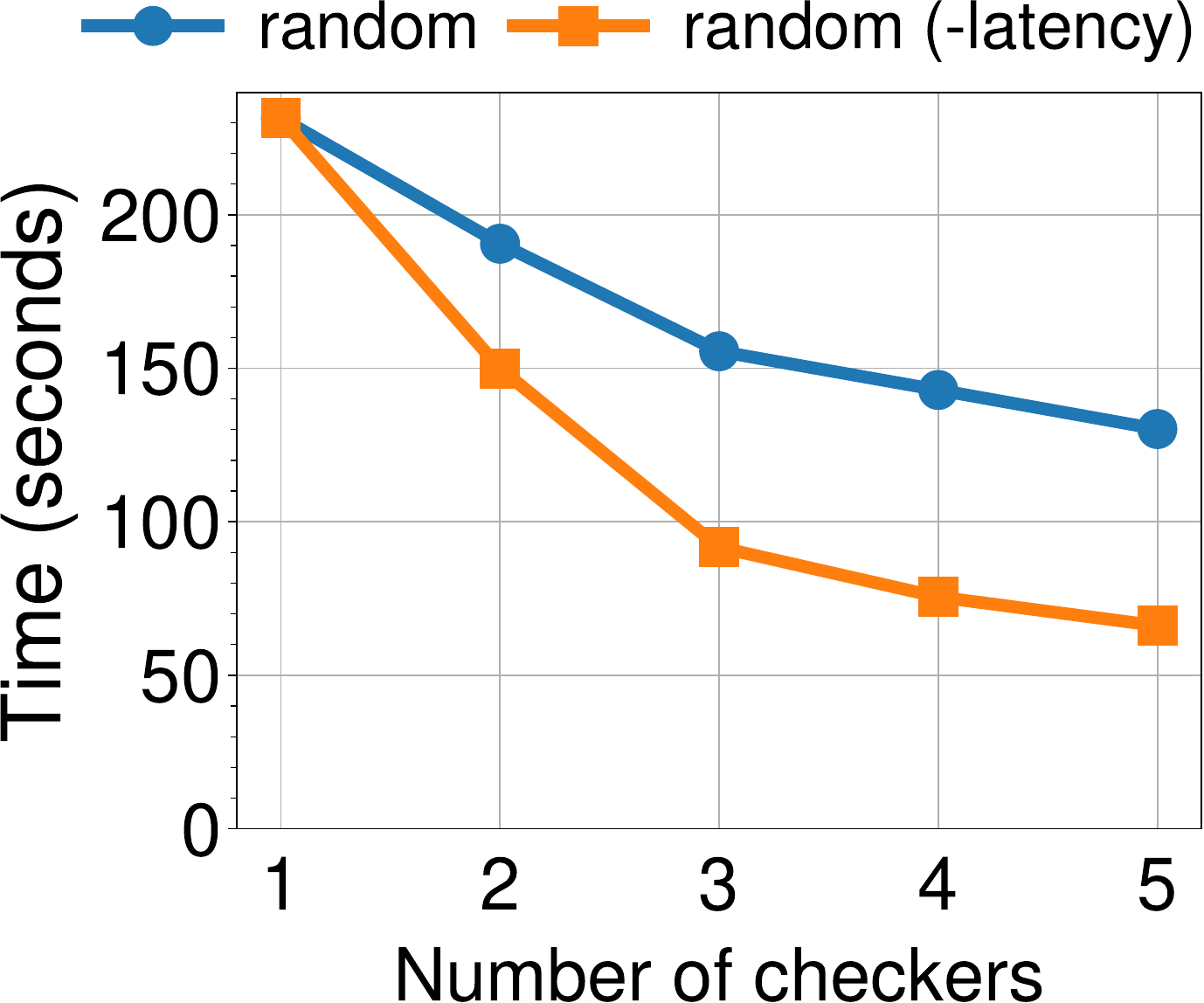}
        \caption{Total time (N4)}
        \label{fig:loop.time.N4}
    \end{subfigure}
    \hfill
    \begin{subfigure}[t]{0.23\linewidth}
        \includegraphics[width=\linewidth]{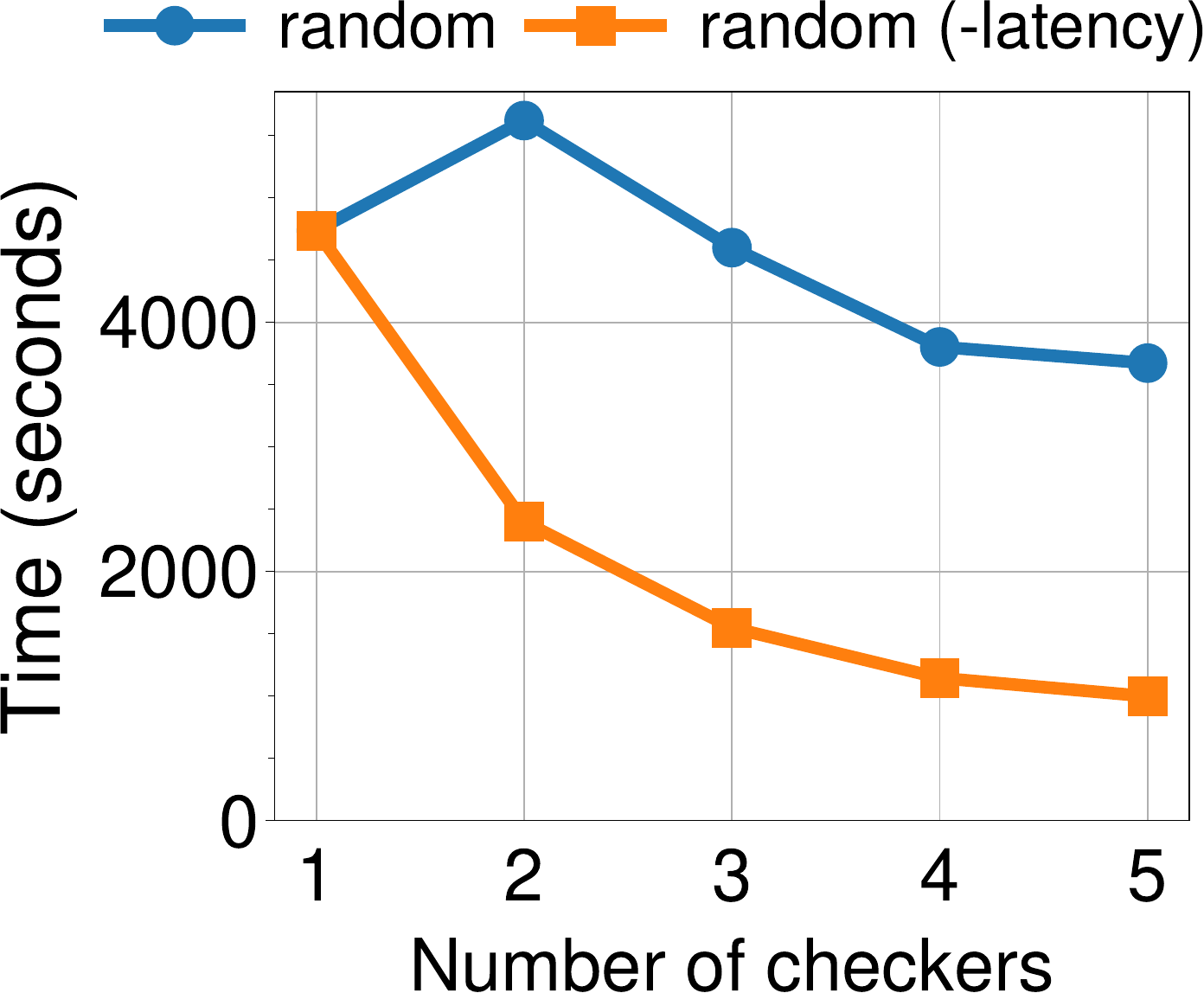}
        \caption{Total time (N5)}
        \label{fig:loop.time.N5}
    \end{subfigure}
    \hfill
    \begin{subfigure}[t]{0.23\linewidth}
        \includegraphics[width=\linewidth]{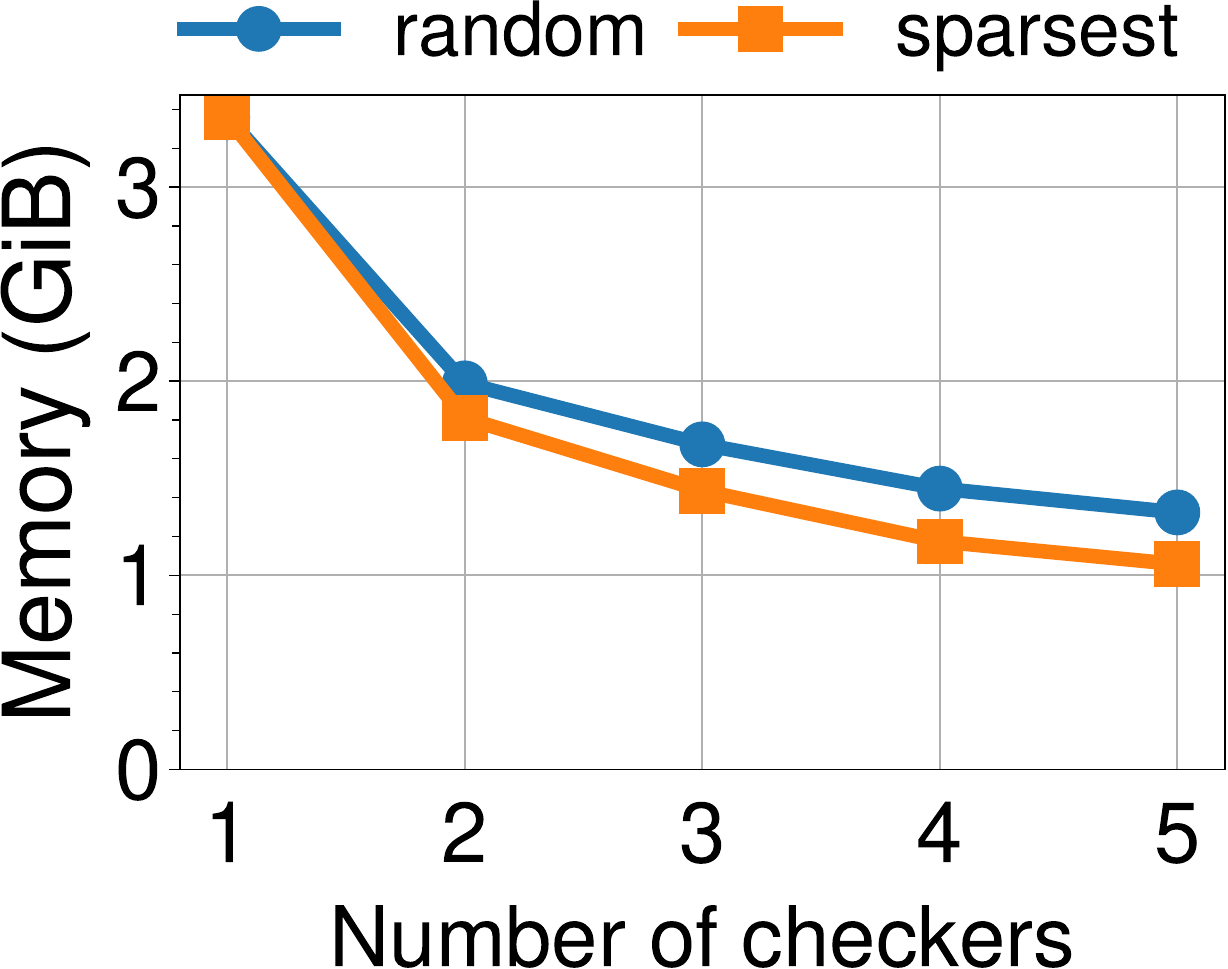}
        \caption{Max memory/checker (N4)}
        \label{fig:loop.memory.N4}
    \end{subfigure}
    \hfill
    \begin{subfigure}[t]{0.23\linewidth}
        \includegraphics[width=\linewidth]{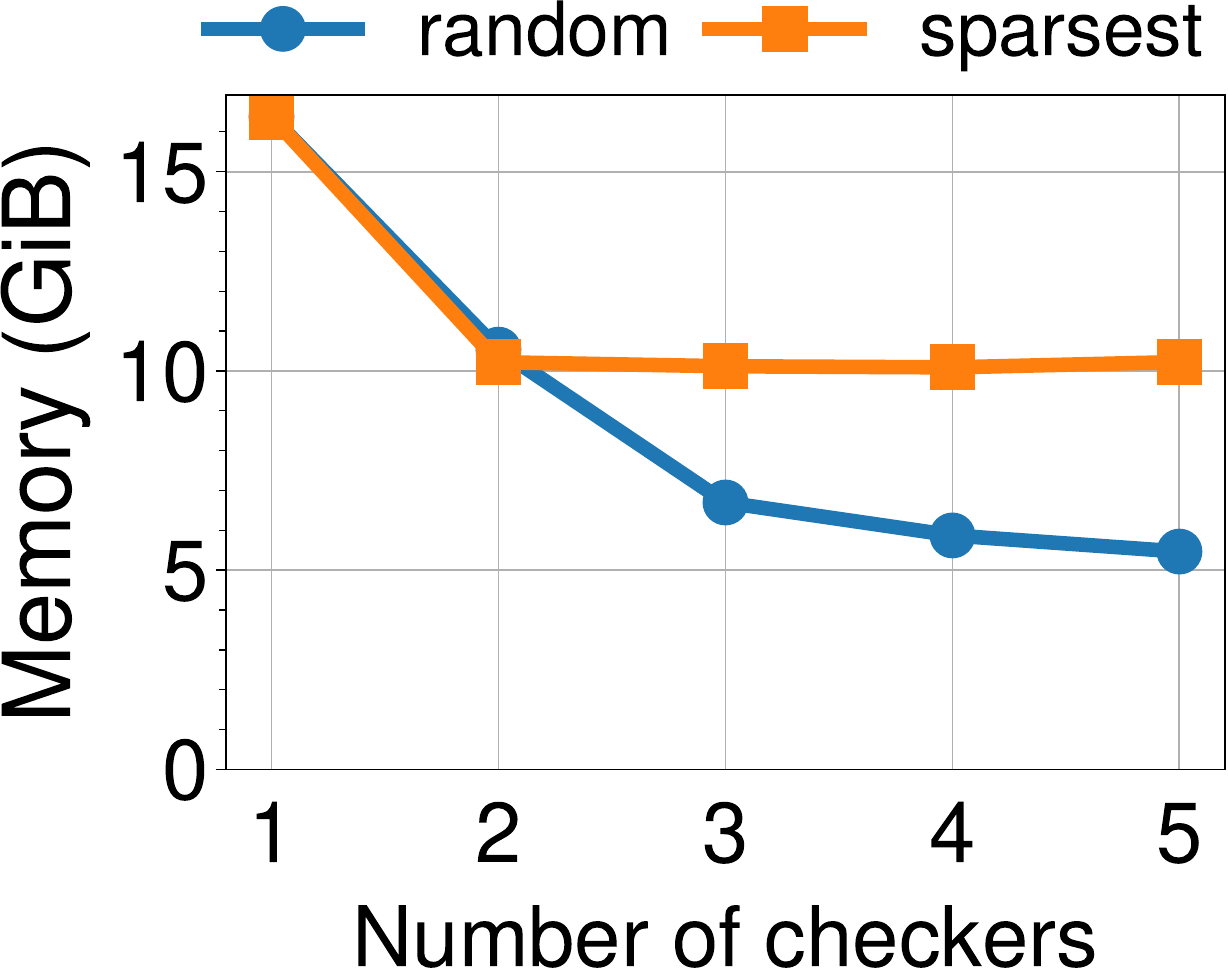}
        \caption{Max memory/checker (N5)}
        \label{fig:loop.memory.N5}
    \end{subfigure}
    \vspace{-0.5em}
    \caption{Performance of distributed loop detection.}
    \label{fig:loop}
    \vspace{-1em}
\end{figure*}

% How is the evaluation conducted.

We now move to the widest intent, loop freedom, which traverses the entire network model. We test \name's distributed loop detection with the larger networks in our dataset (N4, N5). To precisely measure memory and collect intermediate messages between checkers, we first run the algorithm on the standalone server by running each checker sequentially, and then replay the collected messages deterministically in a live cluster to measure the messaging latency.
% More specifically, we sent the messages between checkers at the end of each superstep, which, in effect, is a pessimistic estimate of the latency, because in reality the messages from the same superstep will not be sent out all at the same time, but rather they will be scattered across time based on when each checker finishes traversing their partition for that step. However, the memory evaluation should be truthful in this approach.

Figure~\ref{fig:loop} shows the total time (including model construction and disk I/O) and maximum memory per checker, across varying numbers of checkers and dividing schemes (\S\ref{subsubsec:distributed-loop}). While run time without messaging latency improves significantly, the latency itself is not negligible. With latency factored in, the performance gain from random partitioning is less pronounced.\footnote{In the prototype, the messages are sent in plain text. A more efficient format (e.g., protobuf) may be used to greatly reduce the messaging latency.} We intentionally omit the sparsest cut scheme's time to avoid potentially unfair comparisons, as it found perfectly disconnected segments (due to the nature of the networks) and no messaging is needed between checkers. Full figures can be found in \S\ref{app:subsec:loop}.

The maximum memory reflects how evenly a network model is divided. A more even division would require less memory per checker. Random partitioning generally achieves a more even division than sparsest cut. However, for N4, due to its composition, sparsest cut found a more balanced distribution, hence requiring slightly less memory. For N5, sparsest cut led to a relatively uneven division, where one segment was larger than others, which limited the reduction of maximum memory with more checkers.
In summary, \name can scale loop checks through distributed verification. With 5 checkers and random partitioning, \name reduced total time by 43.7\% (N4) and 22.5\% (N5), while cutting maximum memory usage by 60.6\% (N4) and 66.6\% (N5).

%%%%%%%%%%%%%%%%%%%%%%%%%%%%%%%%%%%%%%%%%%%%%%%%%%%%%%%%%%%%%%%%%%%%%%%%%%%%%%%%

\section{Limitations and Future Work}
\label{sec:limitation}

\para{Incremental traversal}
When a network update affects an intent slice, \name currently recheck the intent by starting a full traversal from the beginning. While this method is quite efficient, as scalability bottlenecks primarily lie in model construction and I/O rather than graph traversal, performance may be further improved by caching previous traversal states and resuming only from the updated vertices.

\vspace{-1em}
\para{Message latency for distributed loop detection}
Currently, a checker sends a separate message for each external path during loop detection. Bundling these paths into a single message could eliminate redundancies like common rules and packets sets shared among paths. The messages may also be encoded in a binary or compressed format instead of plain text JSON to further reduce the latency.

\vspace{-1em}
\para{Integrate loop detection with intent colocation}
While we presented two dividing schemes for distributed loop detection, neither considers other intents. To improve utilization, loop check segments may be divided based on existing intent slices.
% This can enhance resource utilization, although it may introduce increased messaging latency.

%%%%%%%%%%%%%%%%%%%%%%%%%%%%%%%%%%%%%%%%%%%%%%%%%%%%%%%%%%%%%%%%%%%%%%%%%%%%%%%%

\section{Related Work}
\label{sec:related-work}

\para{Distributed data plane verification}
NetPlumber~\cite{2013-KazemianEtAl} and Libra~\cite{2014-ZengEtAl} partition networks by IP subnets to verify intents distributively, but face inefficiencies due to rule duplication across partitions, especially for non-IP rules like MPLS or policy-based routing. Libra also lacks support for header rewrites due to mutually disjoint partitions~\cite{2014-ZengEtAl}.
RCDC~\cite{2019-JayaramanEtAl} partitions networks based on topology by deriving and validating local contracts but only applies to specific reachability intents and structured datacenter networks, limiting its generalizability.
Tulkun~\cite{2023-XiangEtAl} uses on-device verifiers for distributed verification within the verified network, but it is unclear about its impact on live traffic, making it unsuitable for production networks. Moreover, on-device verifiers are generally unavailable in most networks, whereas \name does not require such network resources.
Medusa~\cite{2024-WenEtAl} uses sets of integer tuples for packet sets and distributes workload across cores within a single machine by dividing both the IP space and the topological space. However, like Libra, Medusa struggles with non-IP rules due to its slicing method and the choice of data structure. While Medusa is faster with multithreading on IP-based networks than prior monolithic designs, it still builds the entire network model in memory on a single machine, leading to the same scalability limitations as monolithic designs (\S\ref{sec:motivation}).

\vspace{-1em}
\para{Data plane verification}
Over the past decade, various techniques have been proposed for DPV~\cite{2005-XieEtAl, 2011-MaiEtAl, 2012-KazemianEtAl, 2013-KhurshidEtAl, 2013-YangLam, 2017-HornEtAl, 2020-ZhangEtAl}. Efforts to enhance scalability and performance include more efficient data structures (e.g., atomic predicates~\cite{2013-YangLam}, ddNF~\cite{2016-BjoernerEtAl}, edge-labeled graph~\cite{2017-HornEtAl}, TOBDD~\cite{2023-GuoEtAl}, RANGESET~\cite{2024-WenEtAl}), network transformations~\cite{2016-PlotkinEtAl}, batching~\cite{2022-GuoEtAl}, and minimizing ECs (APKeep~\cite{2020-ZhangEtAl}). However, these approaches still treat the data plane model as monolithic, requiring scaling up rather than out. \name operates on intent-based slices instead of a single monolithic model.

\vspace{-1em}
\para{Formal methods in the network domain}
Formal techniques have been applied to networking with increasing scope and scale. Beyond DPV, network configuration verification~\cite{2015-FogelEtAl, 2017-BeckettEtAl, 2020-PrabhuEtAl, 2020-AbhashkumarEtAl, 2022-ZhangEtAl} enables verifying network configurations before deployment. Stateful network verification~\cite{2017-PandaEtAl, 2020-YousefiEtAl, 2020-YuanEtAl} enables reasoning about data planes with stateful network functions. Automated tools based on formal methods also bring rigor to programmable networks~\cite{2018-KheradmandRosu, 2018-LiuEtAl, 2022-AlbabEtAl, 2023-RenganathanEtAl, 2023-RuffyEtAl}.
% Network function verification~\cite{2019-ZaostrovnykhEtAl, 2020-ZhangEtAla, 2022-PirelliEtAl} verifies stateful network function against given correctness properties.
Other formal techniques have also been explored, such as equivalence checking~\cite{2019-DumitrescuEtAl}, summarization~\cite{2018-BirknerEtAl, 2020-Kheradmand}, differential verification~\cite{2022-ZhangEtAla}, and synthesis~\cite{2018-El-HassanyEtAl}.

%%%%%%%%%%%%%%%%%%%%%%%%%%%%%%%%%%%%%%%%%%%%%%%%%%%%%%%%%%%%%%%%%%%%%%%%%%%%%%%%

\section{Conclusion}
\label{sec:conclusion}

We present \name, a data plane verification method which enables fast verification in a manner that scales out easily, with resource requirements decoupled with the network size. \name operates by creating slices of the network model, which are stored in memory across a cluster and are minimally sufficient for checking pre-configured intents. Evaluation of our implementation shows that \name enables checking thousands of intents on real networks within a short time, and also flattens the resource growth curve for intent checks, allowing significantly larger networks to be supported, while also modeling a rich variety of data plane features.

%%%%%%%%%%%%%%%%%%%%%%%%%%%%%%%%%%%%%%%%%%%%%%%%%%%%%%%%%%%%%%%%%%%%%%%%%%%%%%%%

% \section*{Acknowledgments}
%
% The USENIX latex style is old and very tired, which is why
% there's no \textbackslash{}acks command for you to use when
% acknowledging. Sorry.

%%%%%%%%%%%%%%%%%%%%%%%%%%%%%%%%%%%%%%%%%%%%%%%%%%%%%%%%%%%%%%%%%%%%%%%%%%%%%%%%

% \section*{Availability}
%
% USENIX program committees give extra points to submissions that are
% backed by artifacts that are publicly available. If you made your code
% or data available, it's worth mentioning this fact in a dedicated
% section.

%%%%%%%%%%%%%%%%%%%%%%%%%%%%%%%%%%%%%%%%%%%%%%%%%%%%%%%%%%%%%%%%%%%%%%%%%%%%%%%%

\bibliographystyle{plain}
\bibliography{scylla}

%%%%%%%%%%%%%%%%%%%%%%%%%%%%%%%%%%%%%%%%%%%%%%%%%%%%%%%%%%%%%%%%%%%%%%%%%%%%%%%%

\clearpage
\appendix

\section{Supplementary Examples}

\subsection{Data Plane Abstraction (DPA)}
\label{app:subsec:dpa-examples}

\begin{figure}
    \begin{lstlisting}[basicstyle=\scriptsize\ttfamily]
    {
      "name": "spine2",
      "vendor": "Cisco Systems, Inc.",
      "model": "N9K-C9364C",
      "timestamp": "1648658732157",
      "interfaces": [ ... ],
      "rule_tables": [
        {
          "name": "spine2_::_VLAN_IN_::_Ethernet1/1",
          "rules": [
            {
              "match": {
                "masked_values": [
                  {
                    "field_type": 6, // Match VLAN ID
                    "value": "2",
                    "mask": "",
                    "depth": 0
                  }
                ]
              },
              "actions": [
                {
                  "type": 0,  // Forward
                  "value": "",
                  "outgoing_interface_id": "9365eca7"
                }
              ],
            },
          ],
        },
        {
          "name": "spine2_::_L3_::_infra:overlay-1",
          "rules": [
            {
              "match": {
                "masked_values": [
                  {
                    "field_type": 6, // Match VLAN ID
                    "value": "0",
                    "mask": "0",
                    "depth": 0
                  },
                  {
                    "field_type": 13, // Match IP dst
                    "value": "10.1.0.1",
                    "mask": "255.255.255.255",
                    "depth": 1
                  }
                ]
              },
              "actions": [
                {
                  "type": 104, // Set VLAN ID
                  "value": "2",
                  "outgoing_interface_id": null
                },
                {
                  "type": 102, // Set L2 dst
                  "value": "2c:4f:52:56:c8:67",
                  "outgoing_interface_id": null
                },
                {
                  "type": 0,   // Forward
                  "value": "",
                  "outgoing_interface_id": "f55a0c8f"
                }
              ],
            },
          ],
        },
      ],
    }
    \end{lstlisting}
    \caption{DPA snippet of a Cisco N9K switch in N3.}
    \label{app:fig:dpa-n9k}
\end{figure}

Figure~\ref{app:fig:dpa-n9k} shows a redacted and simplified DPA snippet of a Cisco N9K-C9364C network switch from the network N3 described in \S\ref{subsec:setup}. Two rules are shown in this snippet. The first one, in the \textit{VLAN\_IN} table, matches for the packets tagged with VLAN 2 and forwards them to a specific interface, which is associated with some other rule table. The second rule, in the \textit{L3} table, matches for the packets tagged with VLAN 0 (no VLAN) and IP destination 10.1.0.1, and then applies the following actions: (1) set the VLAN ID to 2, (2) set the Ethernet destination address to 2c:4f:52:56:c8:67, and (3) forward the matched packet to another interface.

\subsection{Intent Specifications}
\label{app:subsec:intent-examples}

\begin{figure}[ht]
    \scriptsize
    \begin{lstlisting}
    {
      "intent_parameters": {
        "bidirectional": false,
        "target_set": [
          {
            "vmName": "DB01",
            "ip": "10.0.3.11",
            "mac": "00:50:56:95:cf:1d",
            "modelKey": "(device ID)"
          },
          (...)
        ],
        "origin_set": [
          {
            "vmName": "App01",
            "ip": "10.0.2.11",
            "mac": "00:50:56:95:32:7c",
            "modelKey": "(device ID)"
          },
          (...)
        ],
        "target_subnet": "[ \"(\", \"^\", [{ \"field_type\": 4, \"depth\": 0, \"mask\": \"\", \"value\": \"00:50:56:95:32:7c\" }], \")\", \"&\", [{ \"field_type\": 12, \"depth\": 1, \"mask\": \"255.255.255.255\", \"value\": \"10.0.2.11\" }], \"&\", [{ \"field_type\": 13, \"depth\": 1, \"mask\": \"255.255.255.255\", \"value\": \"10.0.3.11\" }] ]"
      },
      "id": "(intent ID)",
      "type": 7  // Reachability
    }
    \end{lstlisting}
    \caption{Specification of a reachability intent.}
    \label{app:fig:reachability-spec}
\end{figure}

\begin{figure}[ht]
    \scriptsize
    \begin{lstlisting}
    {
      "intent_parameters": {},
      "id": "(intent ID)",
      "type": 1  // Loop freedom
    }
    \end{lstlisting}
    \caption{Specification of the loop freedom intent.}
    \label{app:fig:loop-spec}
\end{figure}

Figure~\ref{app:fig:reachability-spec} and \ref{app:fig:loop-spec} show redacted examples of intent specifications of a reachability intent and the loop freedom intent. The specifications of other intent types have similar syntax as the reachability intent and are thus omitted. In Figure~\ref{app:fig:reachability-spec}, \textit{target\_subnet} does not actually specify a subnet, but the initial packet set for the symbolic traversal. The parameter name is called \textit{target\_subnet} due to historical reasons. In this case, it specifies the destination Ethernet address to be 00:50:56:95:32:7c, the source IP address to be 10.0.2.11, and the destination IP address to be 10.0.3.11. On the other hand, the loop freedom intent (Figure~\ref{app:fig:loop-spec}) requires no parameter.

\subsection{Algorithms for Verifying Intents}
\label{app:subsec:verify-intents}

\begin{algorithm}[ht]
    \small
    \caption{Get the next violating path.}
    \label{app:alg:traversal}
    \Fn{GetNextPath ()}{
        \KwData{rt: rule table.}
        \KwData{ps: packet set.}
        \KwData{path: the current path; a list of (rt,~ps) pairs.}
        \KwData{frontier: a stack of path.}
        \KwData{status: the violation status.}
        path $\gets \varnothing$\;
        status $\gets 0$\;
        \Do{status $= 0$}{
            \Do{status $= -1$}{
                \If{frontier.size $= 0$}{
                    \Return{$\varnothing$}\;
                }
                path $\gets$ frontier.pop ()\;
                status $\gets$ \textit{IntentPredicate}~(path, frontier)\;
            }
            \If{status $\neq 1$}{
                nexthops $\gets$ \textit{GetNextHops}~(path.last)\;
                \ForEach{$(rt, ps) \in$ nexthops}{
                    new\_path $\gets$ path + (rt, ps)\;
                    frontier.push (new\_path)\;
                }
            }
        }
        \Return{path}\;
    }
\end{algorithm}

Algorithm~\ref{app:alg:traversal} illustrates the pseudocode of the symbolic traversal as described in \S\ref{subsec:traversal}. However, to begin with, we need to first define the \emph{violation status} of the traversal state. The violation status is $1$ if the current traversal state constitutes a violation of the given intent. The violation status is $-1$ if, based on the current traversal state, going down the current path would never lead to a violation, in which case the traversal algorithm will backtrack to the previous branch and continue exploring other paths. The violation status is $0$ if it is yet inconclusive, meaning that it may lead to a violation in the future, and the symbolic traversal should continue.

\begin{algorithm}[ht]
    \small
    \caption{Compute the next hops.}
    \label{app:alg:nexthops}
    \Fn{GetNextHops (rt, ps)}{
        \KwData{rt: rule table.}
        \KwData{ps: packet set.}
        \KwData{fn: flow node.}
        nexthops $\gets \varnothing$\;
        \If{$\nexists$ rt}{
            Fetch and load the DPA containing rt\;
            $\forall$ rt $\in$ DPA. rt.universe $\gets \varnothing$\;
        }
        \If{ps $\nsubseteq$ rt.universe}{
            rt.universe $\gets$ rt.universe $\cup$ ps\;
            rt.build\_flow\_nodes (rt.universe)\;
        }
        \ForEach{fn $\in$ rt.overlapping\_flow\_nodes(ps)}{
            (rt', ps') $\gets$ fn.apply\_actions (ps)\;
            nexthops $\gets$ nexthops $\cup$ (rt', ps')\;
        }
        \Return{nexthops}\;
    }
\end{algorithm}

\begin{algorithm}[ht]
    \small
    \caption{Black hole freedom as an example.}
    \label{app:alg:blackhole}
    \Fn{BlackHole::IntentPredicate (path, frontier)}{
        \KwData{path: the current path; a list of (rt,~ps) pairs.}
        \KwData{frontier: a stack of path.}
        \If{path contains a loop or leaves the network}{
            \Return{-1}\;
        }
        \If{path ends with packets being accepted}{
            \Return{-1}\;
        }
        \If{path ends with packets being dropped}{
            \Return{1}\;
        }
        \Return{0}\;
    }
\end{algorithm}

At each hop, before the path ends, the \emph{GetNextHops} function computes the set of successor $(R_i,P_i)$ pairs, as described in \S\ref{subsec:traversal} and builds up the intent slice as needed for the traversal. Algorithm~\ref{app:alg:nexthops} shows the pseudocode. The \emph{IntentPredicate} function is specific to the intent types (\S\ref{subsubsec:intent-types}). It determines when the traversal state constitutes a violation, which governs the exploration. We provide the pseudocode in Algorithm~\ref{app:alg:blackhole} for verifying black hole freedom as an example.

\section{Supplementary Experiment Results}
\label{app:sec:additional-results}

\subsection{Rule-level partitioning}
\label{app:subsec:granularity}

\begin{figure}
    \centering
    \begin{subfigure}[t]{0.48\linewidth}
        \includegraphics[width=\linewidth]{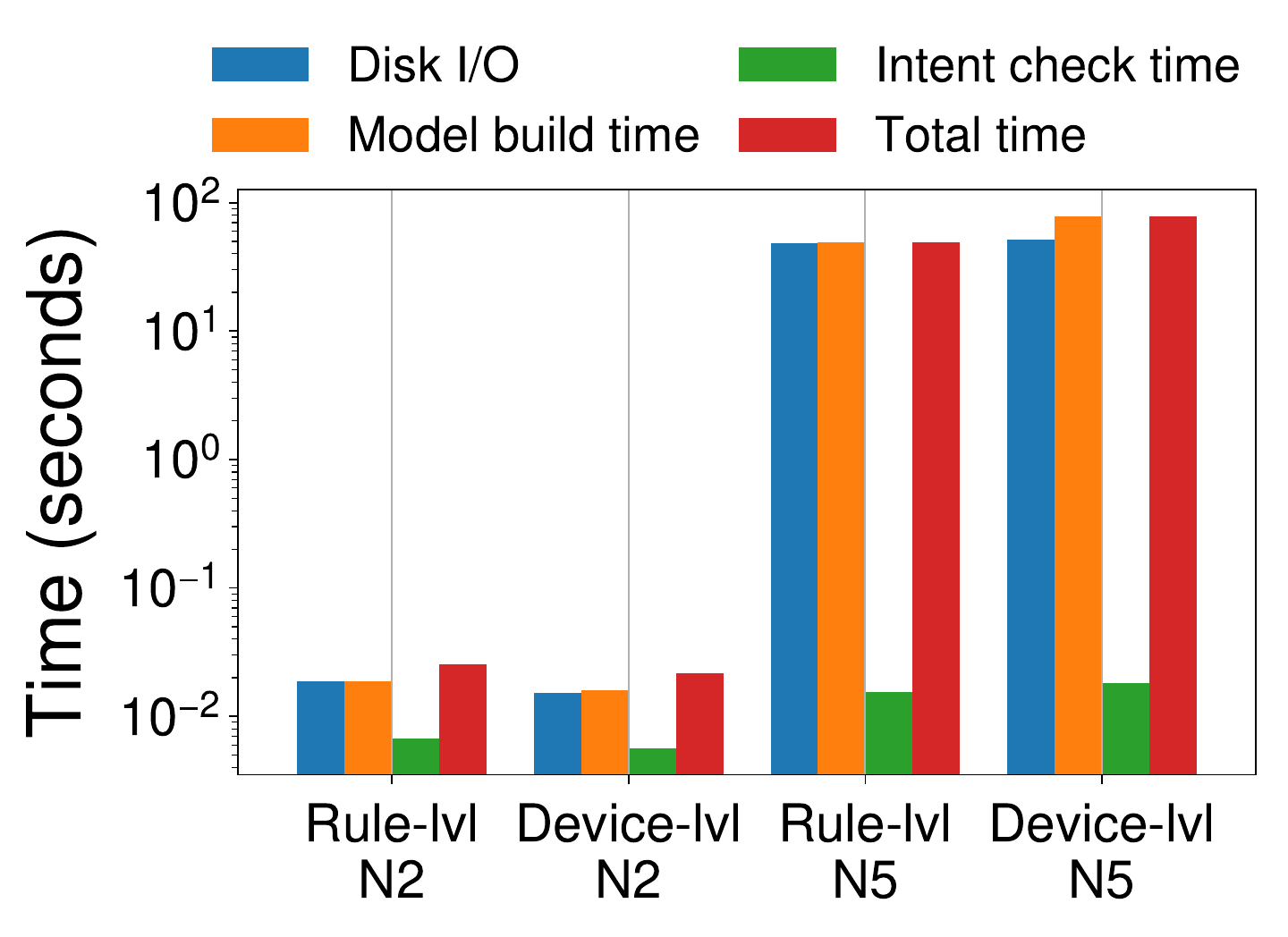}
        \caption{Time (N2, N5) (log scale)}
    \end{subfigure}
    \hfill
    \begin{subfigure}[t]{0.48\linewidth}
        \includegraphics[width=\linewidth]{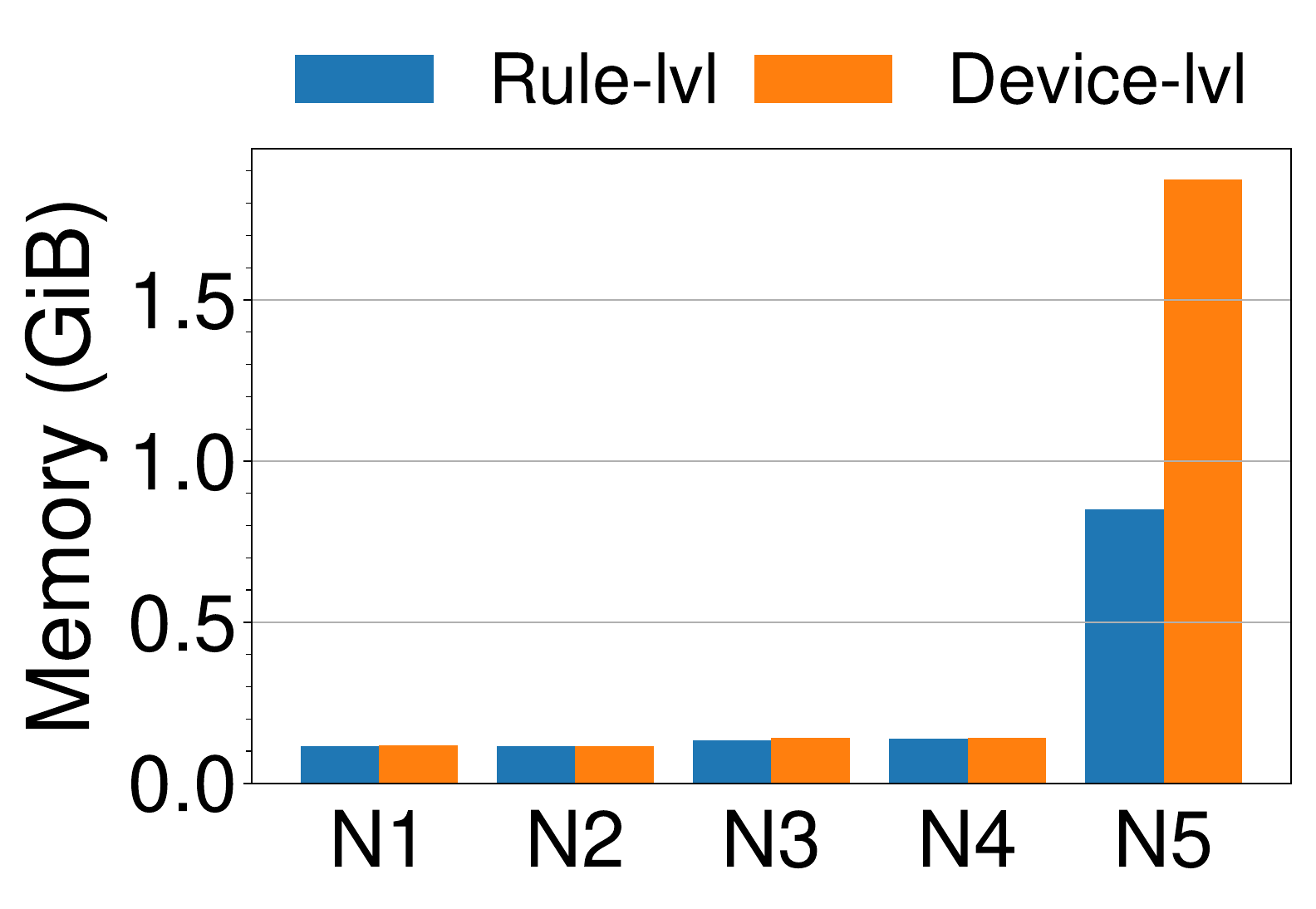}
        \caption{Memory (all networks)}
    \end{subfigure}
    \caption{Performance comparison between device-level vs. rule-level granularity to verify a new intent from scratch.}
    \label{app:fig:granularity.first-run}
\end{figure}

At an earlier stage of the project, we experimented with device-level granularity, meaning for each device, either everything is built into the model or none at all. Later on we implemented rule-level slicing, and here we evaluate how much benefit it brings by comparing the two with single-intent microbenchmarks, as shown in Figure~\ref{app:fig:granularity.first-run}.

\begin{figure*}[ht]
    \centering
    \begin{subfigure}[t]{0.48\linewidth}
        \includegraphics[width=\linewidth]{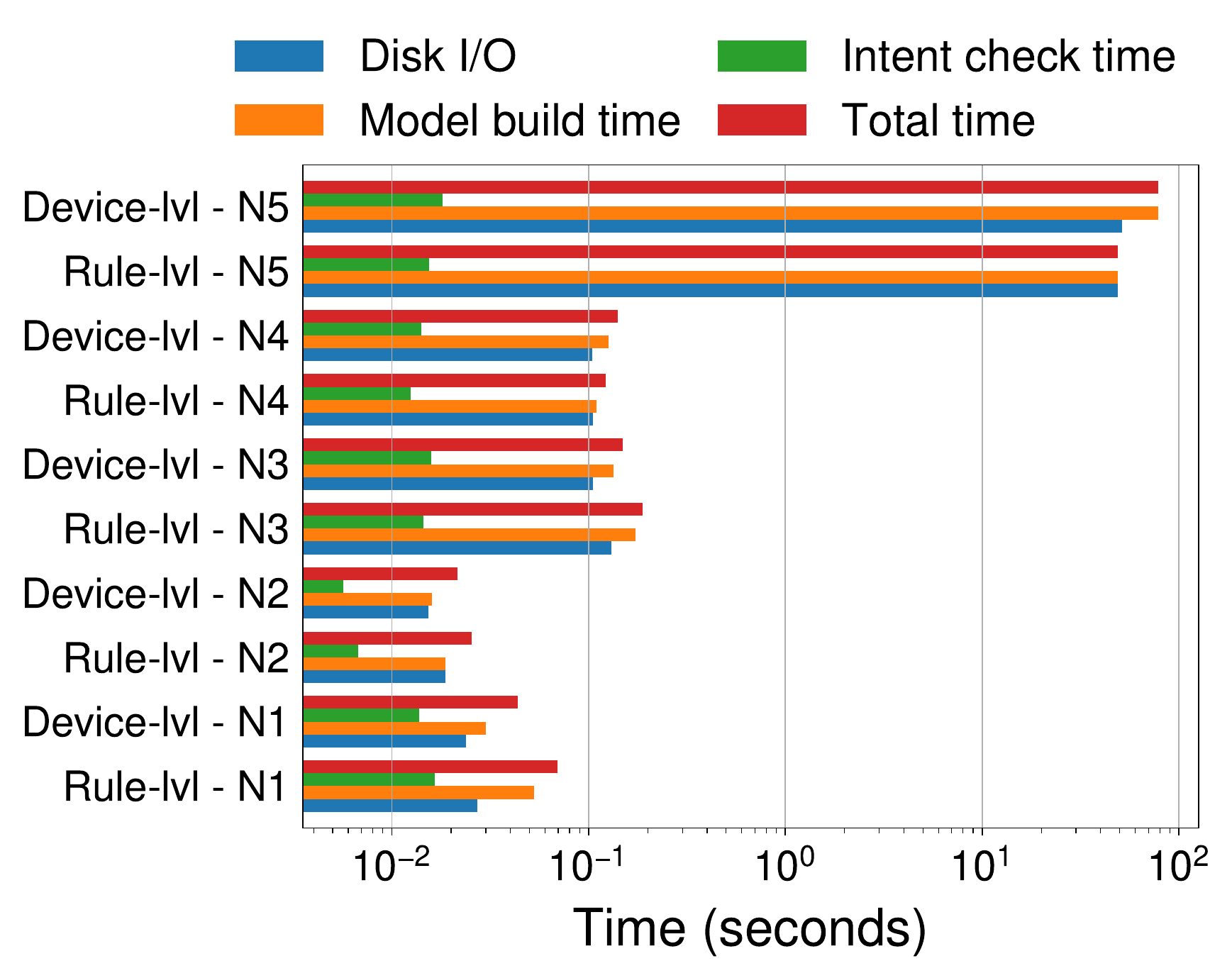}
        \caption{New intent from scratch.}
        % \label{}
    \end{subfigure}
    \hfill
    \begin{subfigure}[t]{0.48\linewidth}
        \includegraphics[width=\linewidth]{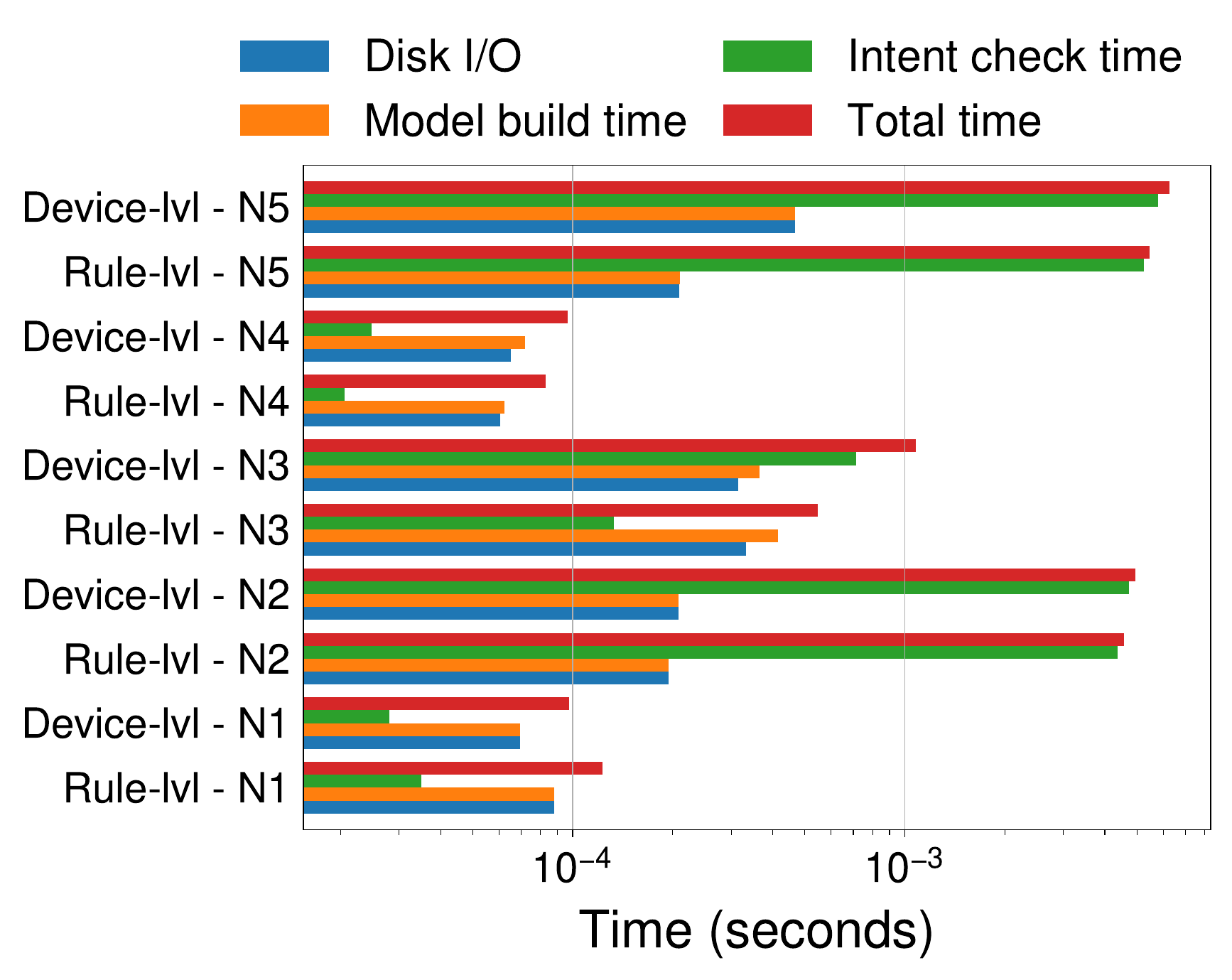}
        \caption{Maintaining existing intent.}
        % \label{}
    \end{subfigure}
    \caption{Run time microbenchmarks to verify one intent with device-level vs. rule-level granularity (log scale).}
    \label{app:fig:granularity.time}
\end{figure*}

\begin{figure}[ht]
    \centering
    \begin{subfigure}[t]{0.48\linewidth}
        \includegraphics[width=\linewidth]{single-intent/granularity.memory.reachability.first-run.pdf}
        \caption{New intent from scratch.}
        % \label{}
    \end{subfigure}
    \hfill
    \begin{subfigure}[t]{0.48\linewidth}
        \includegraphics[width=\linewidth]{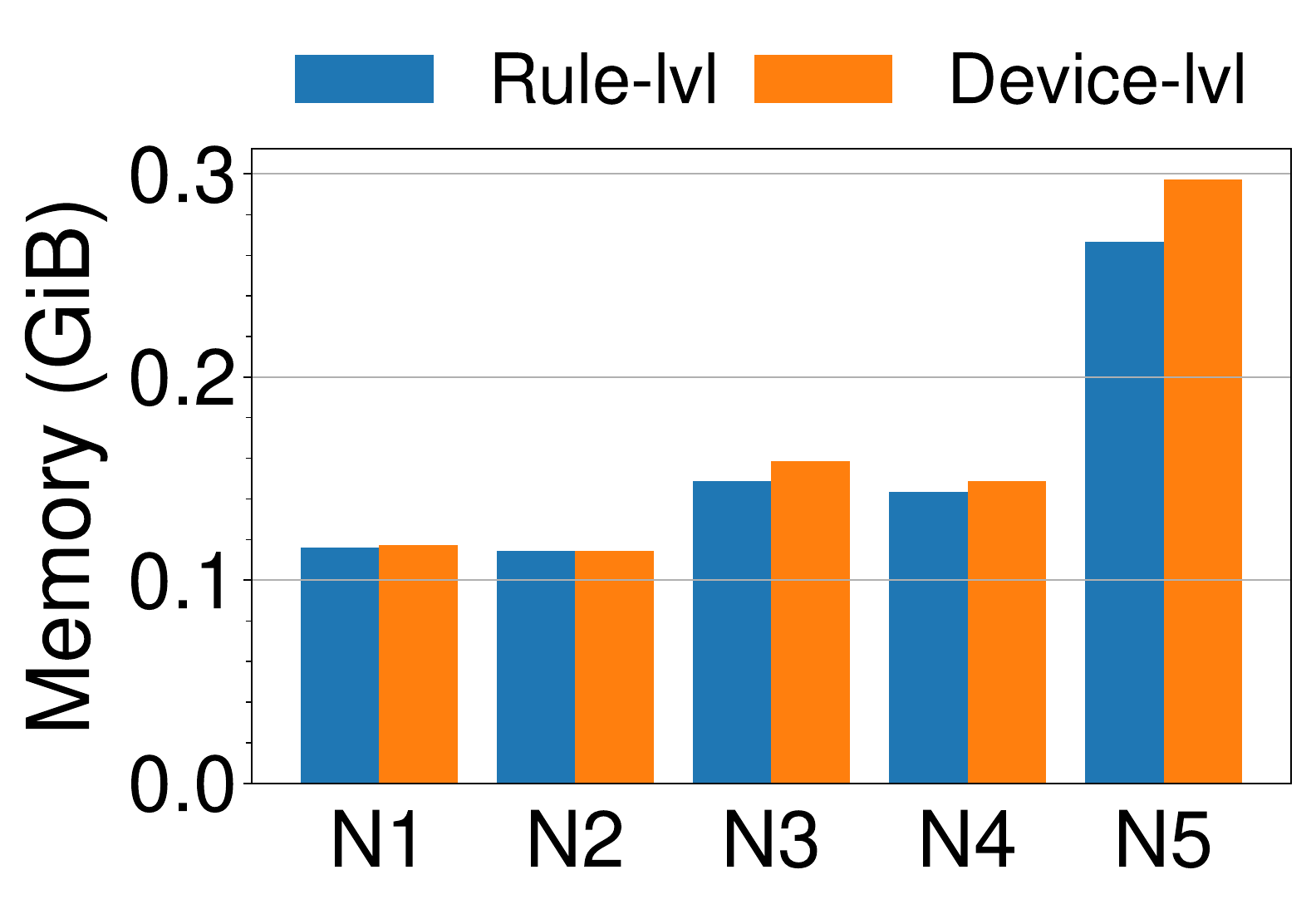}
        \caption{Maintain existing intent.}
        % \label{}
    \end{subfigure}
    \caption{Memory comparison between device-level vs. rule-level granularity for a single intent.}
    \label{app:fig:granularity.memory}
\end{figure}

For N2, the device-level granularity was faster, because the size of a device is small enough that building only a slice of it turns out to be slightly slower due to the additional overhead for intent overlap calculations. However, for N5, with large devices, we can clearly see the benefit of rule-level slicing both in time and memory. The intent processing time is 38\% faster and requires only 45\% of the memory compared to the device-level implementation. For networks with larger devices, finer granularity is in favor. In addition, with these smaller intent slices, it also means we can distribute the workload more easily with finer granularity. Figure~\ref{app:fig:granularity.time} and \ref{app:fig:granularity.memory} presents the full results of the granularity comparison.

\subsection{Single-intent local microbenchmarks}
\label{app:subsec:single-intent}

\begin{figure*}[ht]
    \centering
    \begin{subfigure}[t]{0.32\linewidth}
        \includegraphics[width=\linewidth]{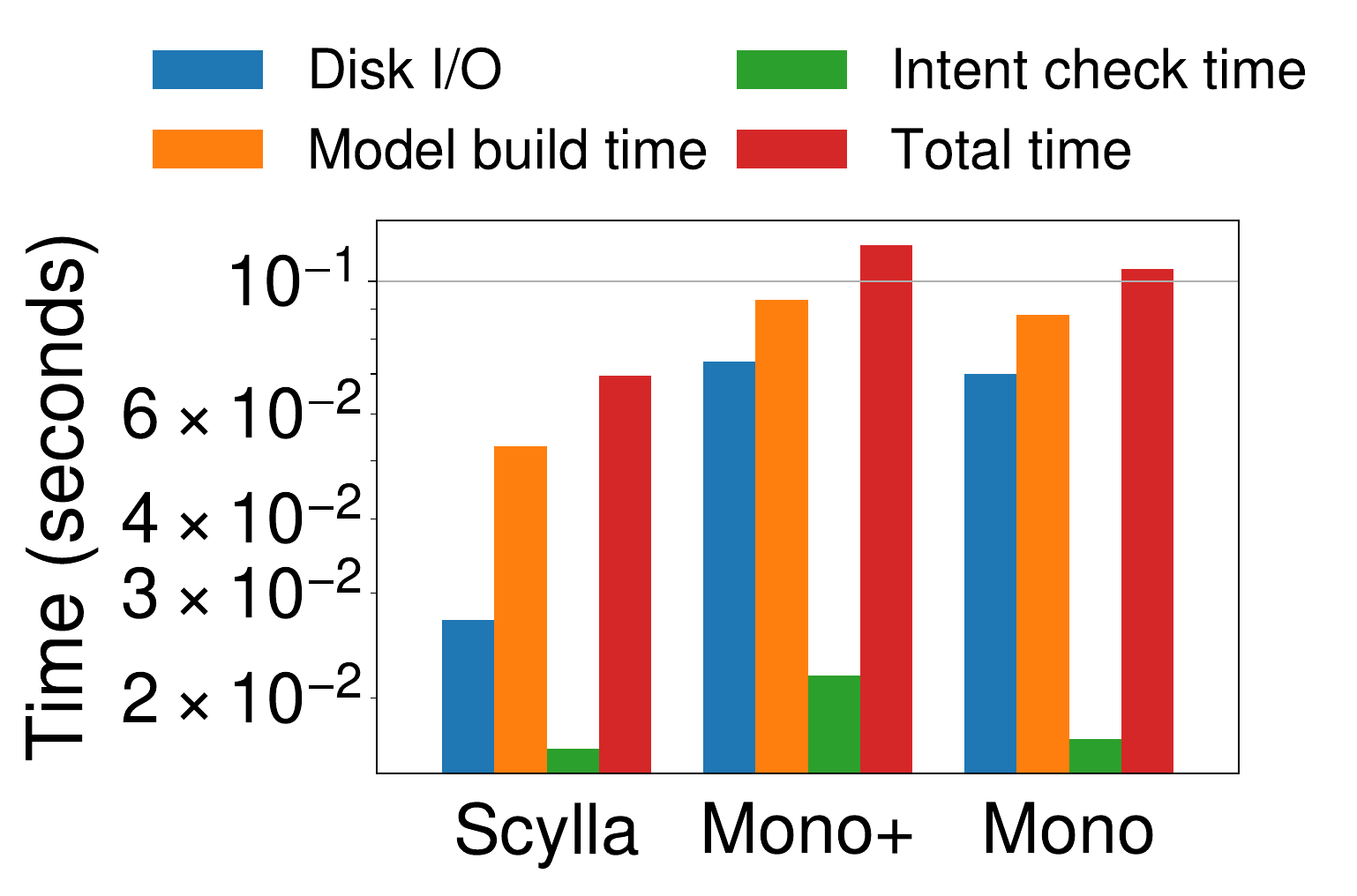}
        \caption{Time (N1)}
    \end{subfigure}
    \hfill
    \begin{subfigure}[t]{0.32\linewidth}
        \includegraphics[width=\linewidth]{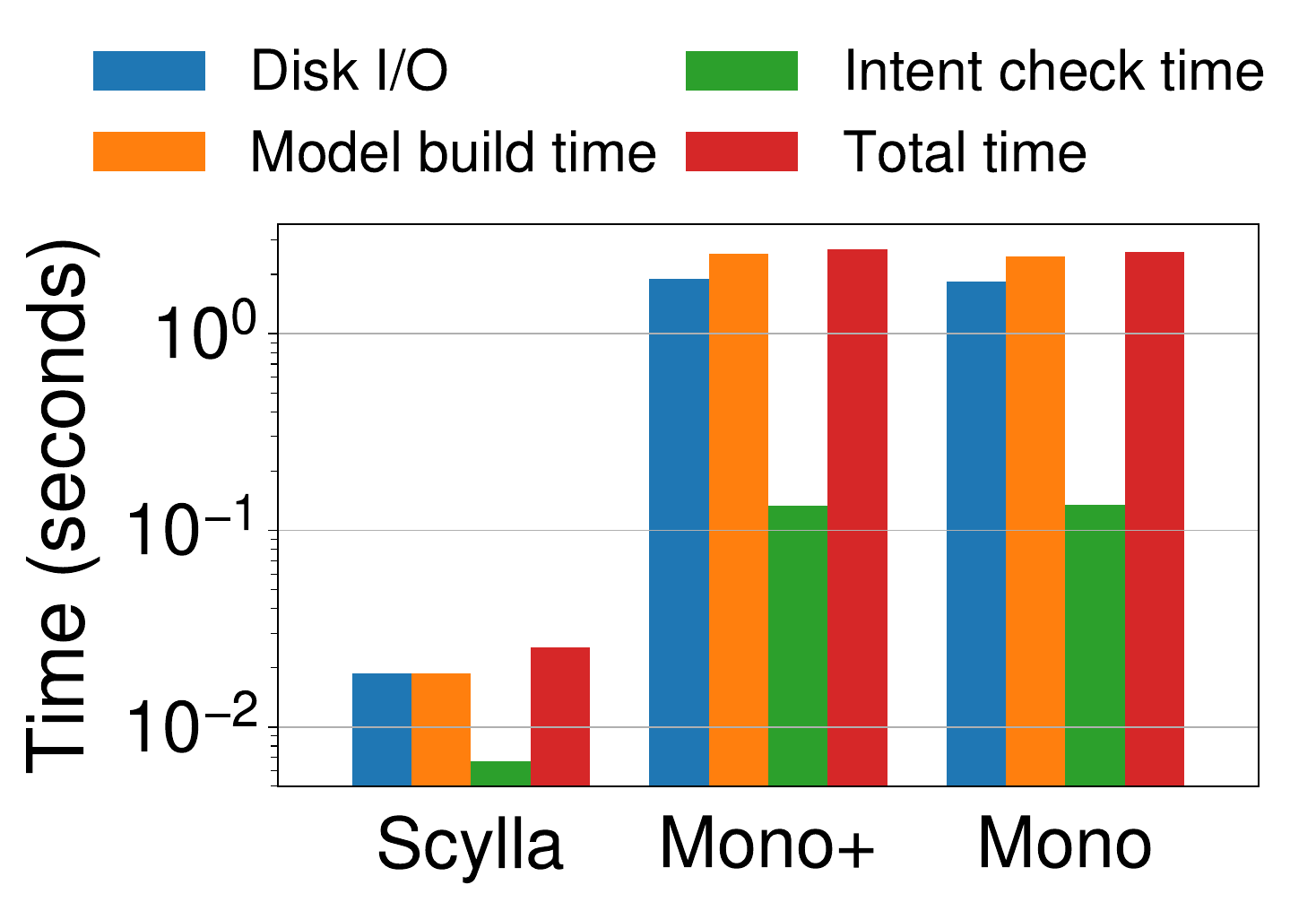}
        \caption{Time (N2)}
    \end{subfigure}
    \hfill
    \begin{subfigure}[t]{0.32\linewidth}
        \includegraphics[width=\linewidth]{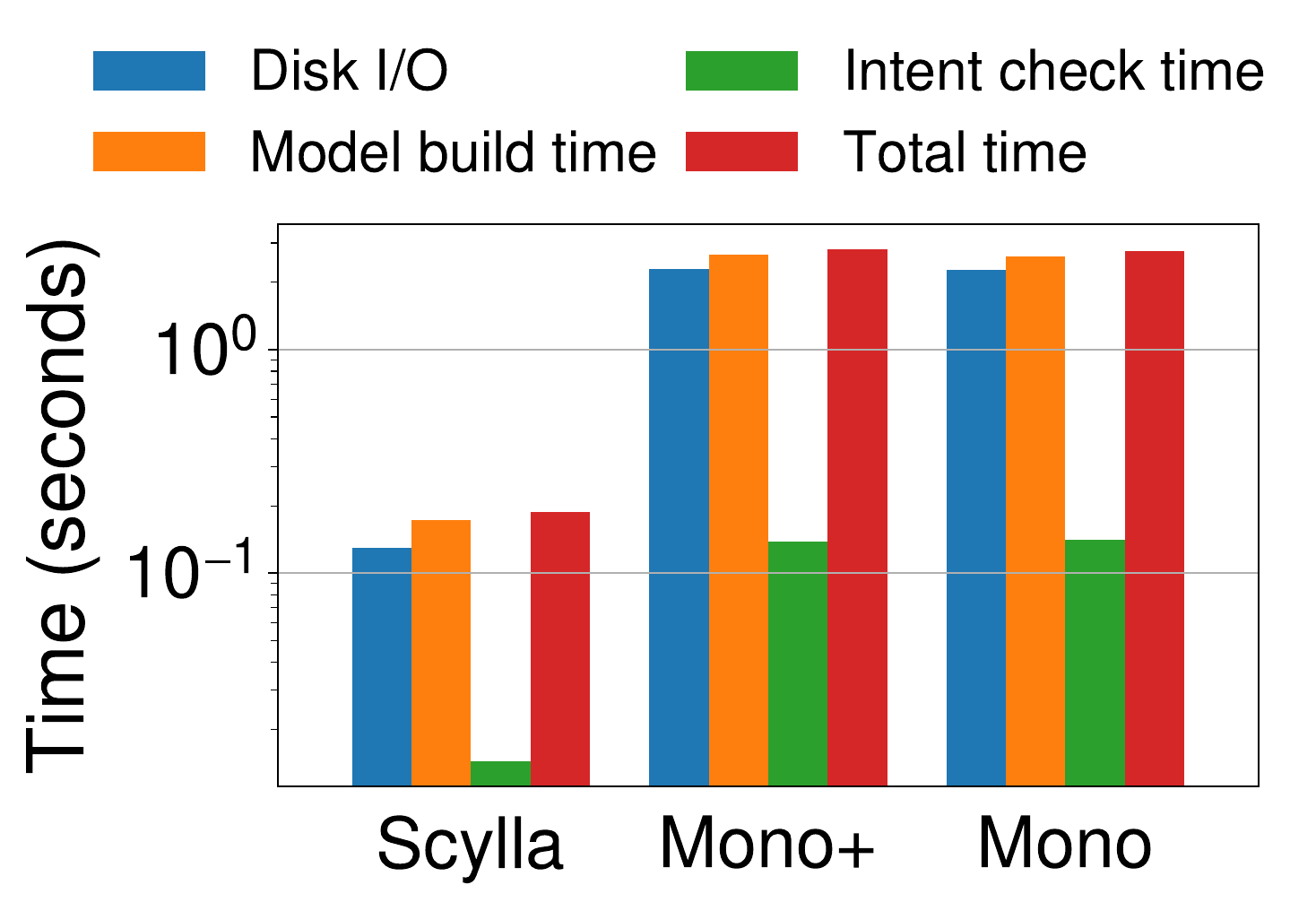}
        \caption{Time (N3)}
    \end{subfigure}
    \hfill
    \begin{subfigure}[t]{0.32\linewidth}
        \includegraphics[width=\linewidth]{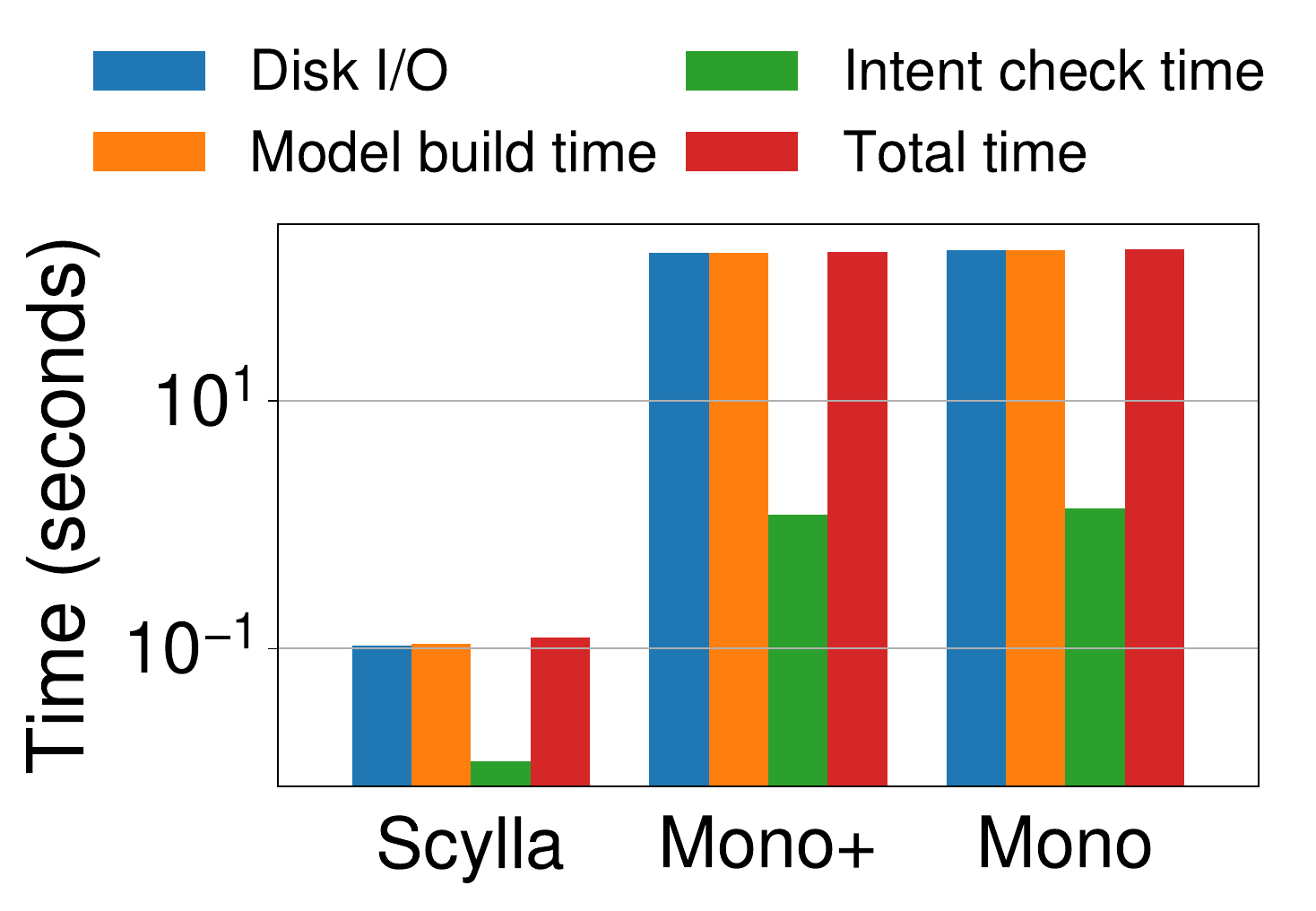}
        \caption{Time (N4)}
    \end{subfigure}
    \hfill
    \begin{subfigure}[t]{0.32\linewidth}
        \includegraphics[width=\linewidth]{single-intent/single-node.time.reachability.first-run.N5.pdf}
        \caption{Time (N5)}
    \end{subfigure}
    \hfill
    \begin{subfigure}[t]{0.32\linewidth}
        \includegraphics[width=\linewidth]{single-intent/single-node.memory.reachability.first-run.pdf}
        \caption{Memory (all networks)}
    \end{subfigure}
    \caption{Performance for verifying a new intent from scratch (log scale).}
    \label{app:fig:single-node.reachability.first-run}
\end{figure*}

\begin{figure*}[ht]
    \centering
    \begin{subfigure}[t]{0.32\linewidth}
        \includegraphics[width=\linewidth]{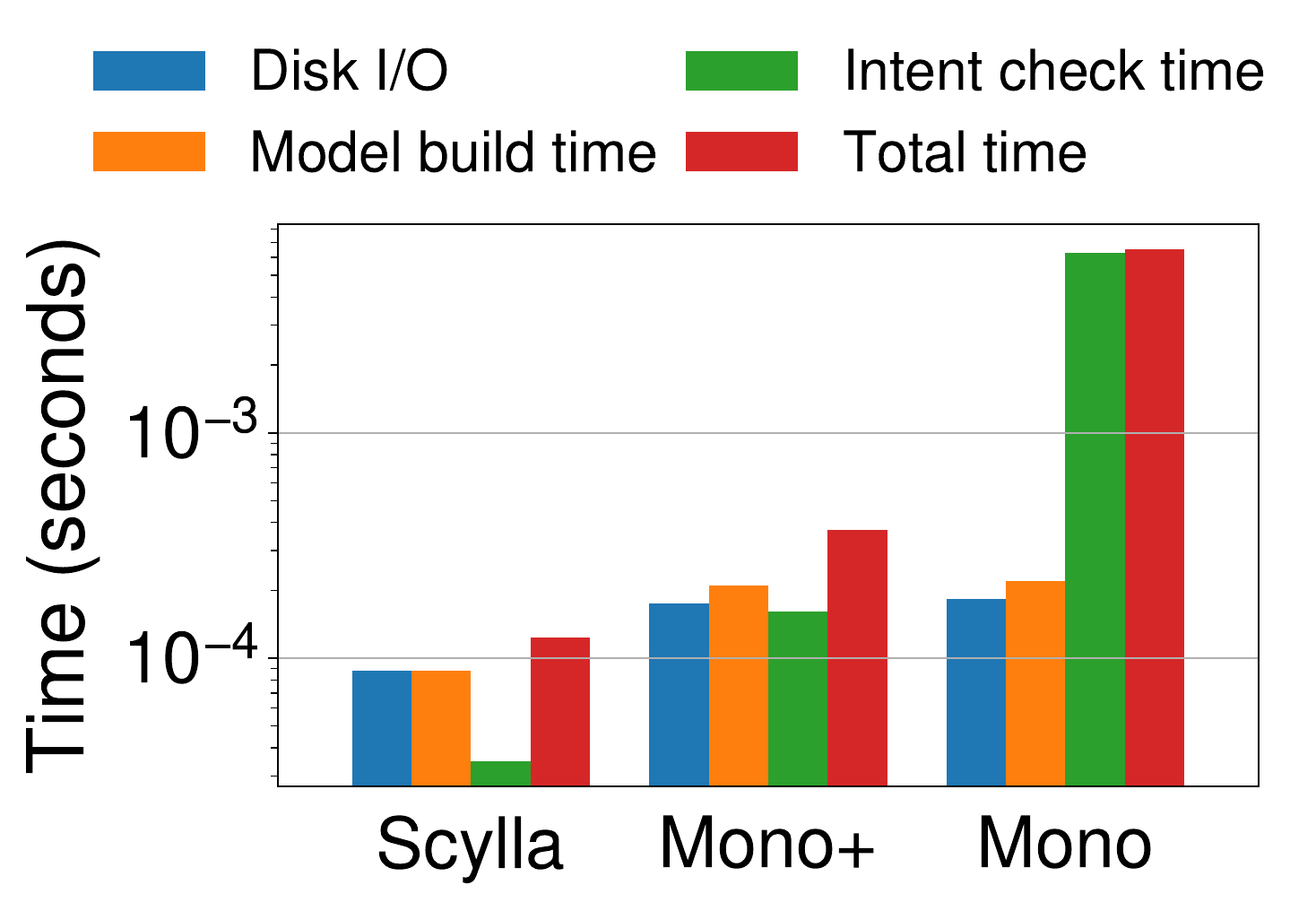}
        \caption{Time (N1)}
    \end{subfigure}
    \hfill
    \begin{subfigure}[t]{0.32\linewidth}
        \includegraphics[width=\linewidth]{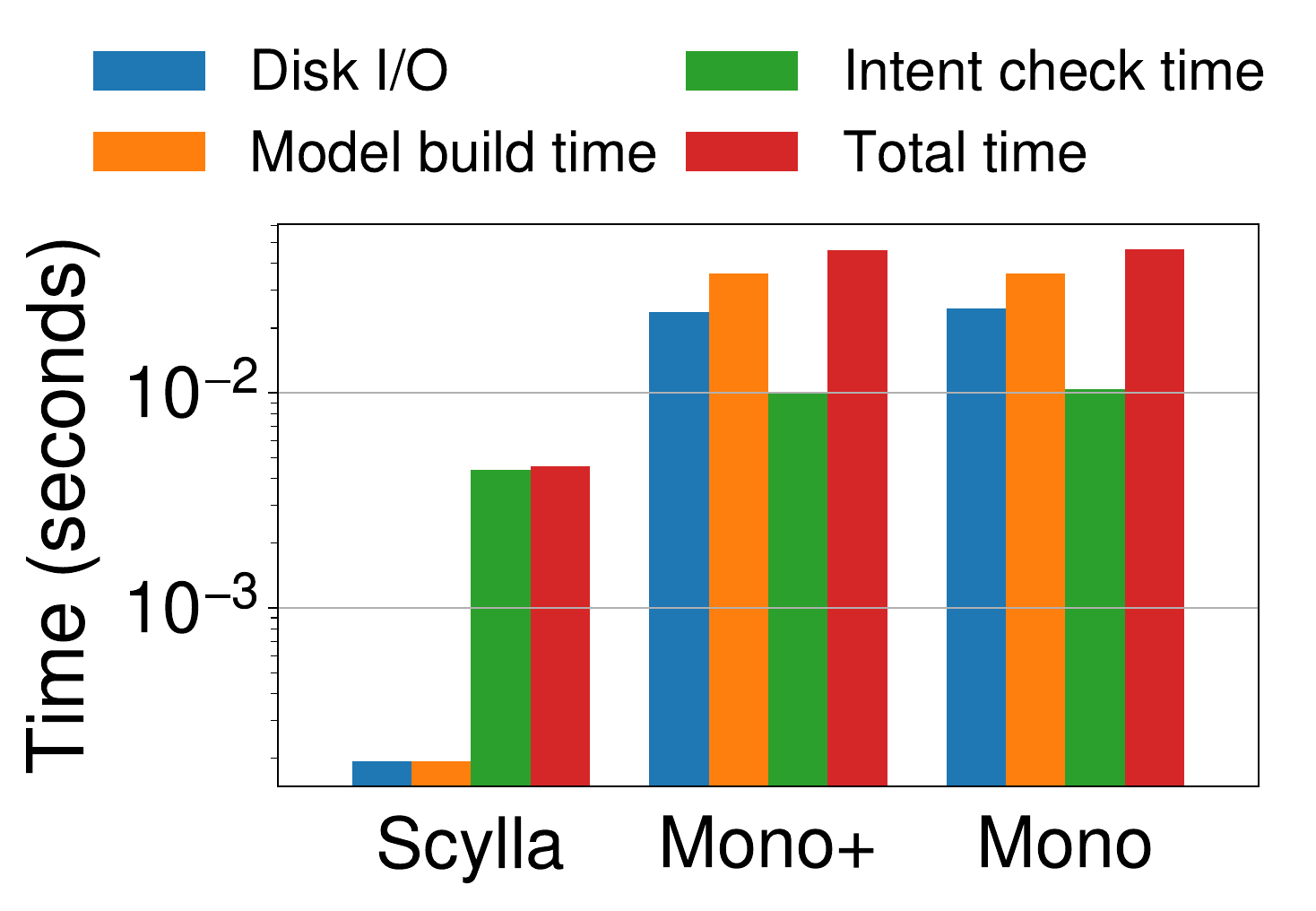}
        \caption{Time (N2)}
    \end{subfigure}
    \hfill
    \begin{subfigure}[t]{0.32\linewidth}
        \includegraphics[width=\linewidth]{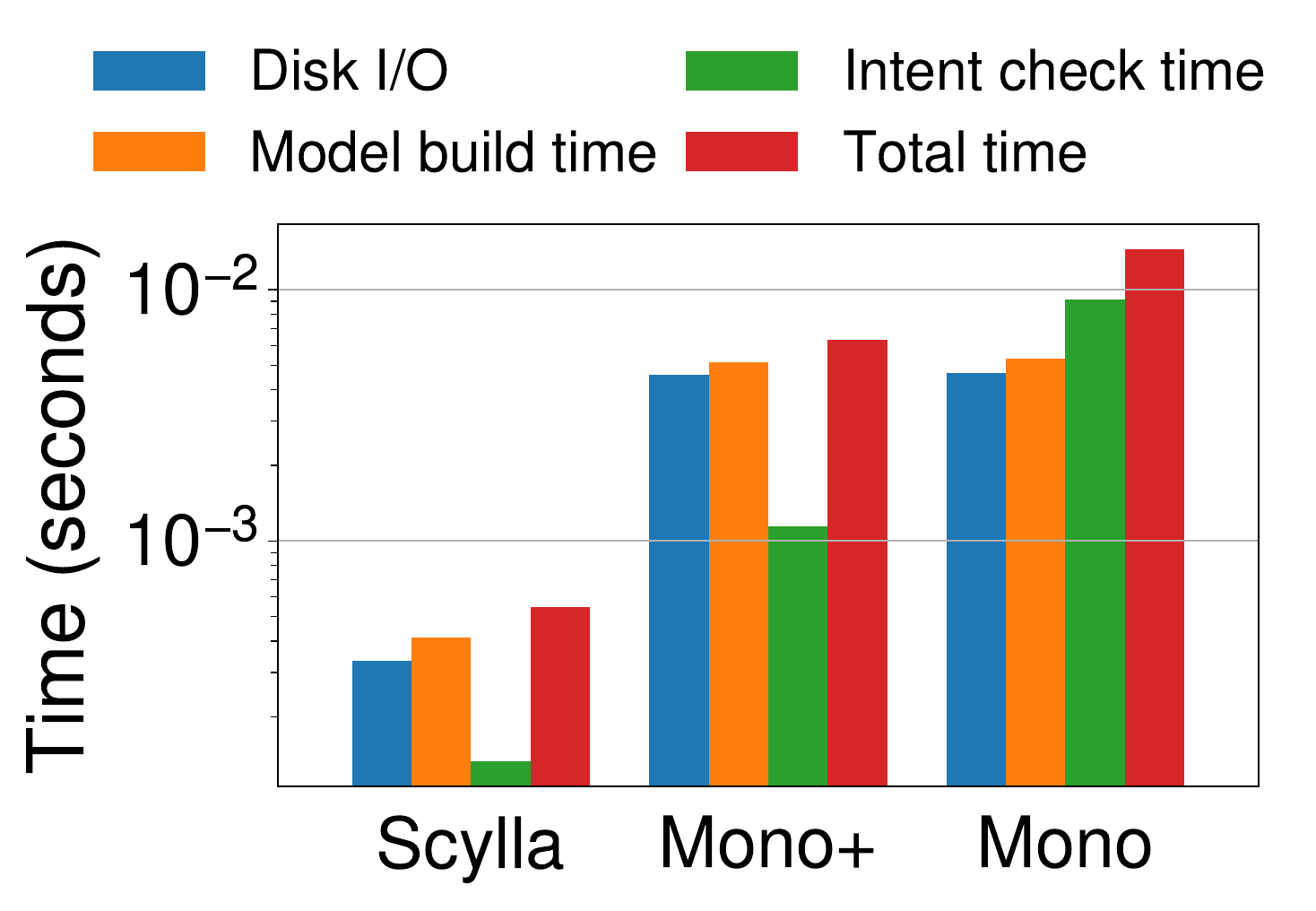}
        \caption{Time (N3)}
    \end{subfigure}
    \hfill
    \begin{subfigure}[t]{0.32\linewidth}
        \includegraphics[width=\linewidth]{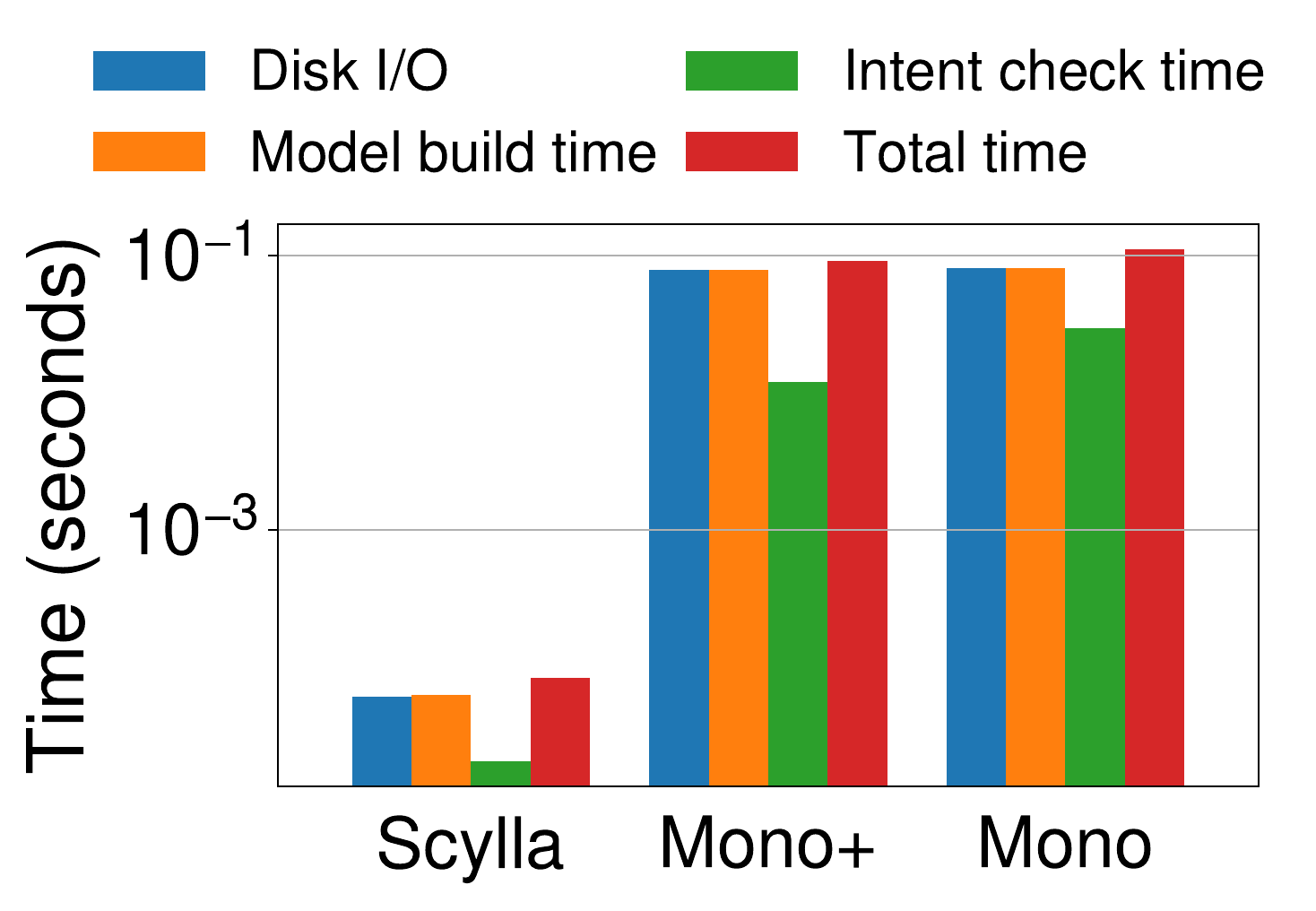}
        \caption{Time (N4)}
    \end{subfigure}
    \hfill
    \begin{subfigure}[t]{0.32\linewidth}
        \includegraphics[width=\linewidth]{single-intent/single-node.time.reachability.update.N5.pdf}
        \caption{Time (N5)}
    \end{subfigure}
    \hfill
    \begin{subfigure}[t]{0.32\linewidth}
        \includegraphics[width=\linewidth]{single-intent/single-node.memory.reachability.update.pdf}
        \caption{Memory (all networks)}
    \end{subfigure}
    \caption{Performance for verifying an existing intent with network updates (log scale).}
    \label{app:fig:single-node.reachability.update}
\end{figure*}

Figure~\ref{app:fig:single-node.reachability.first-run} and
\ref{app:fig:single-node.reachability.update} present the full results for
\S\ref{subsubsec:single-intent}.

\subsection{Multi-intent local microbenchmarks}
\label{app:subsec:multi-intent}

\begin{figure*}[ht]
    \centering
    \begin{subfigure}[t]{0.195\linewidth}
        \includegraphics[width=\linewidth]{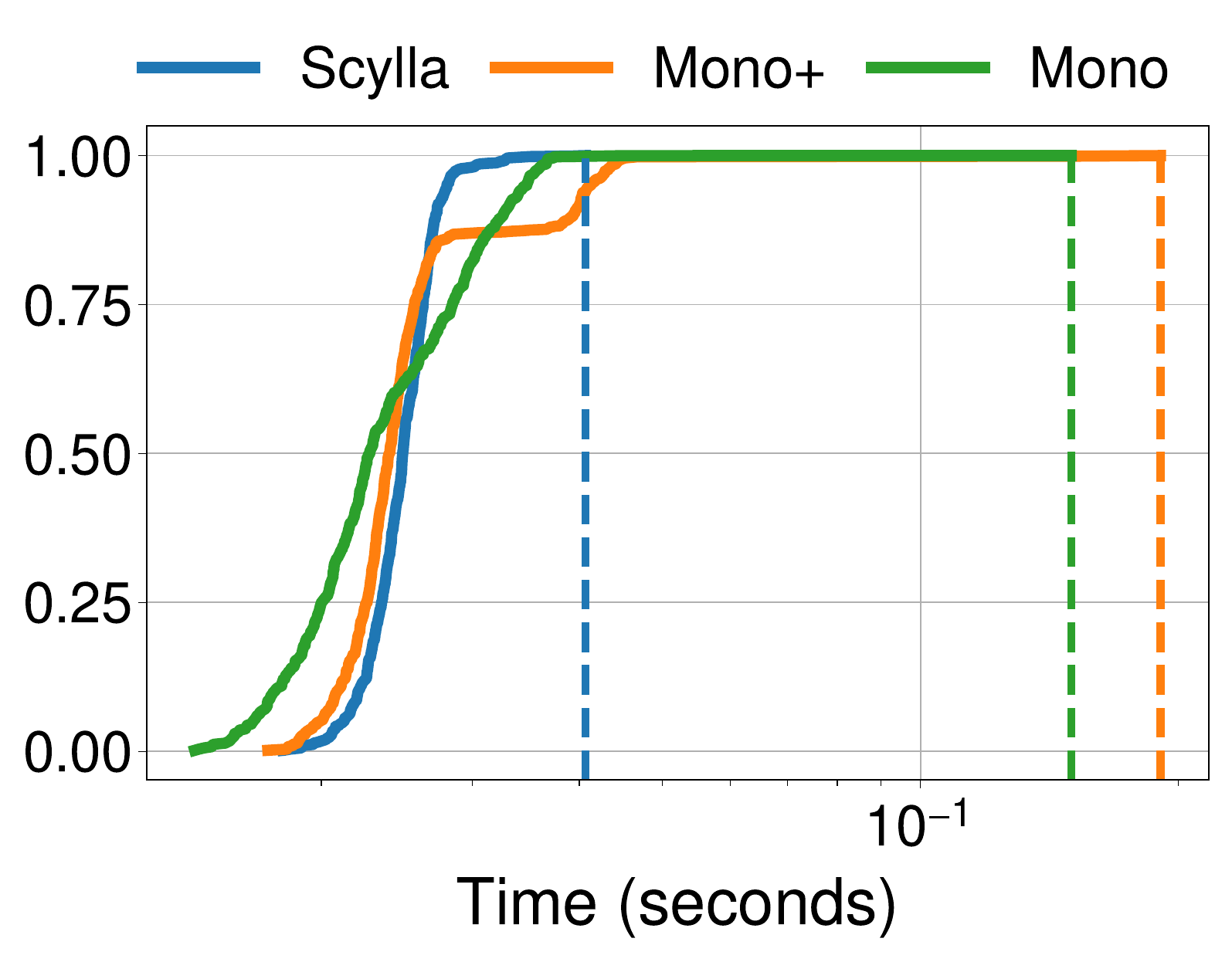}
        \caption{Time CDF (N1)}
    \end{subfigure}
    \hfill
    \begin{subfigure}[t]{0.195\linewidth}
        \includegraphics[width=\linewidth]{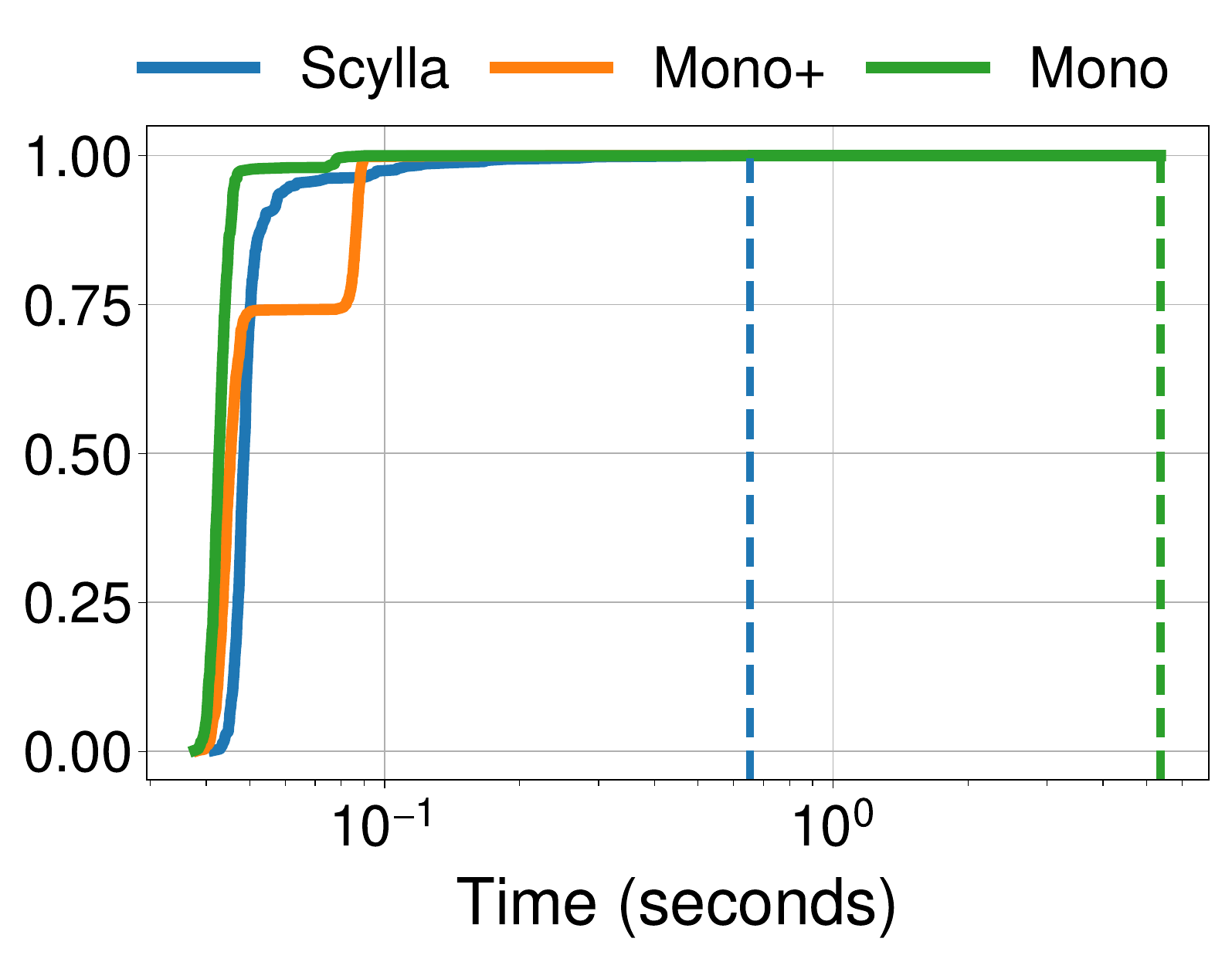}
        \caption{Time CDF (N2)}
    \end{subfigure}
    \hfill
    \begin{subfigure}[t]{0.195\linewidth}
        \includegraphics[width=\linewidth]{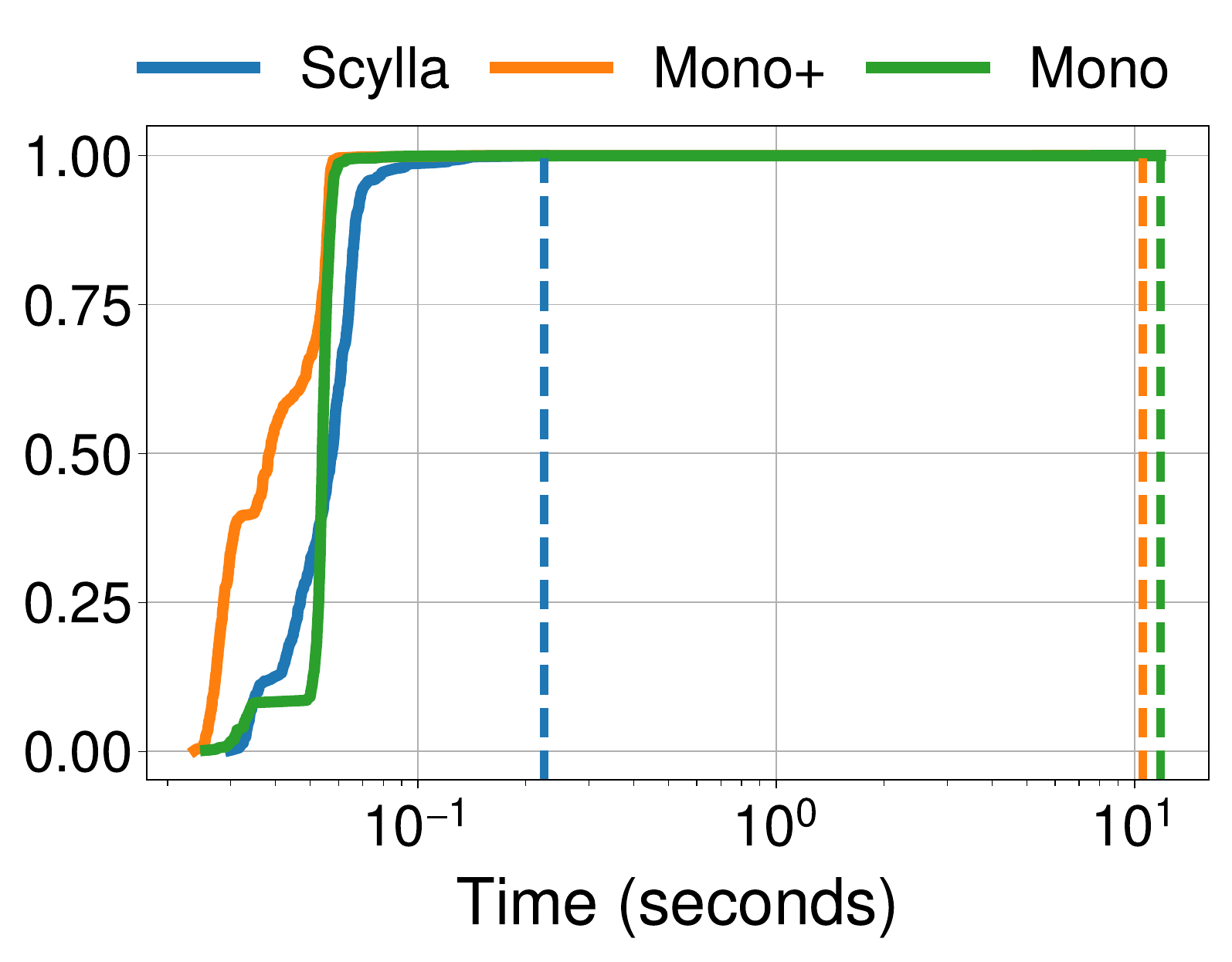}
        \caption{Time CDF (N3)}
    \end{subfigure}
    \hfill
    \begin{subfigure}[t]{0.195\linewidth}
        \includegraphics[width=\linewidth]{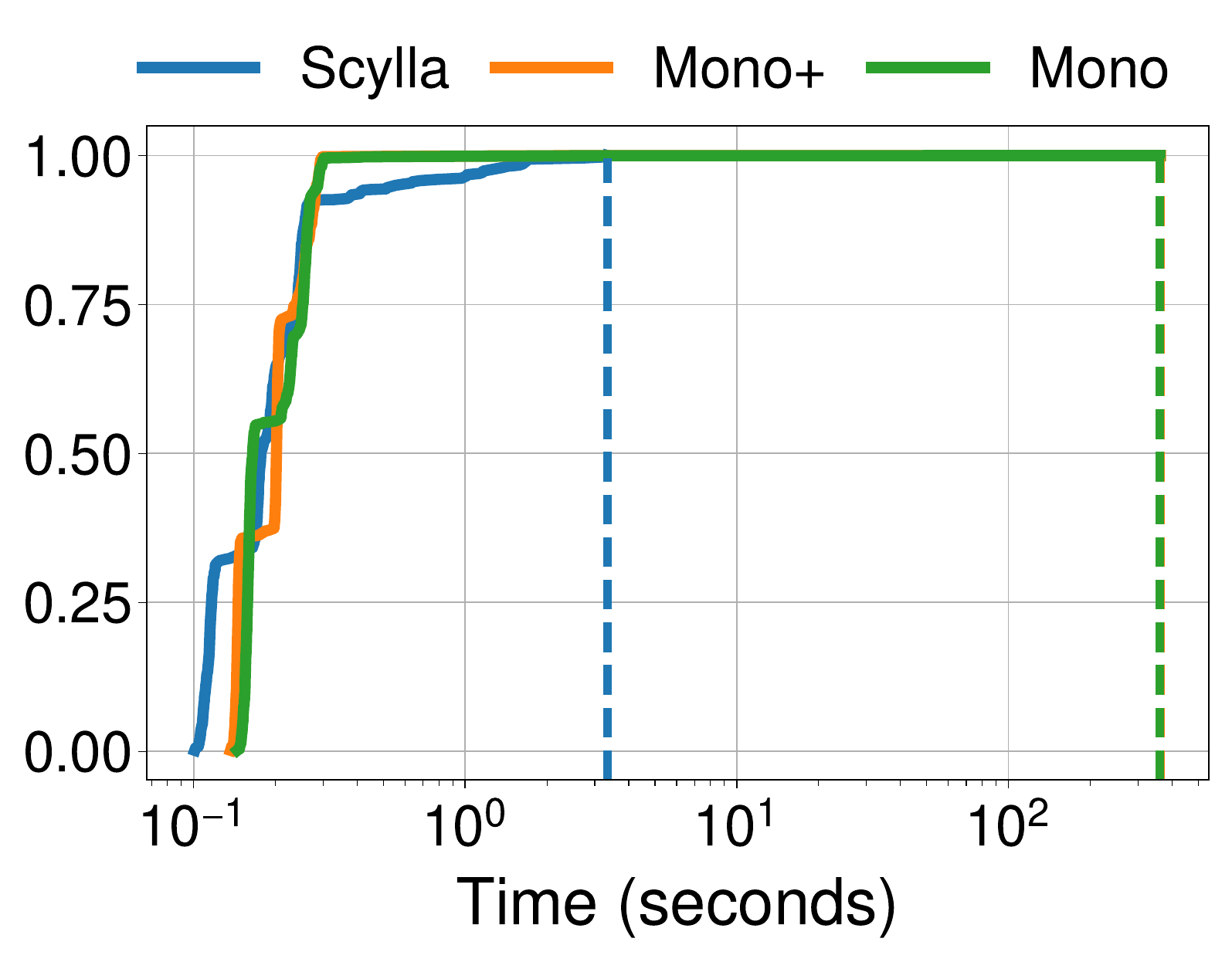}
        \caption{Time CDF (N4)}
    \end{subfigure}
    \hfill
    \begin{subfigure}[t]{0.195\linewidth}
        \includegraphics[width=\linewidth]{multi-intent/time.first-run.N5.pdf}
        \caption{Time CDF (N5)}
    \end{subfigure}
    \hfill
    \begin{subfigure}[t]{0.195\linewidth}
        \includegraphics[width=\linewidth]{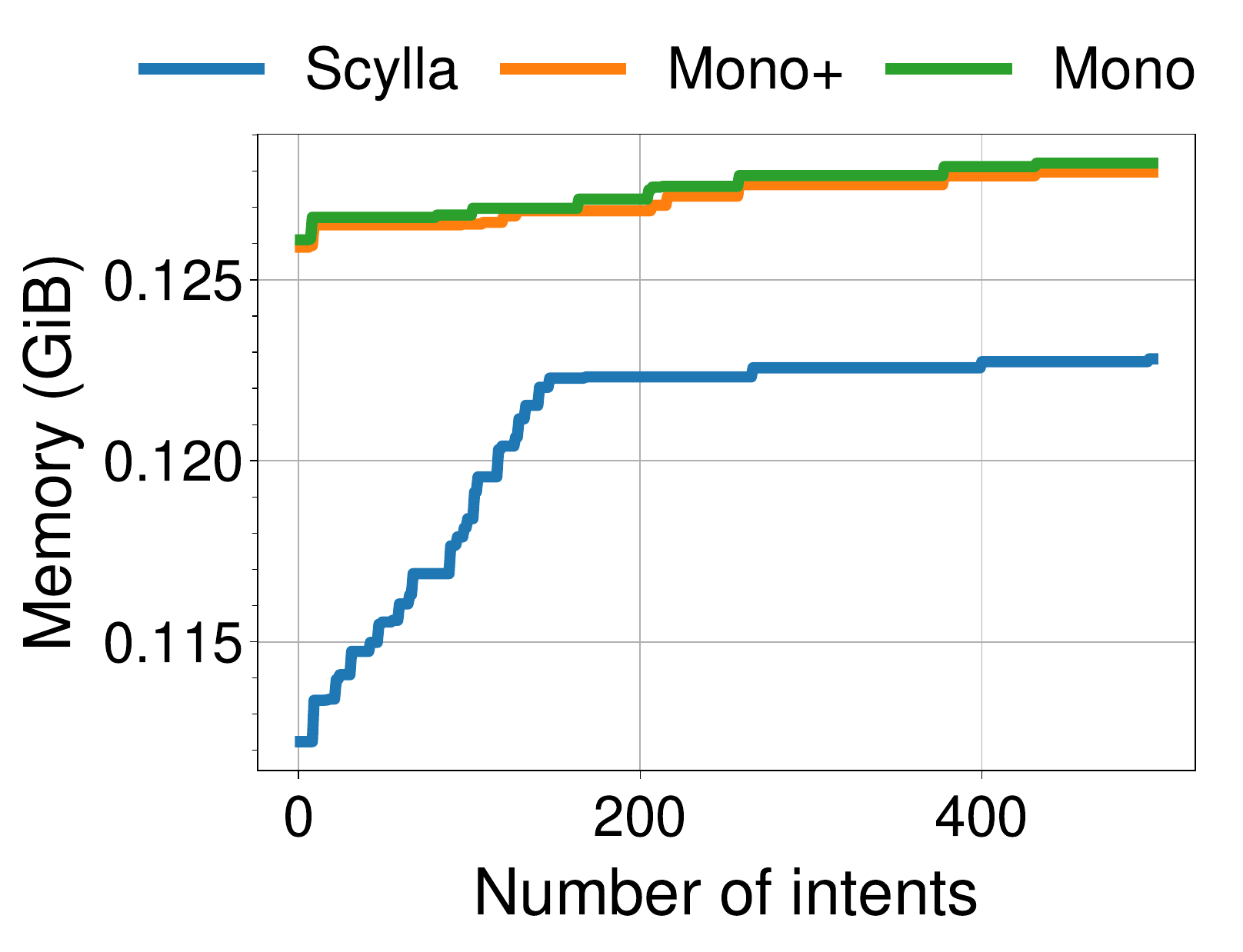}
        \caption{Memory (N1)}
    \end{subfigure}
    \hfill
    \begin{subfigure}[t]{0.195\linewidth}
        \includegraphics[width=\linewidth]{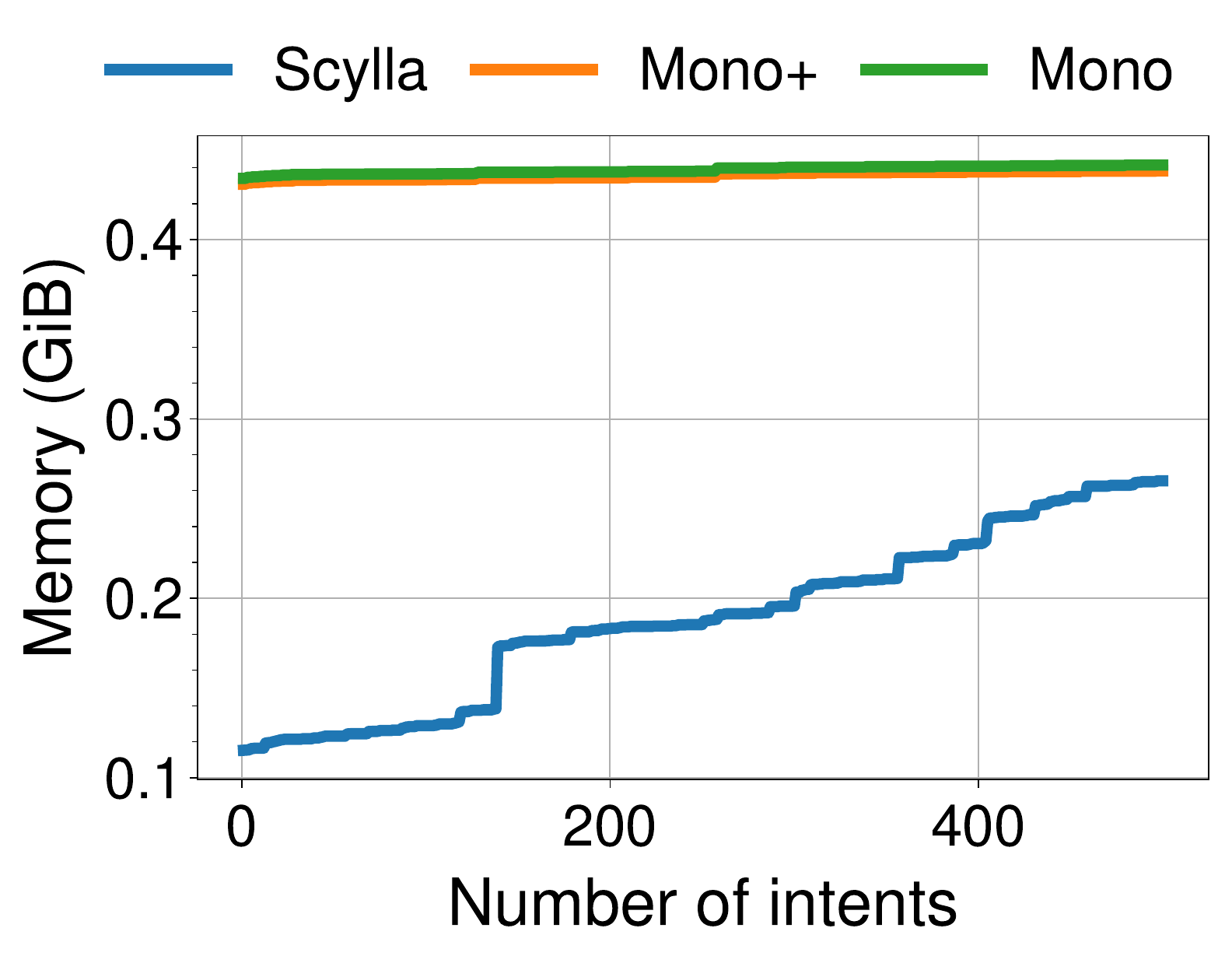}
        \caption{Memory (N2)}
    \end{subfigure}
    \hfill
    \begin{subfigure}[t]{0.195\linewidth}
        \includegraphics[width=\linewidth]{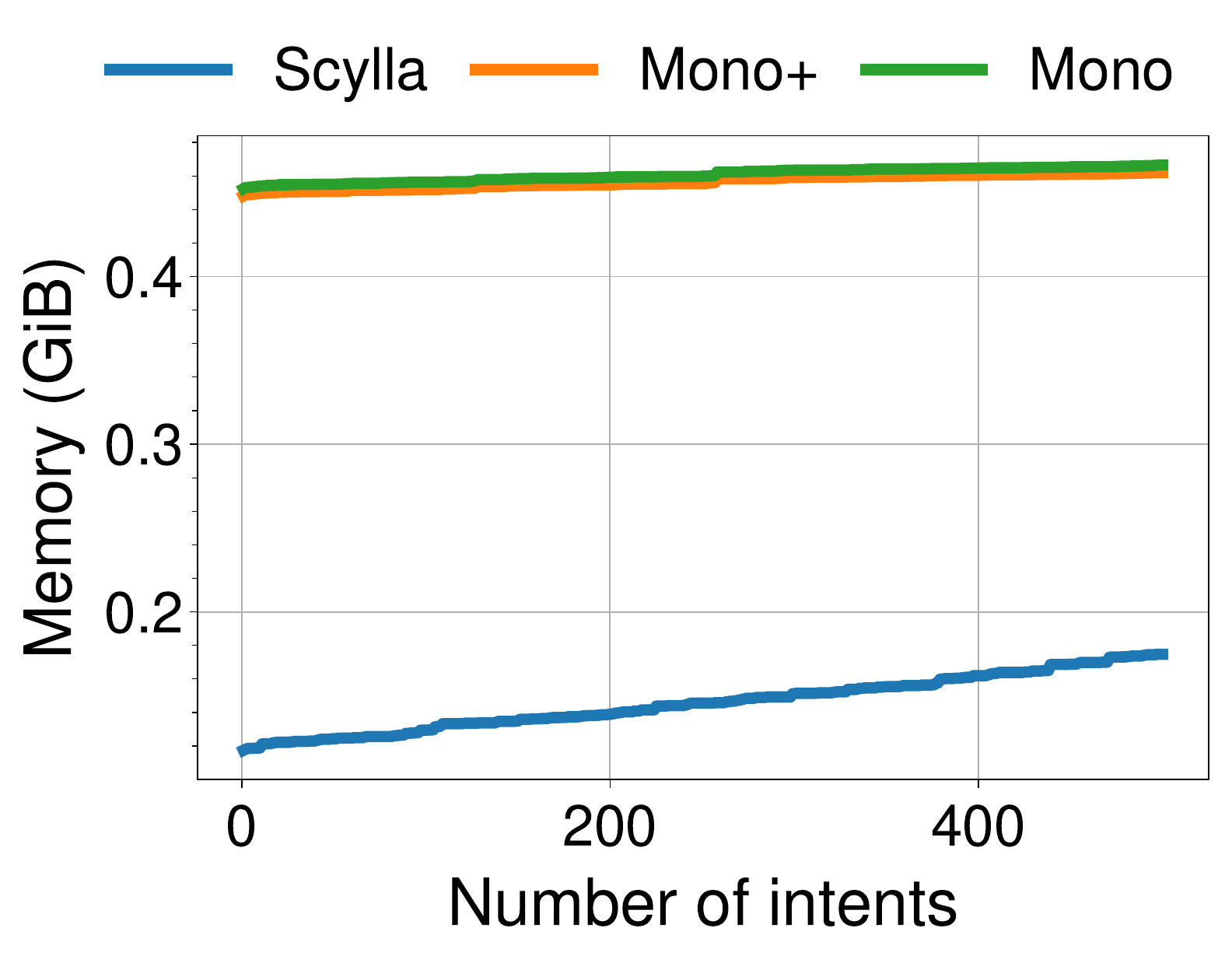}
        \caption{Memory (N3)}
    \end{subfigure}
    \hfill
    \begin{subfigure}[t]{0.195\linewidth}
        \includegraphics[width=\linewidth]{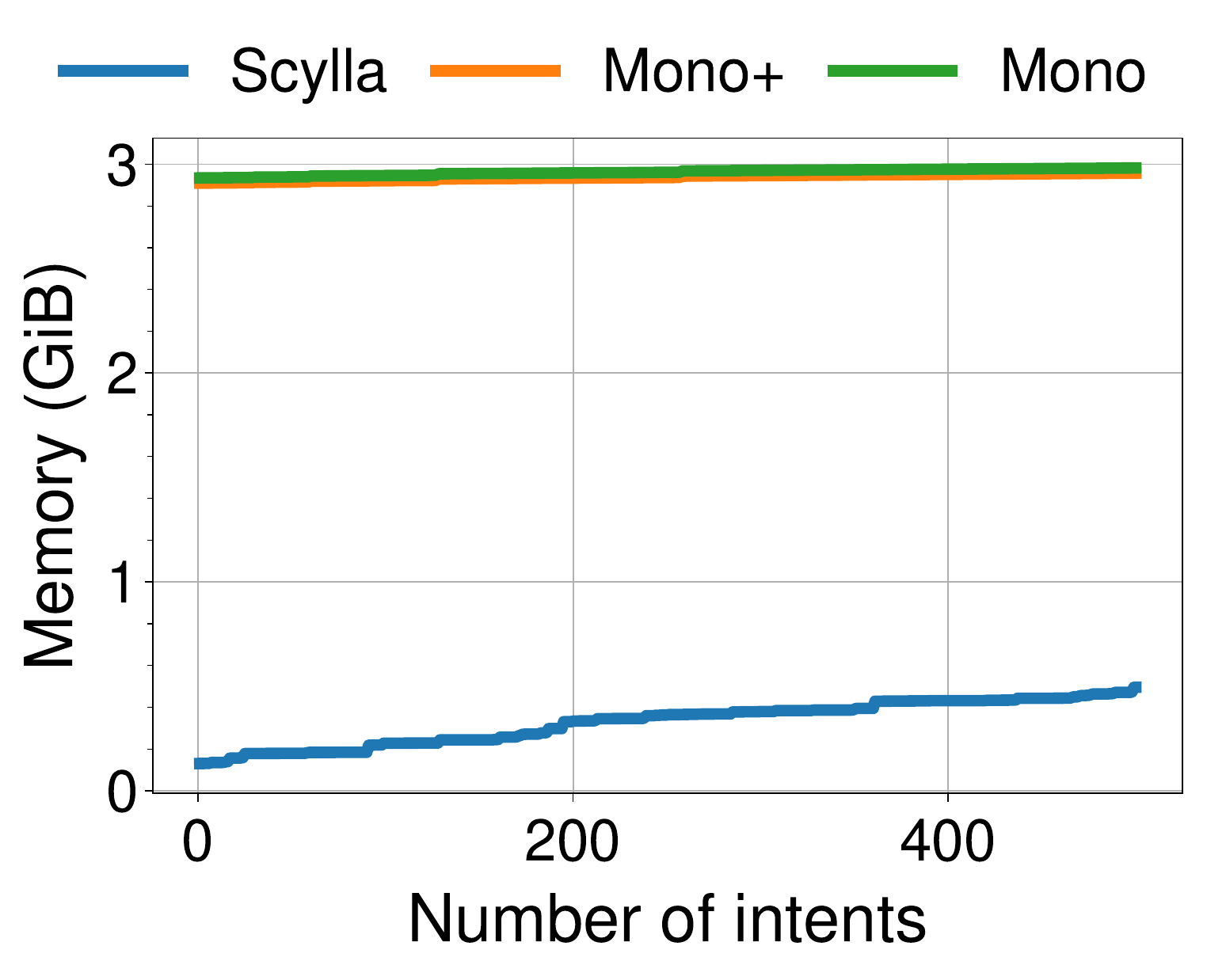}
        \caption{Memory (N4)}
    \end{subfigure}
    \hfill
    \begin{subfigure}[t]{0.195\linewidth}
        \includegraphics[width=\linewidth]{multi-intent/memory.first-run.N5.pdf}
        \caption{Memory (N5)}
    \end{subfigure}
    \caption{Performance for verifying 500 new intents from scratch (time in log scale).}
    \label{app:fig:multi-intent.first-run}
\end{figure*}

\begin{figure*}[ht]
    \centering
    \begin{subfigure}[t]{0.195\linewidth}
        \includegraphics[width=\linewidth]{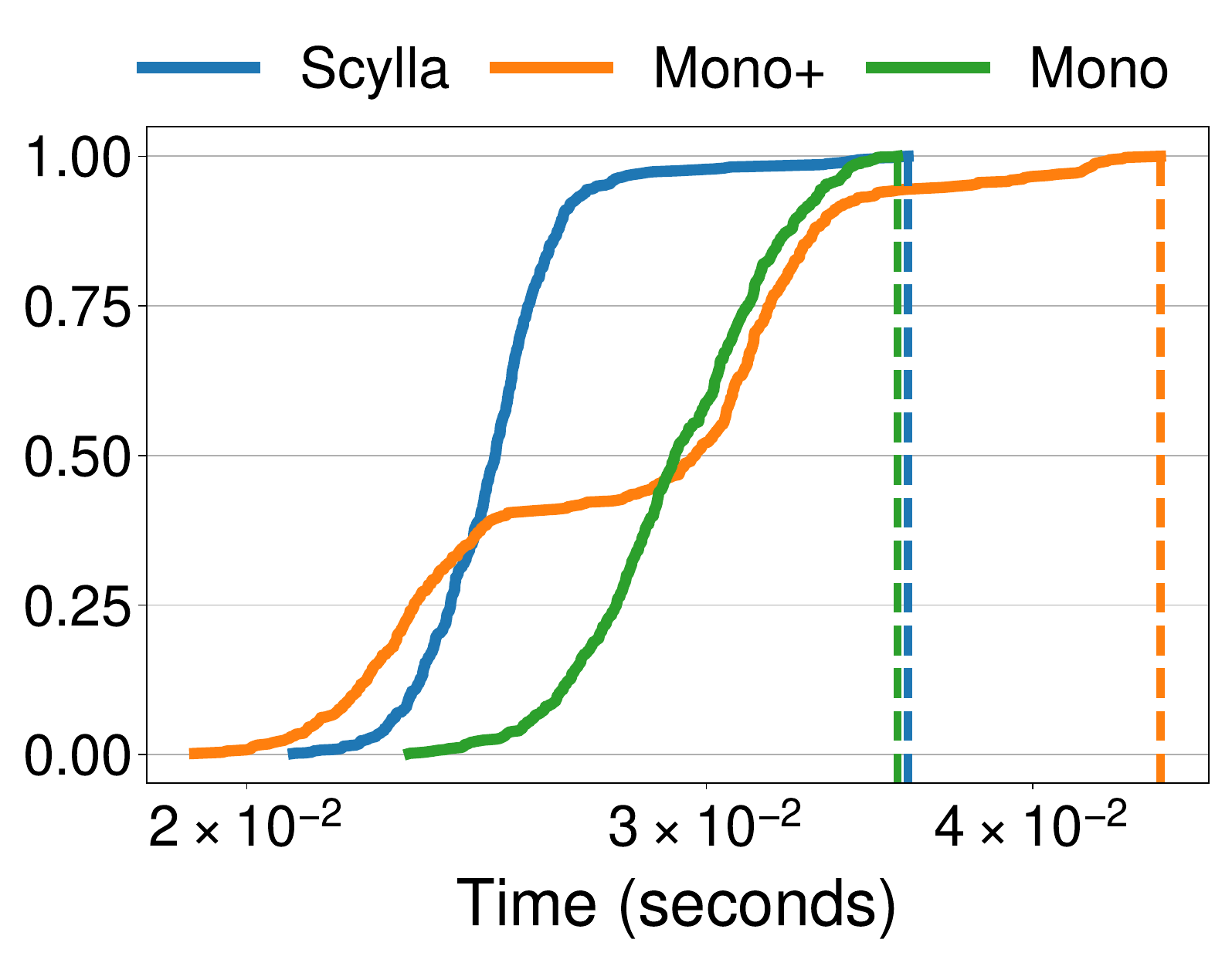}
        \caption{Time CDF (N1)}
    \end{subfigure}
    \hfill
    \begin{subfigure}[t]{0.195\linewidth}
        \includegraphics[width=\linewidth]{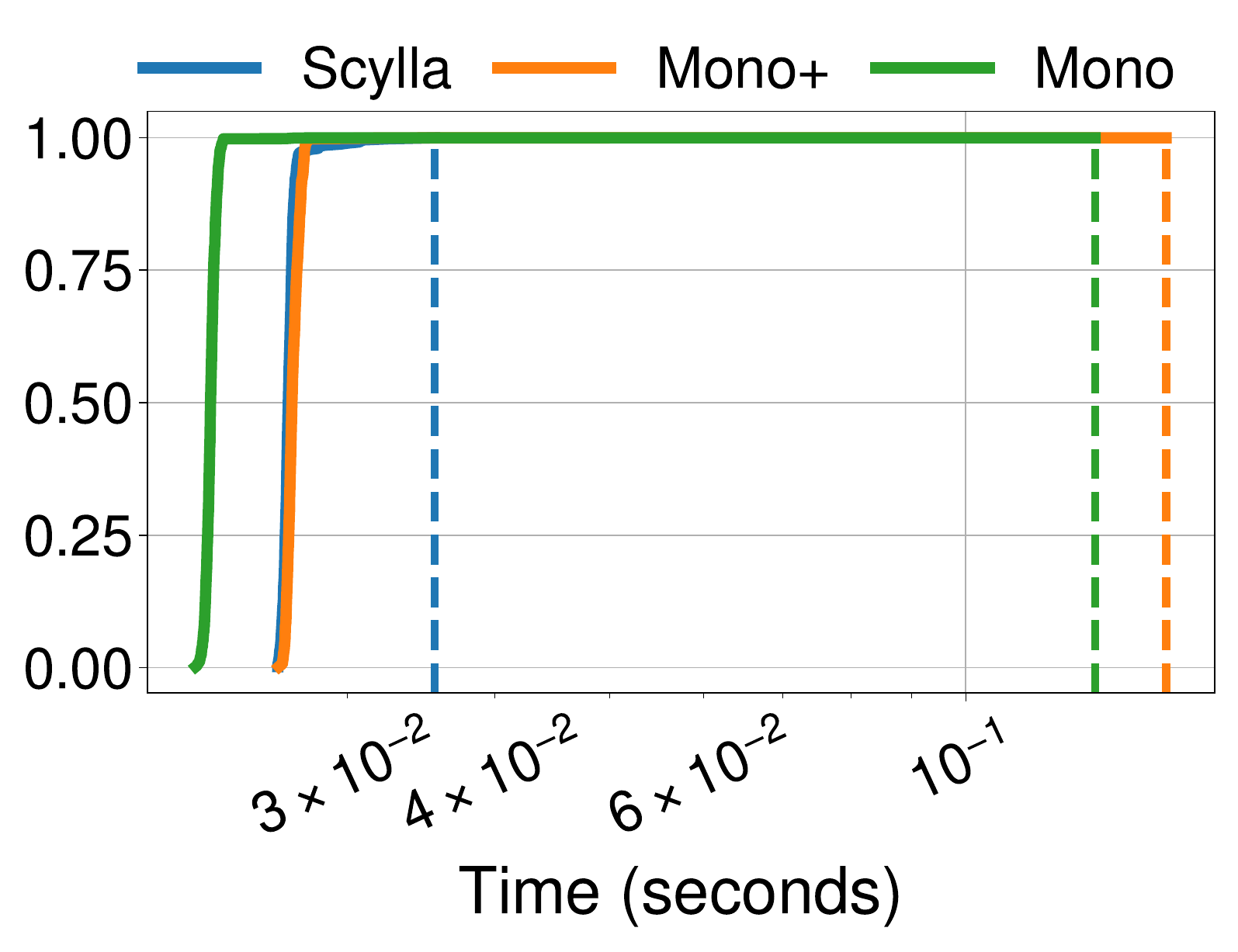}
        \caption{Time CDF (N2)}
    \end{subfigure}
    \hfill
    \begin{subfigure}[t]{0.195\linewidth}
        \includegraphics[width=\linewidth]{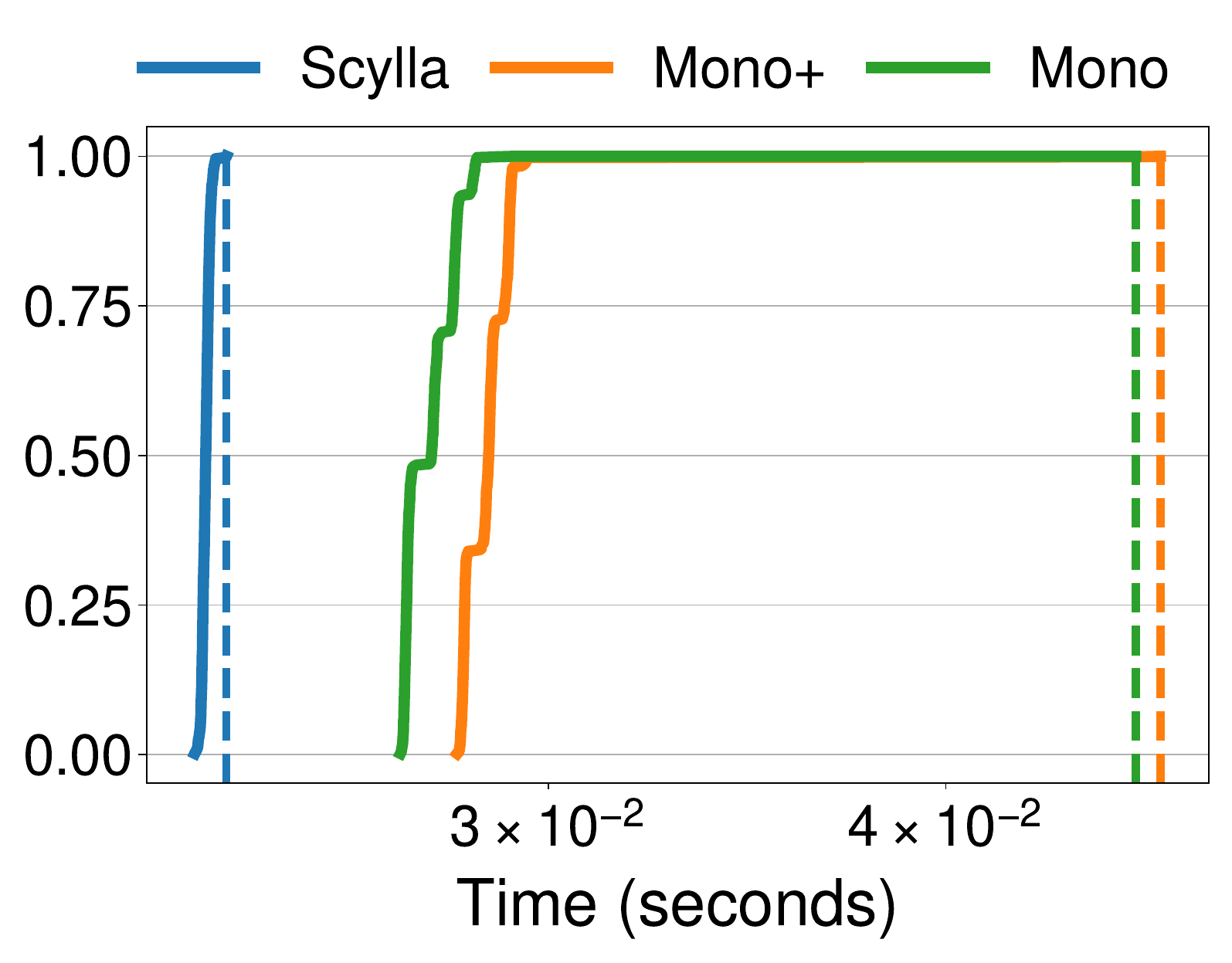}
        \caption{Time CDF (N3)}
    \end{subfigure}
    \hfill
    \begin{subfigure}[t]{0.195\linewidth}
        \includegraphics[width=\linewidth]{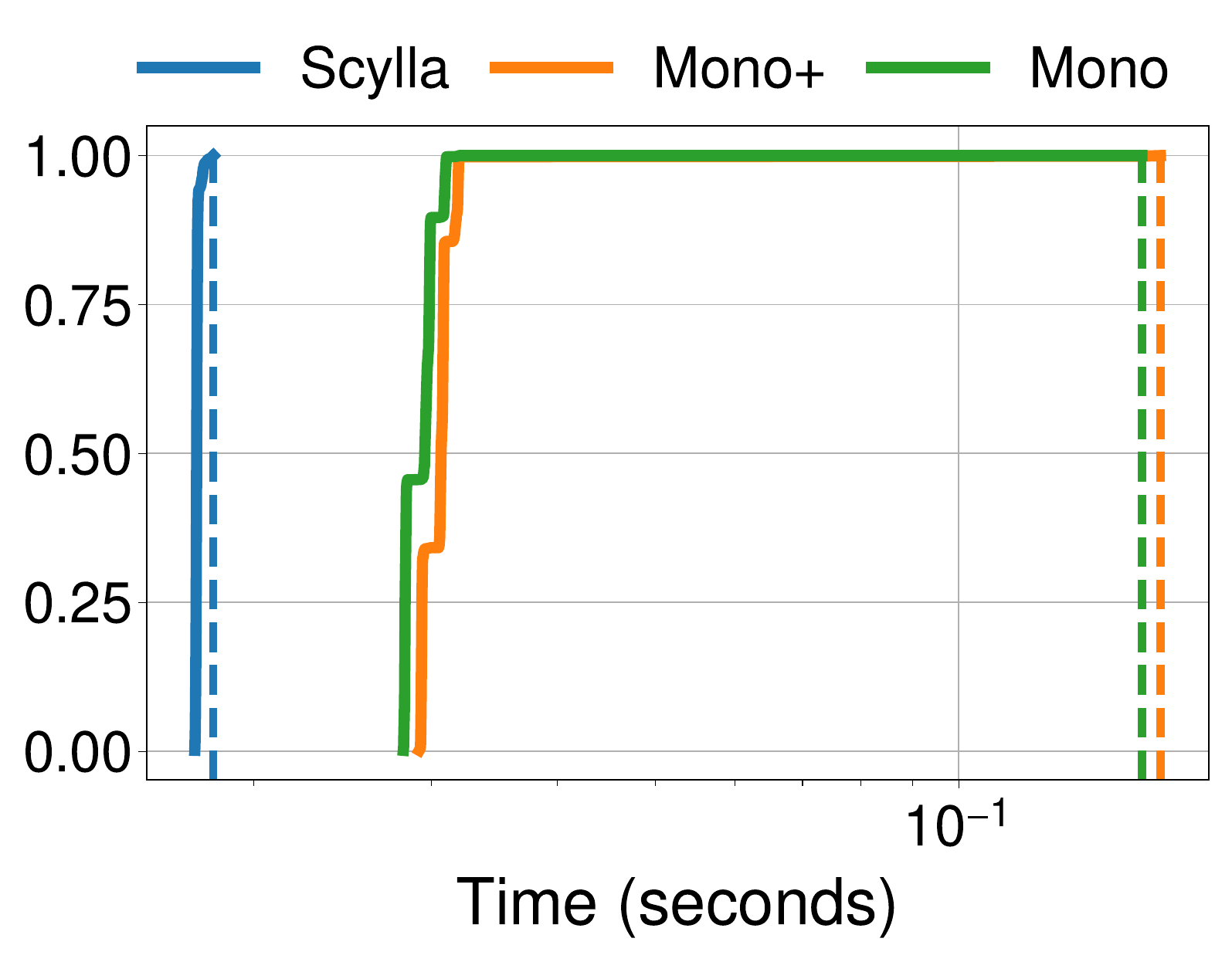}
        \caption{Time CDF (N4)}
    \end{subfigure}
    \hfill
    \begin{subfigure}[t]{0.195\linewidth}
        \includegraphics[width=\linewidth]{multi-intent/time.update.N5.pdf}
        \caption{Time CDF (N5)}
    \end{subfigure}
    \hfill
    \begin{subfigure}[t]{0.195\linewidth}
        \includegraphics[width=\linewidth]{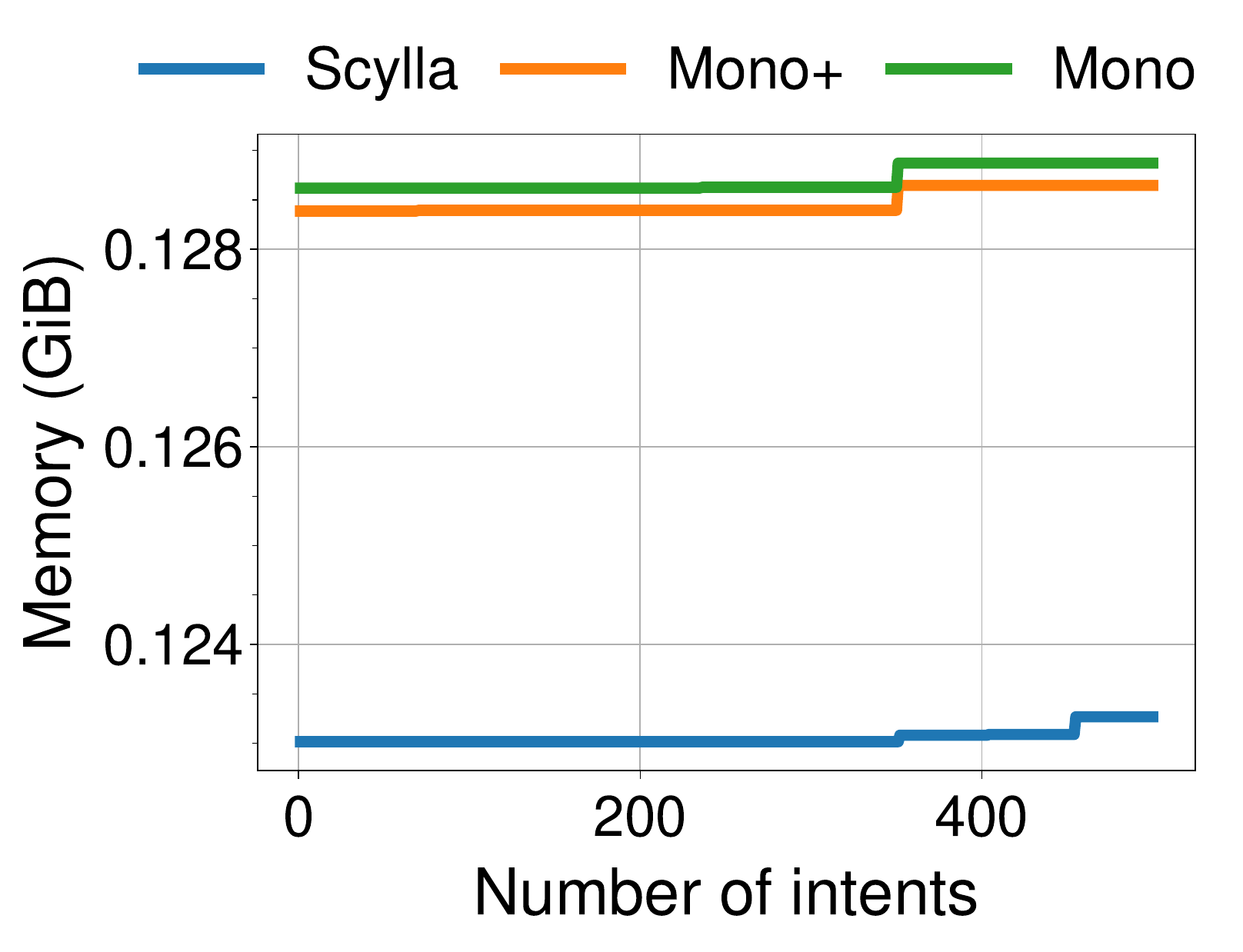}
        \caption{Memory (N1)}
    \end{subfigure}
    \hfill
    \begin{subfigure}[t]{0.195\linewidth}
        \includegraphics[width=\linewidth]{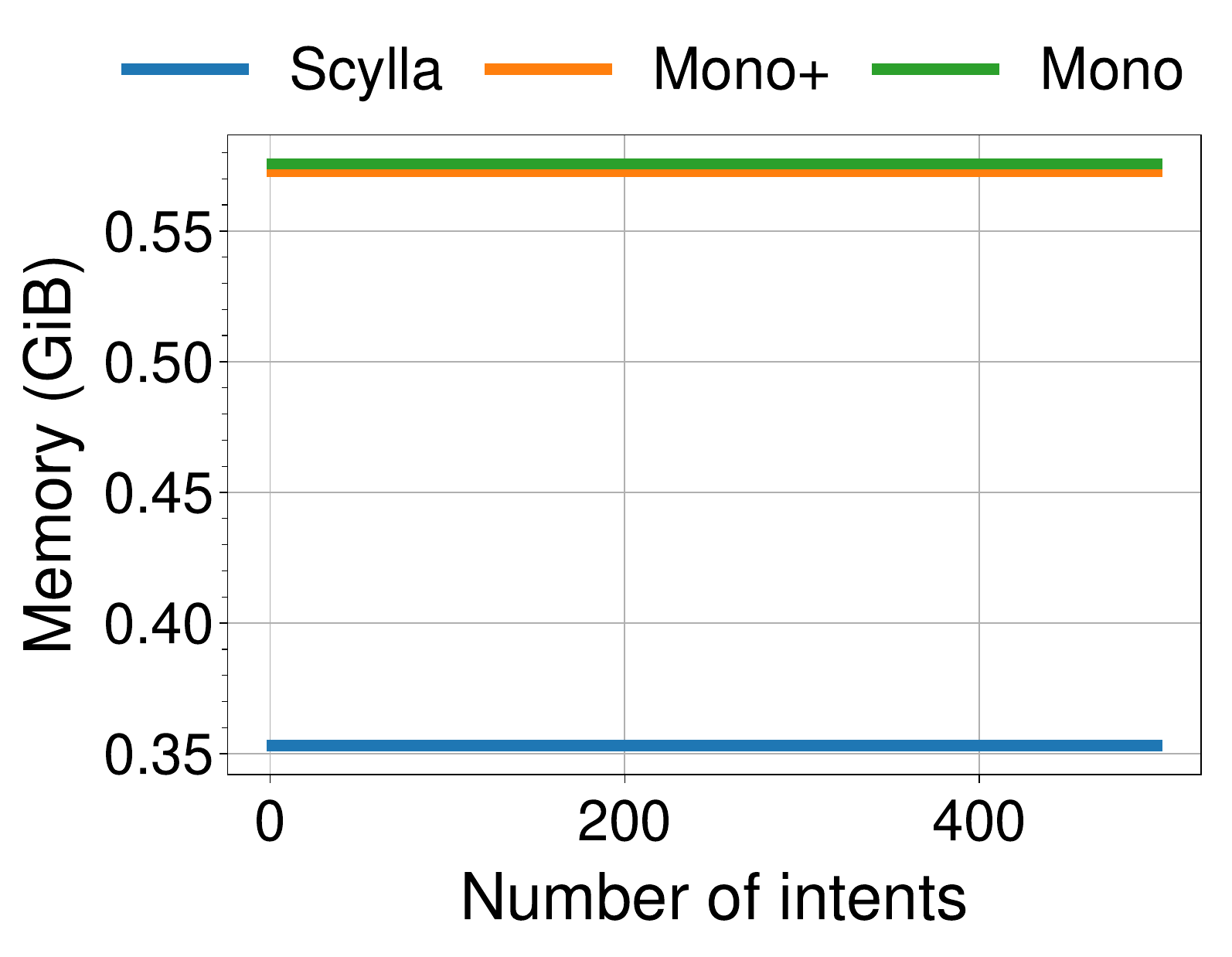}
        \caption{Memory (N2)}
    \end{subfigure}
    \hfill
    \begin{subfigure}[t]{0.195\linewidth}
        \includegraphics[width=\linewidth]{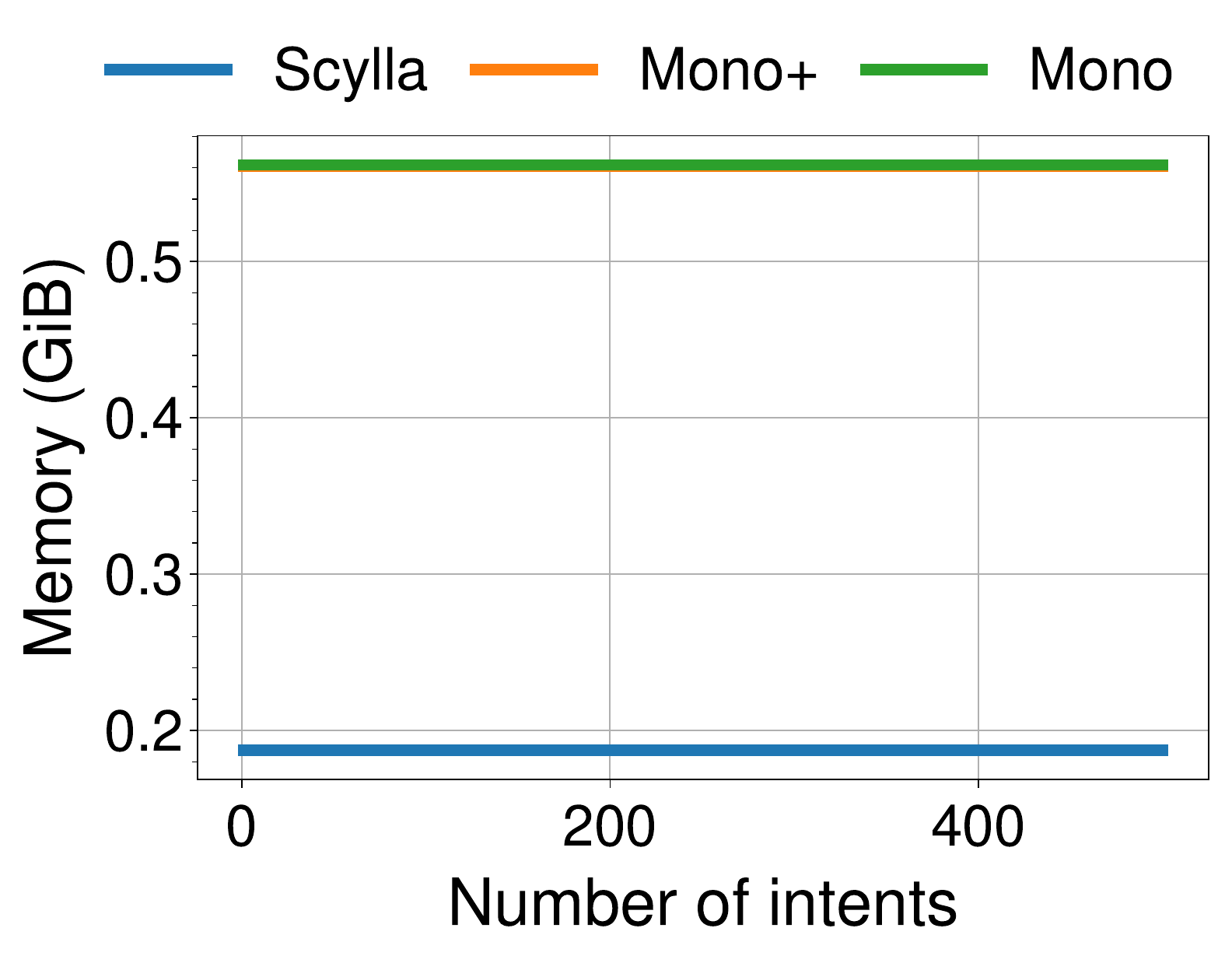}
        \caption{Memory (N3)}
    \end{subfigure}
    \hfill
    \begin{subfigure}[t]{0.195\linewidth}
        \includegraphics[width=\linewidth]{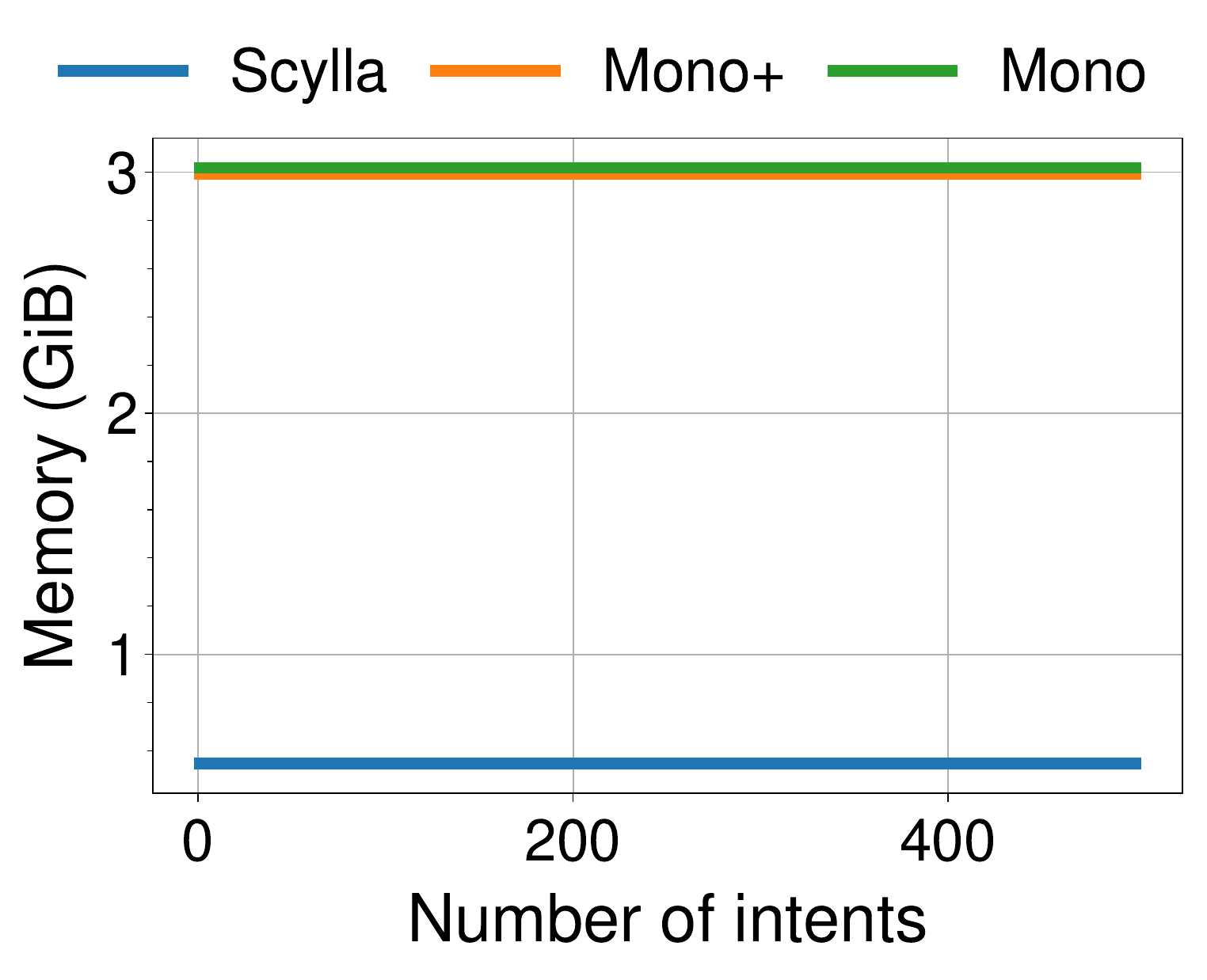}
        \caption{Memory (N4)}
    \end{subfigure}
    \hfill
    \begin{subfigure}[t]{0.195\linewidth}
        \includegraphics[width=\linewidth]{multi-intent/memory.update.N5.pdf}
        \caption{Memory (N5)}
    \end{subfigure}
    \caption{Performance for maintaining 500 intents across network updates (time in log scale).}
    \label{app:fig:multi-intent.update}
\end{figure*}

Figure~\ref{app:fig:multi-intent.first-run} and
\ref{app:fig:multi-intent.update} present the full results for
\S\ref{subsubsec:multi-intent}.

\subsection{Libra comparison}
\label{app:subsec:libra-compare}

\begin{figure*}[ht]
    \centering
    \begin{subfigure}[t]{0.195\linewidth}
        \includegraphics[width=\linewidth]{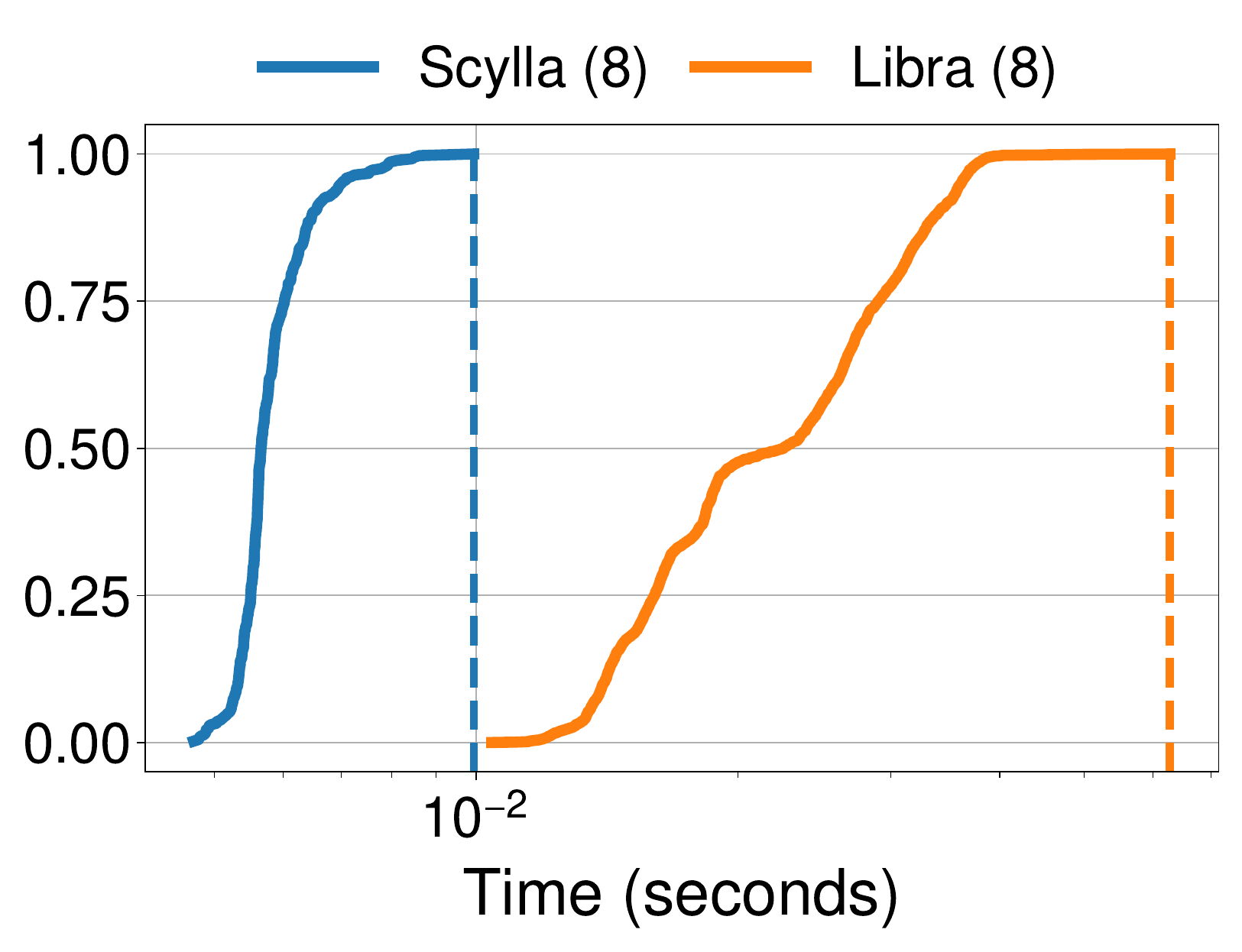}
        \caption{Time CDF (N1 underlay)}
    \end{subfigure}
    \hfill
    \begin{subfigure}[t]{0.195\linewidth}
        \includegraphics[width=\linewidth]{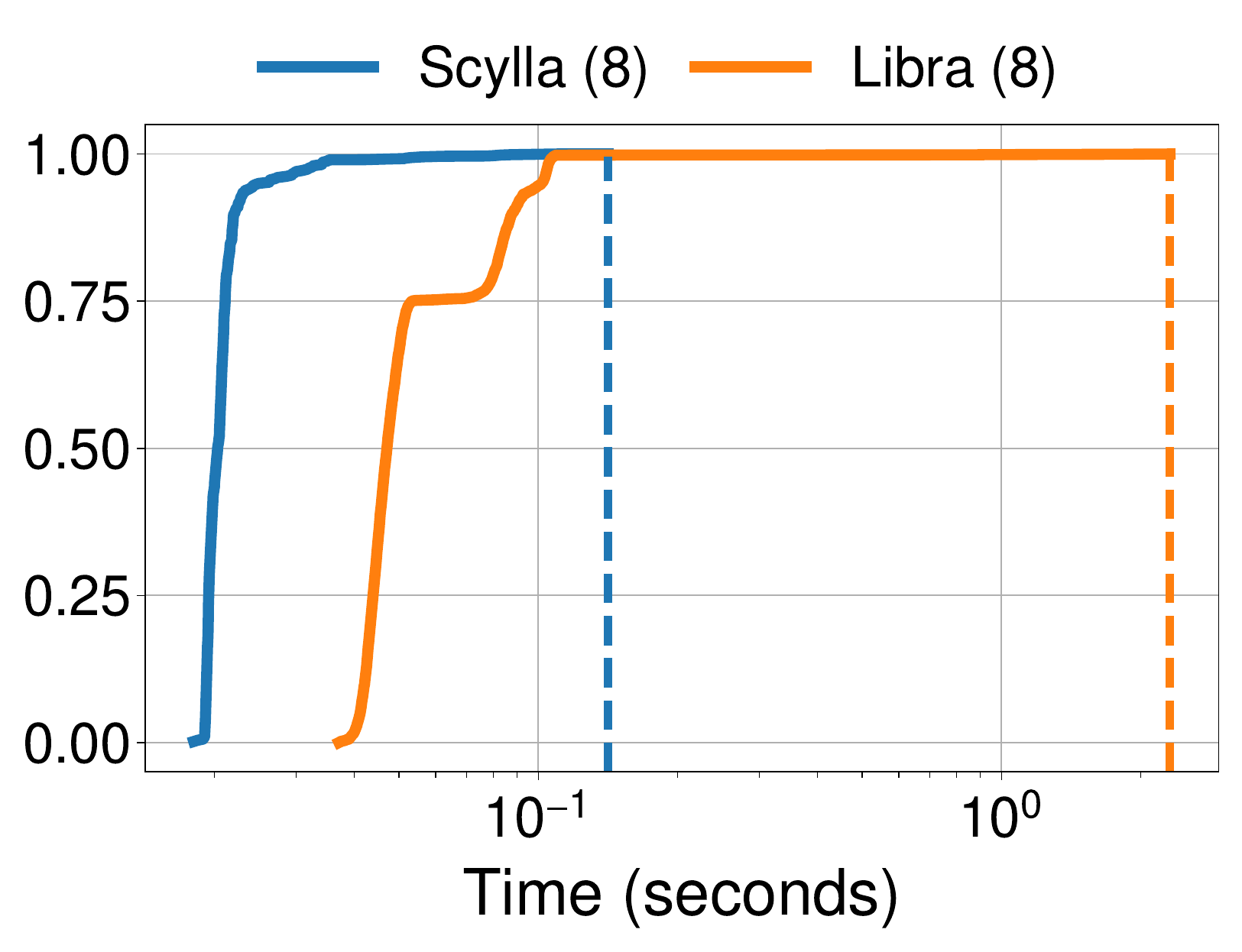}
        \caption{Time CDF (N2 underlay)}
    \end{subfigure}
    \hfill
    \begin{subfigure}[t]{0.195\linewidth}
        \includegraphics[width=\linewidth]{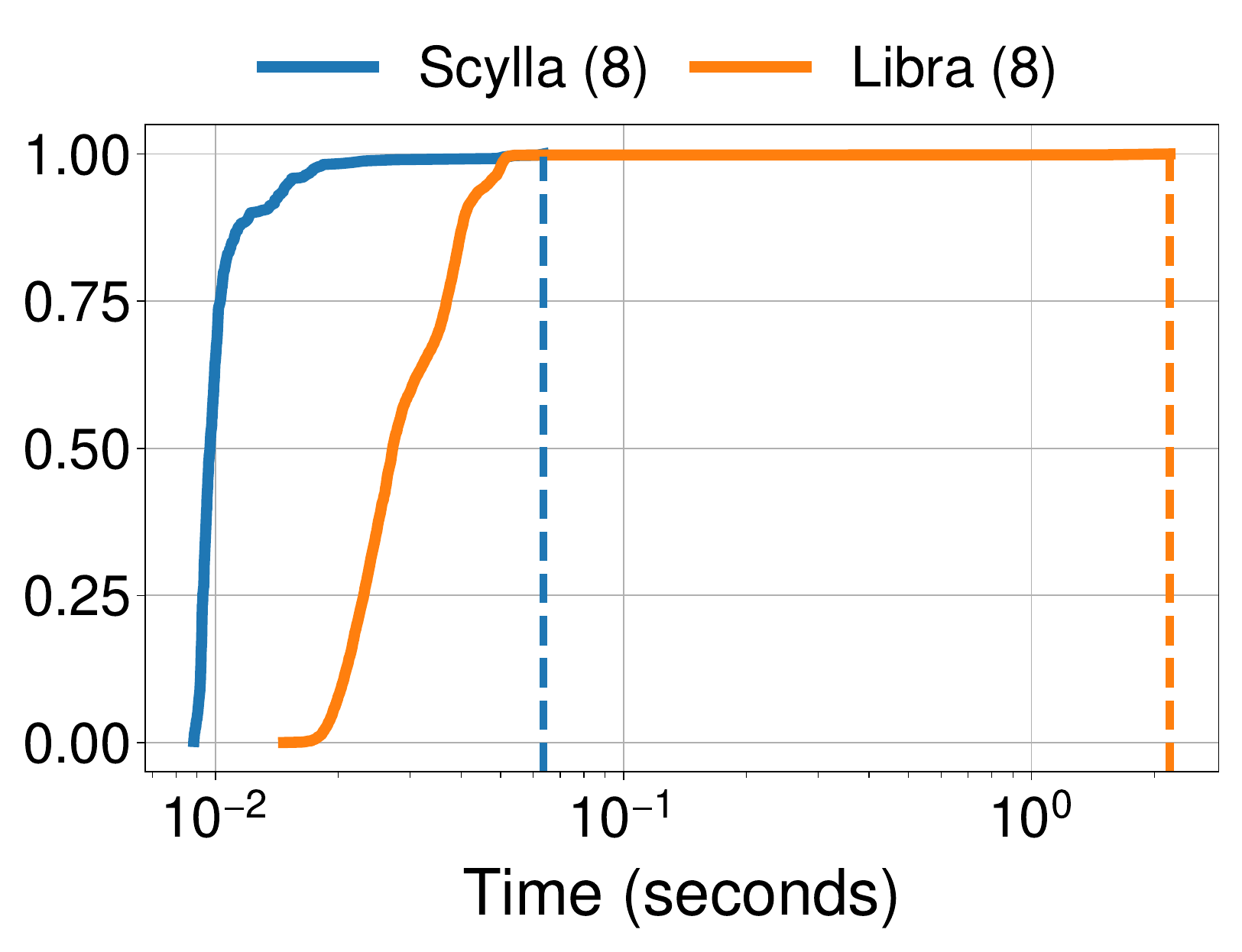}
        \caption{Time CDF (N3 underlay)}
    \end{subfigure}
    \hfill
    \begin{subfigure}[t]{0.195\linewidth}
        \includegraphics[width=\linewidth]{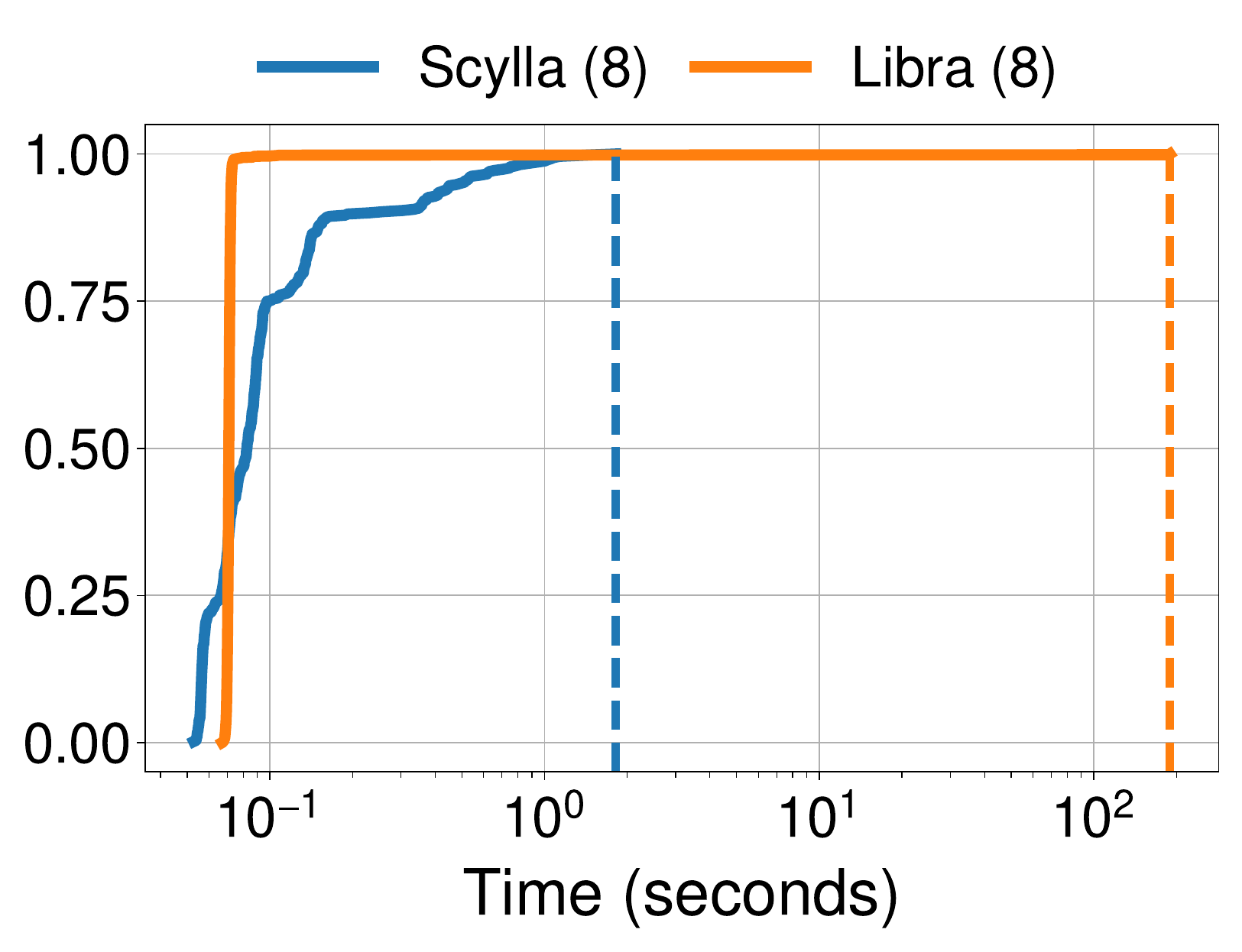}
        \caption{Time CDF (N4 underlay)}
    \end{subfigure}
    \hfill
    \begin{subfigure}[t]{0.195\linewidth}
        \includegraphics[width=\linewidth]{multi-intent/time.first-run.N5-underlay.pdf}
        \caption{Time CDF (N5 underlay)}
    \end{subfigure}
    \hfill
    \begin{subfigure}[t]{0.195\linewidth}
        \includegraphics[width=\linewidth]{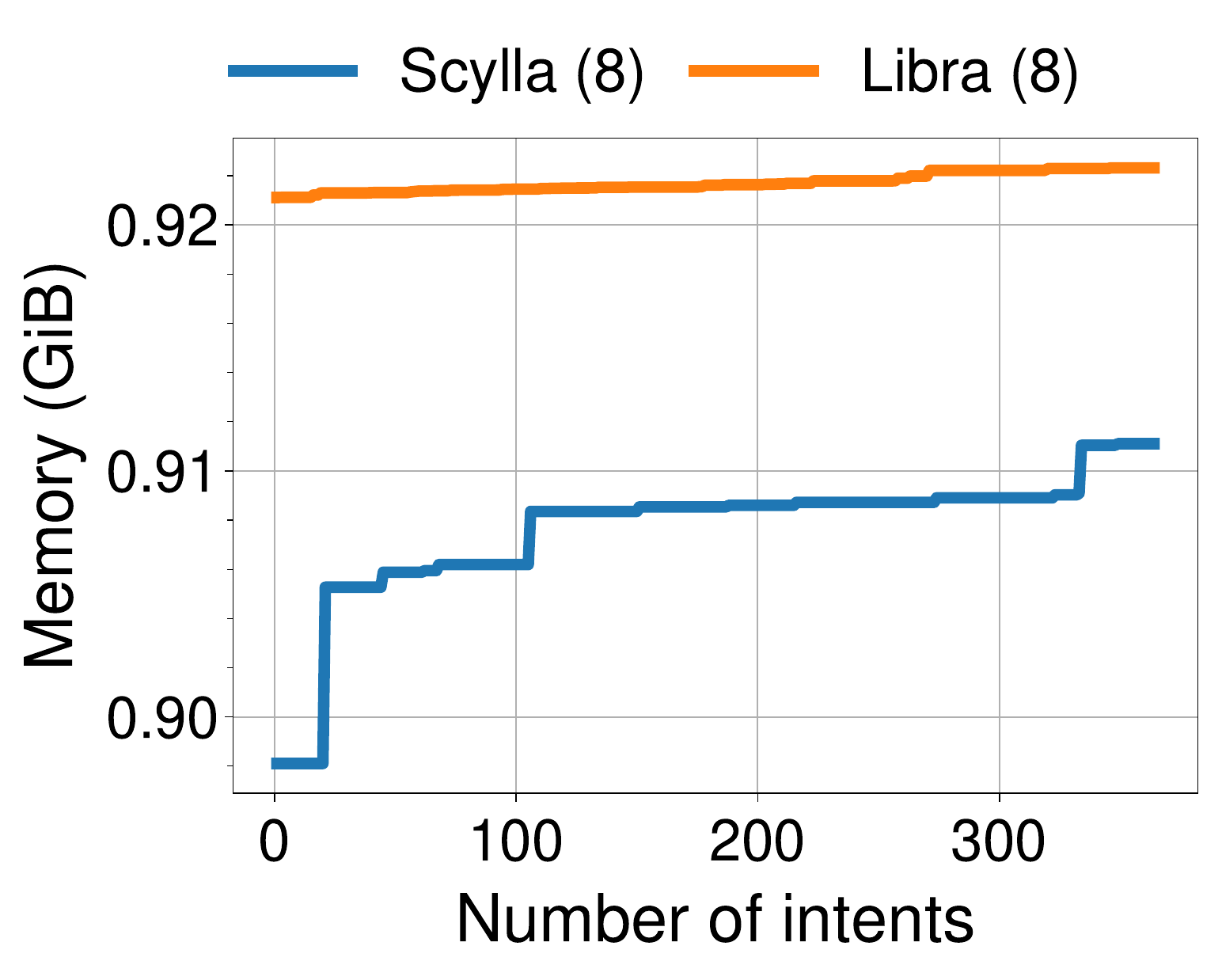}
        \caption{Memory (N1 underlay)}
    \end{subfigure}
    \hfill
    \begin{subfigure}[t]{0.195\linewidth}
        \includegraphics[width=\linewidth]{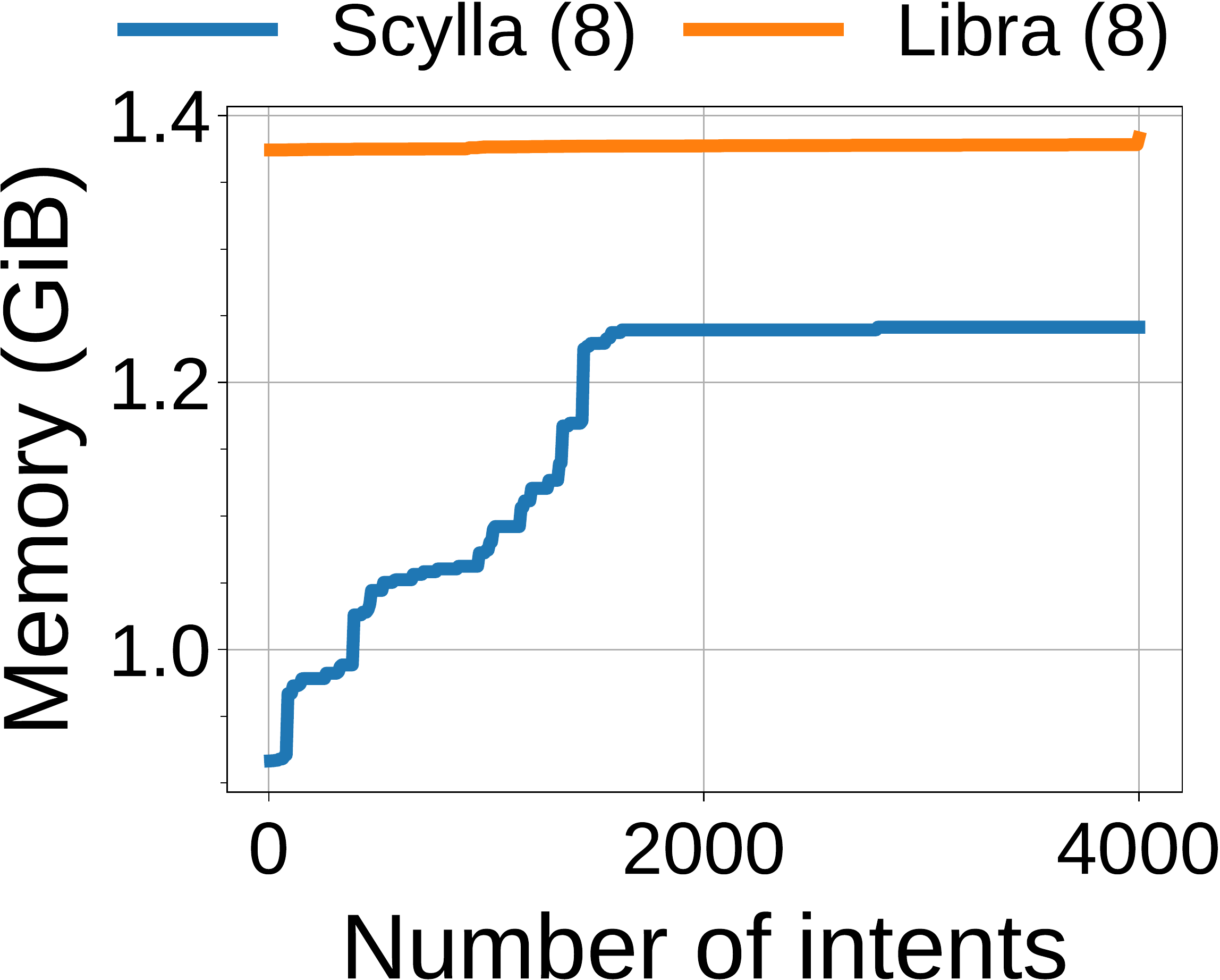}
        \caption{Memory (N2 underlay)}
    \end{subfigure}
    \hfill
    \begin{subfigure}[t]{0.195\linewidth}
        \includegraphics[width=\linewidth]{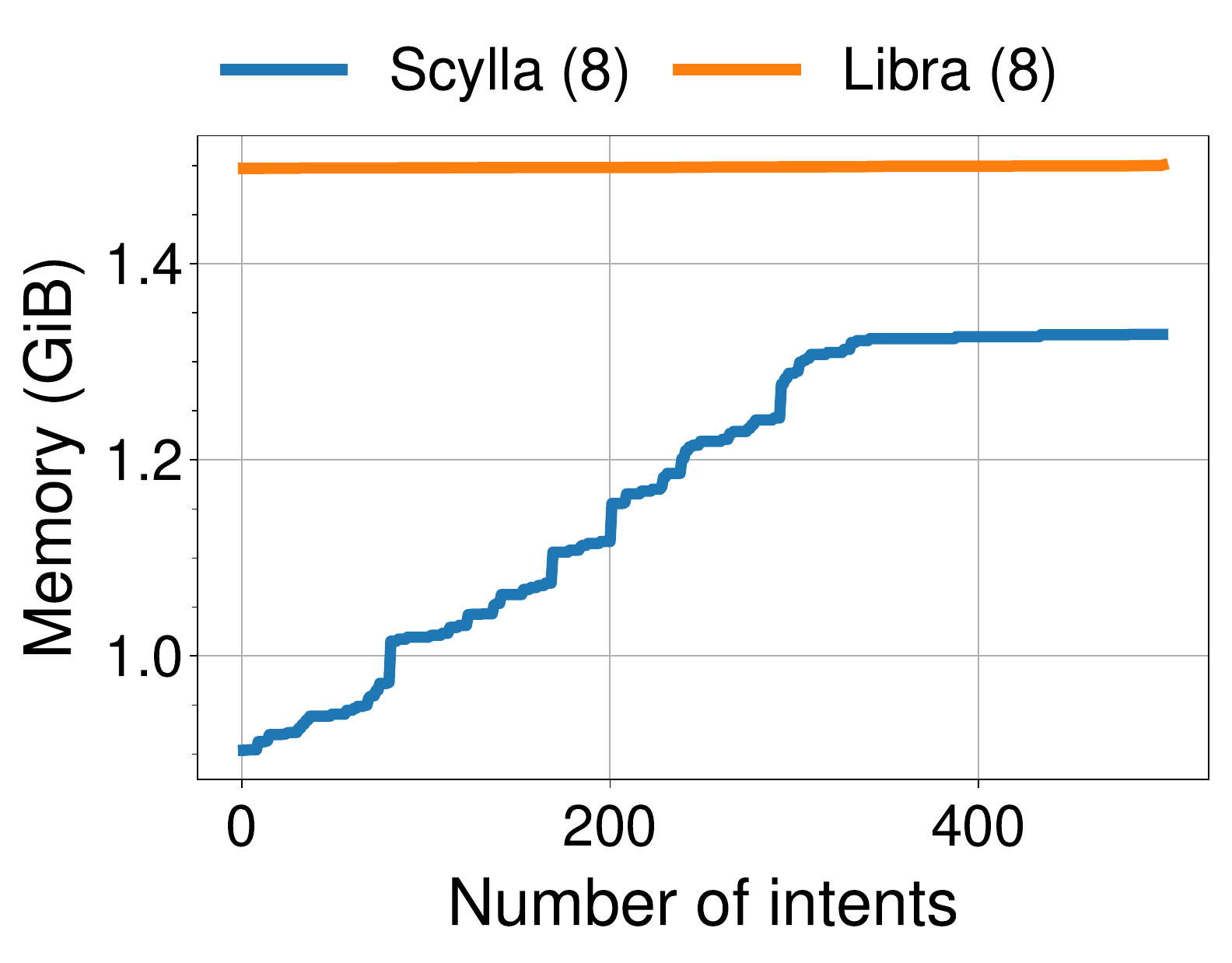}
        \caption{Memory (N3 underlay)}
    \end{subfigure}
    \hfill
    \begin{subfigure}[t]{0.195\linewidth}
        \includegraphics[width=\linewidth]{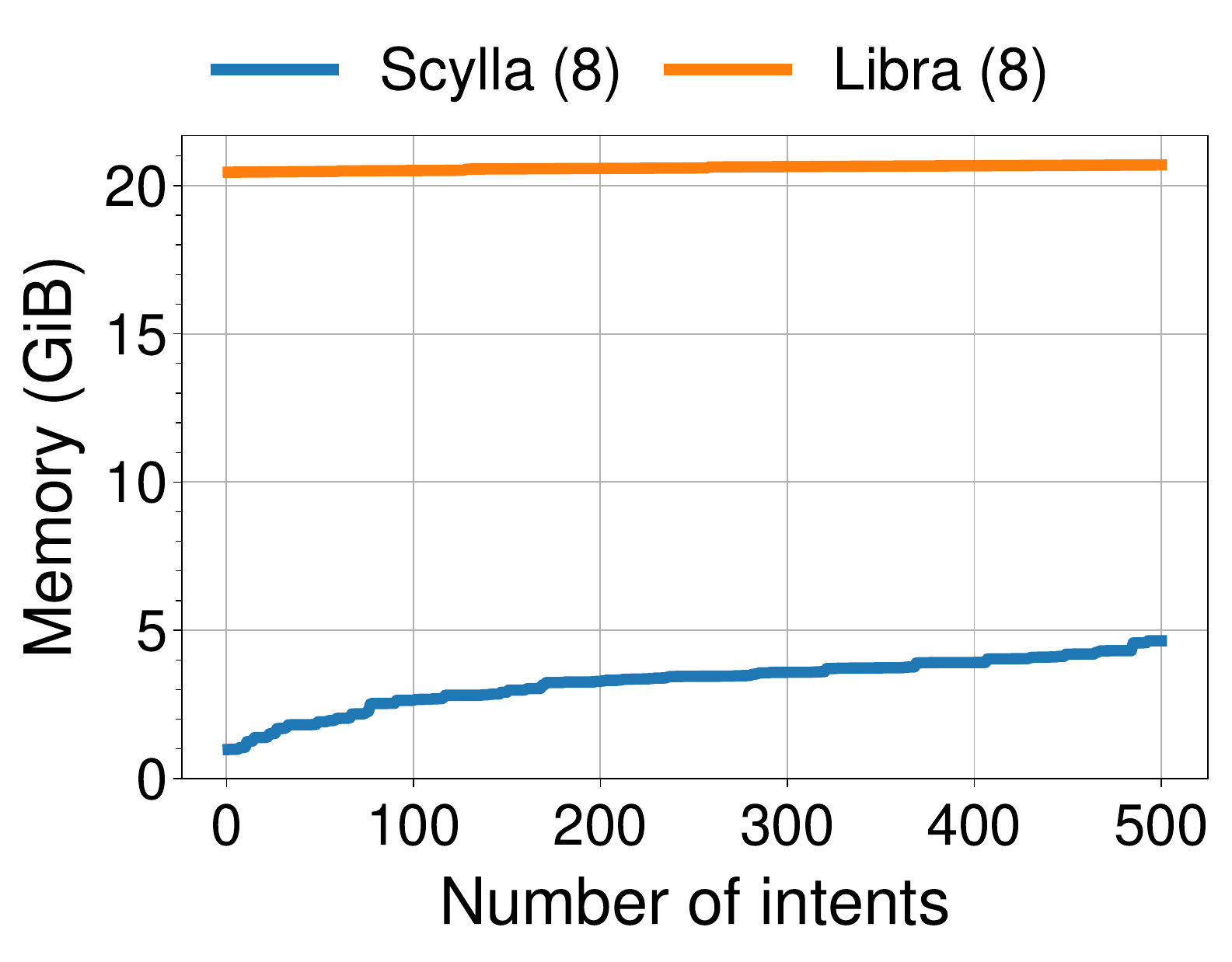}
        \caption{Memory (N4 underlay)}
    \end{subfigure}
    \hfill
    \begin{subfigure}[t]{0.195\linewidth}
        \includegraphics[width=\linewidth]{multi-intent/memory.first-run.N5-underlay.pdf}
        \caption{Memory (N5 underlay)}
    \end{subfigure}
    \caption{Comparison with Libra for verifying 500 intents from scratch (time in log scale).}
    \label{app:fig:multi-intent.libra.first-run}
\end{figure*}

\begin{figure*}[ht]
    \centering
    \begin{subfigure}[t]{0.195\linewidth}
        \includegraphics[width=\linewidth]{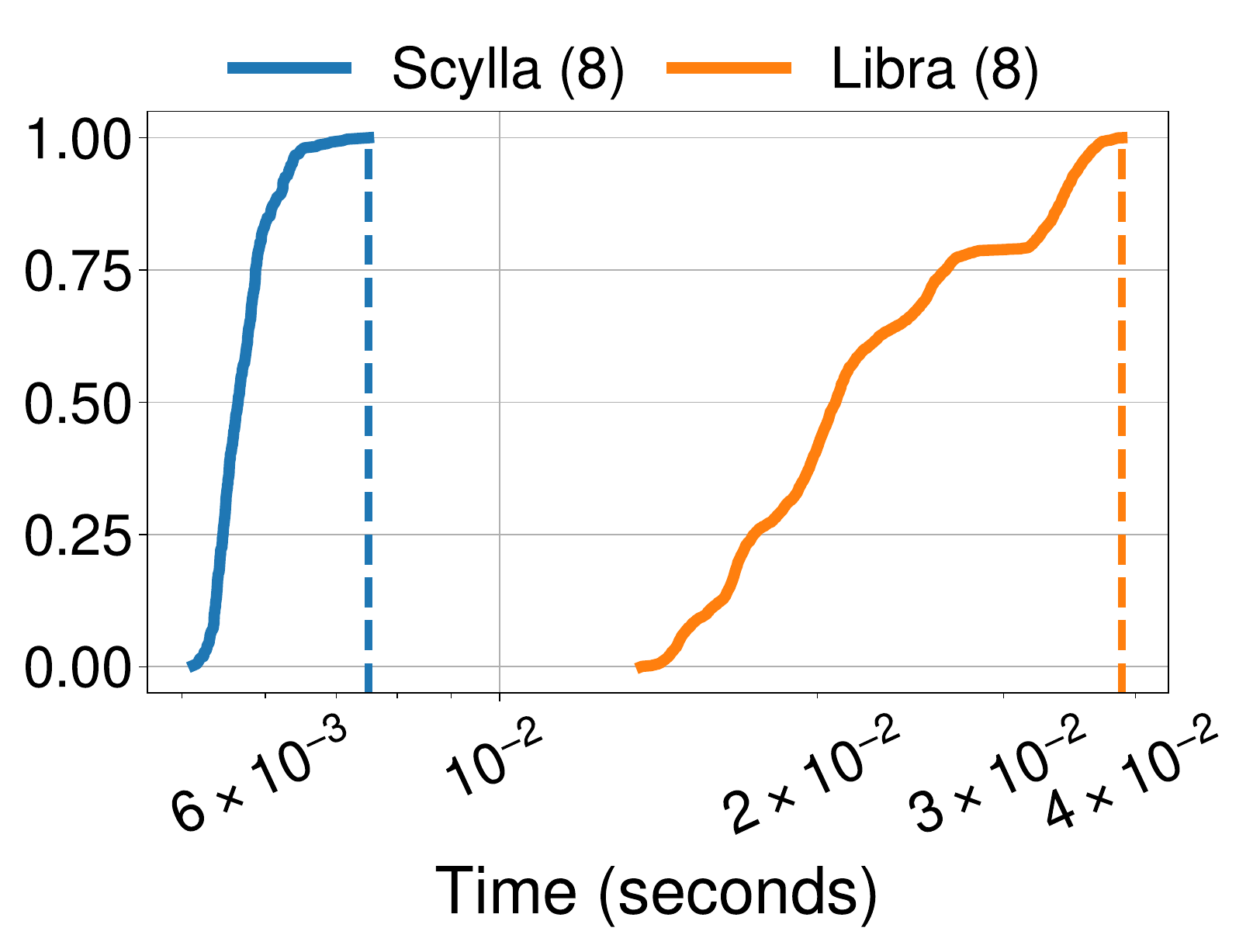}
        \caption{Time CDF (N1 underlay)}
    \end{subfigure}
    \hfill
    \begin{subfigure}[t]{0.195\linewidth}
        \includegraphics[width=\linewidth]{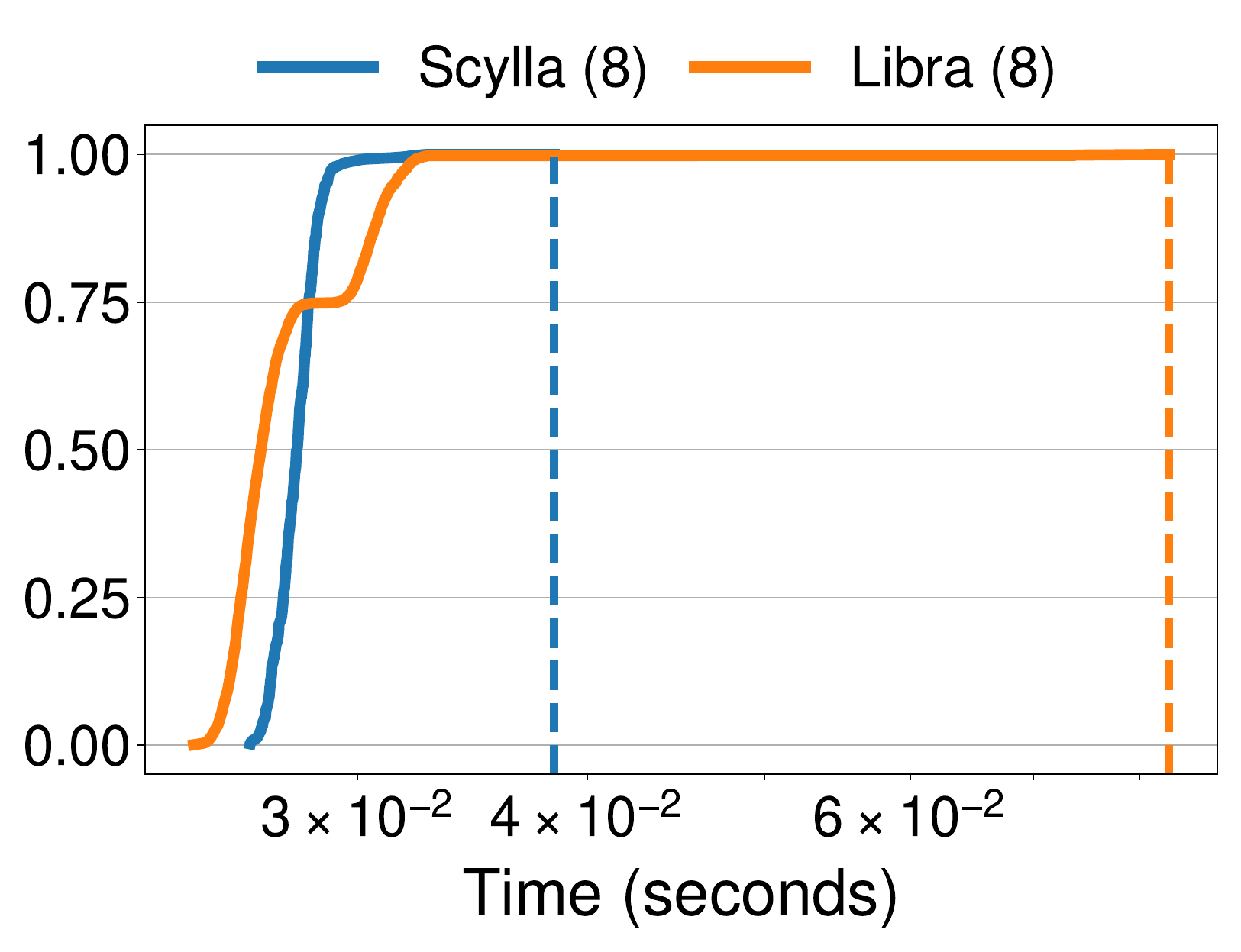}
        \caption{Time CDF (N2 underlay)}
    \end{subfigure}
    \hfill
    \begin{subfigure}[t]{0.195\linewidth}
        \includegraphics[width=\linewidth]{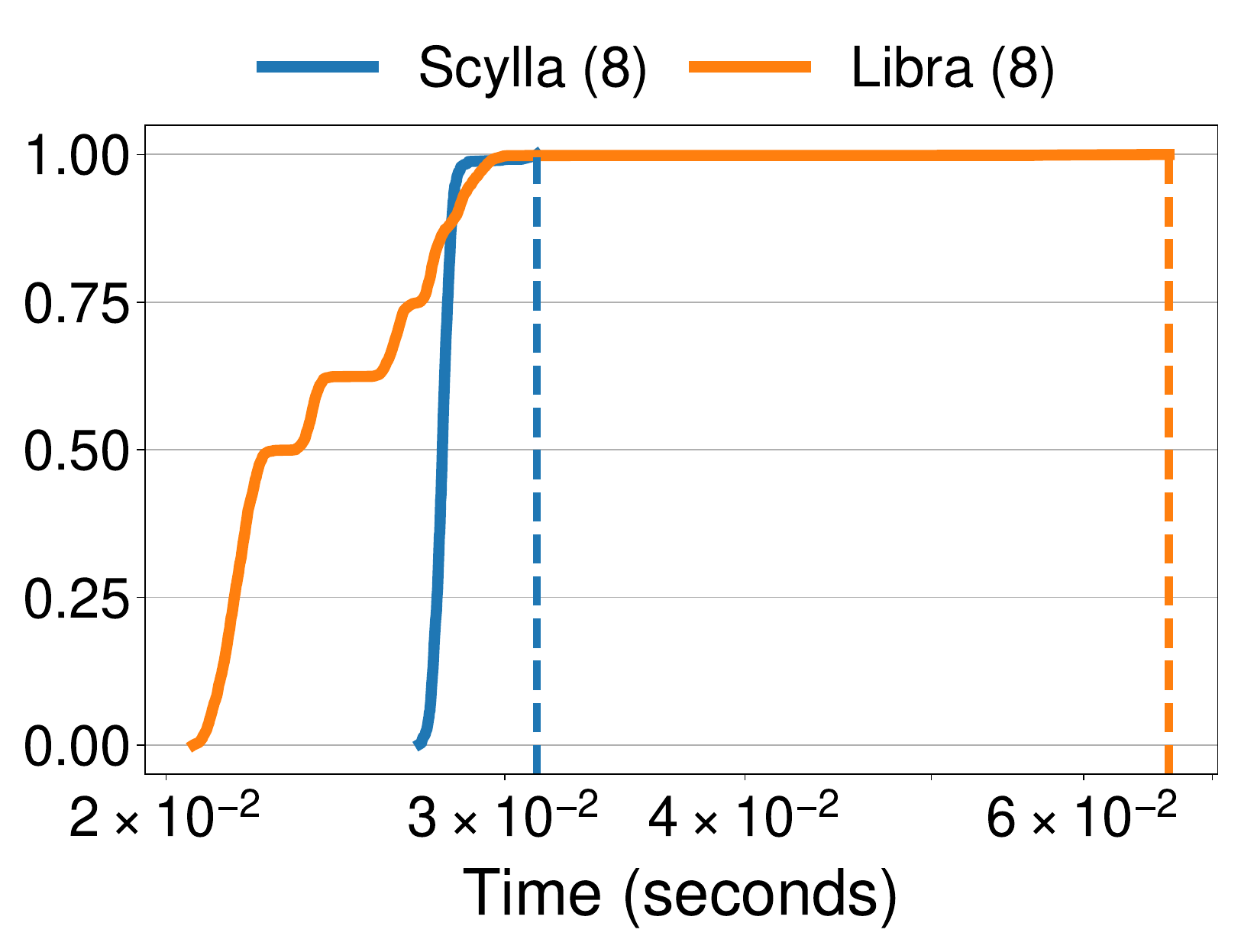}
        \caption{Time CDF (N3 underlay)}
    \end{subfigure}
    \hfill
    \begin{subfigure}[t]{0.195\linewidth}
        \includegraphics[width=\linewidth]{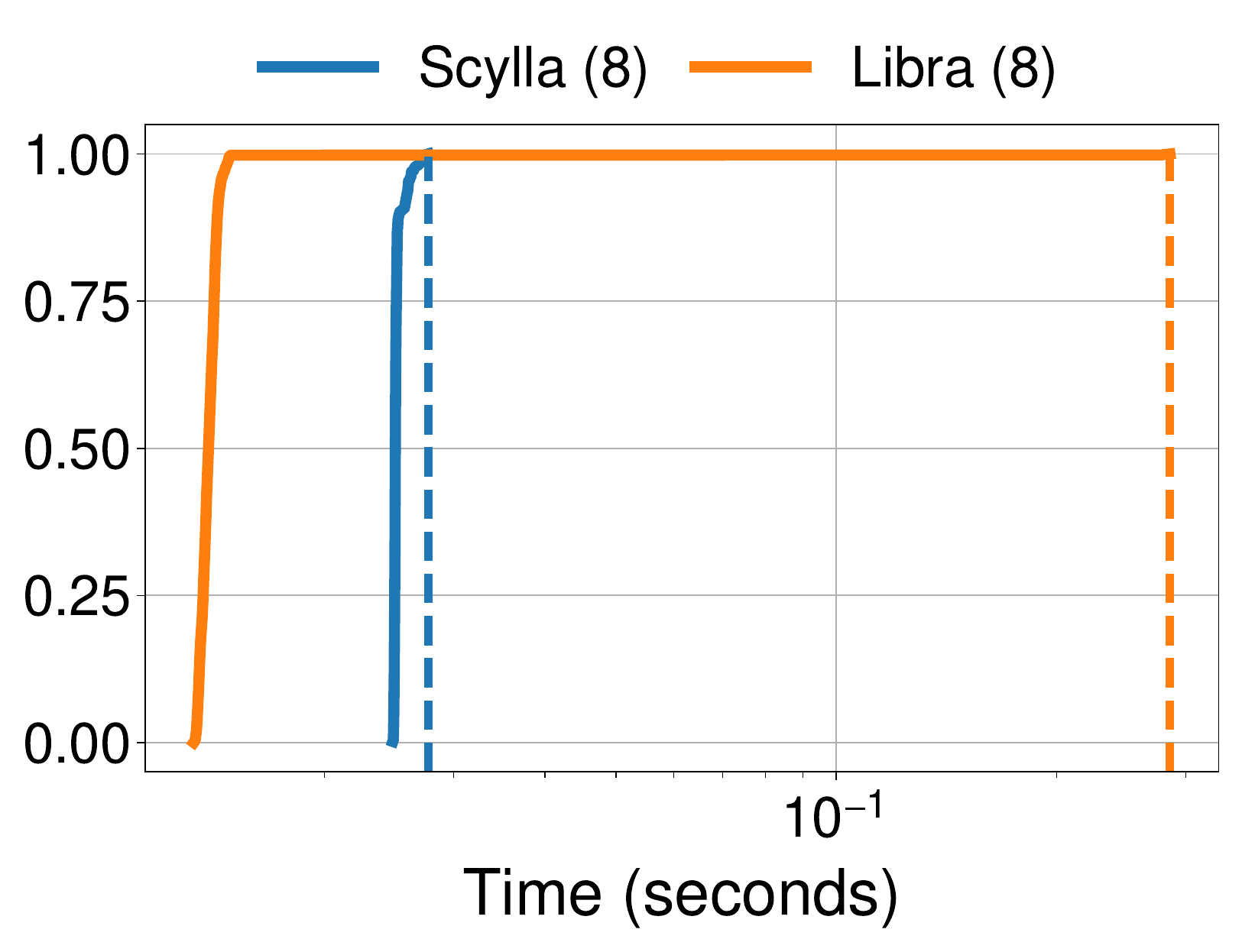}
        \caption{Time CDF (N4 underlay)}
    \end{subfigure}
    \hfill
    \begin{subfigure}[t]{0.195\linewidth}
        \includegraphics[width=\linewidth]{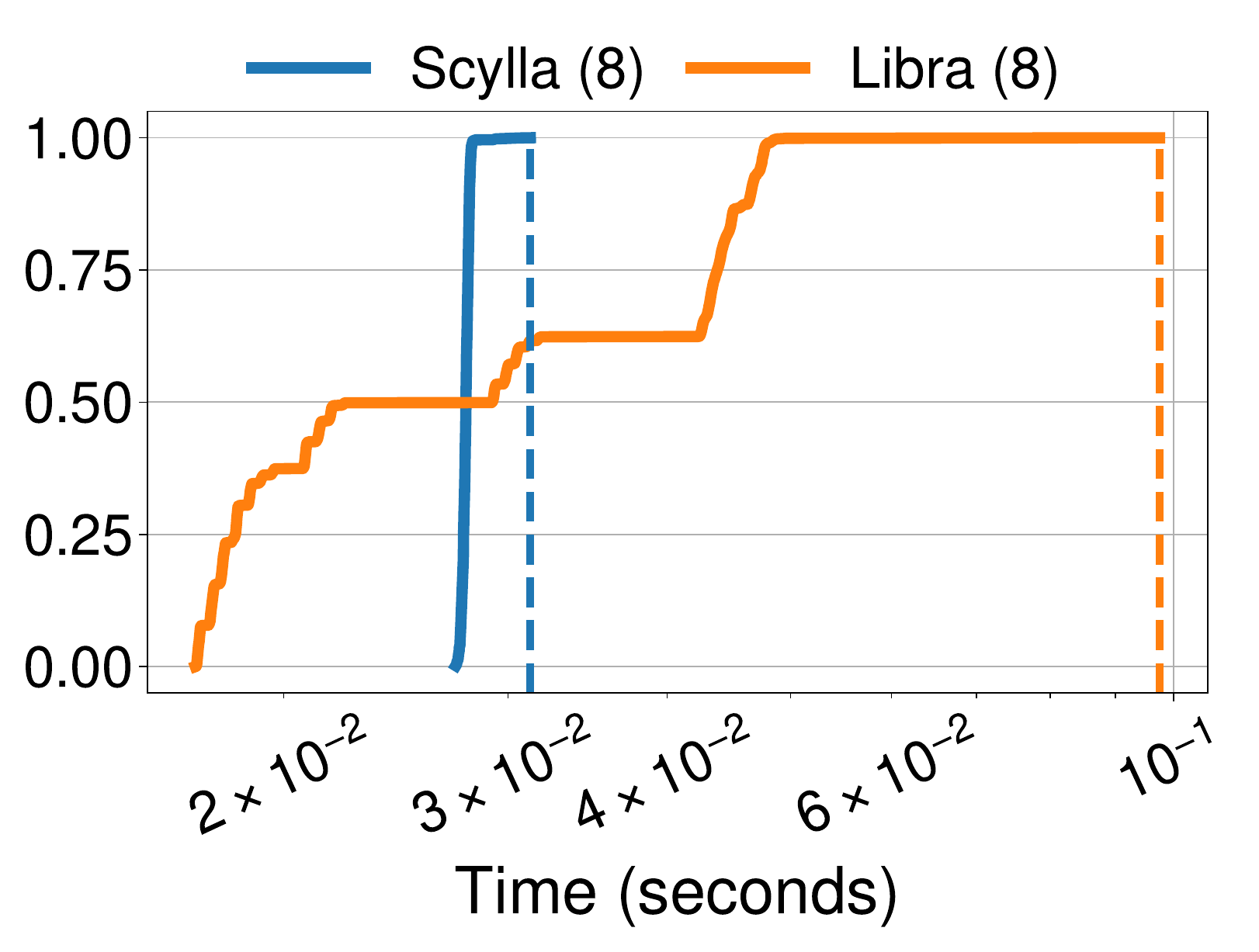}
        \caption{Time CDF (N5 underlay)}
    \end{subfigure}
    \hfill
    \begin{subfigure}[t]{0.195\linewidth}
        \includegraphics[width=\linewidth]{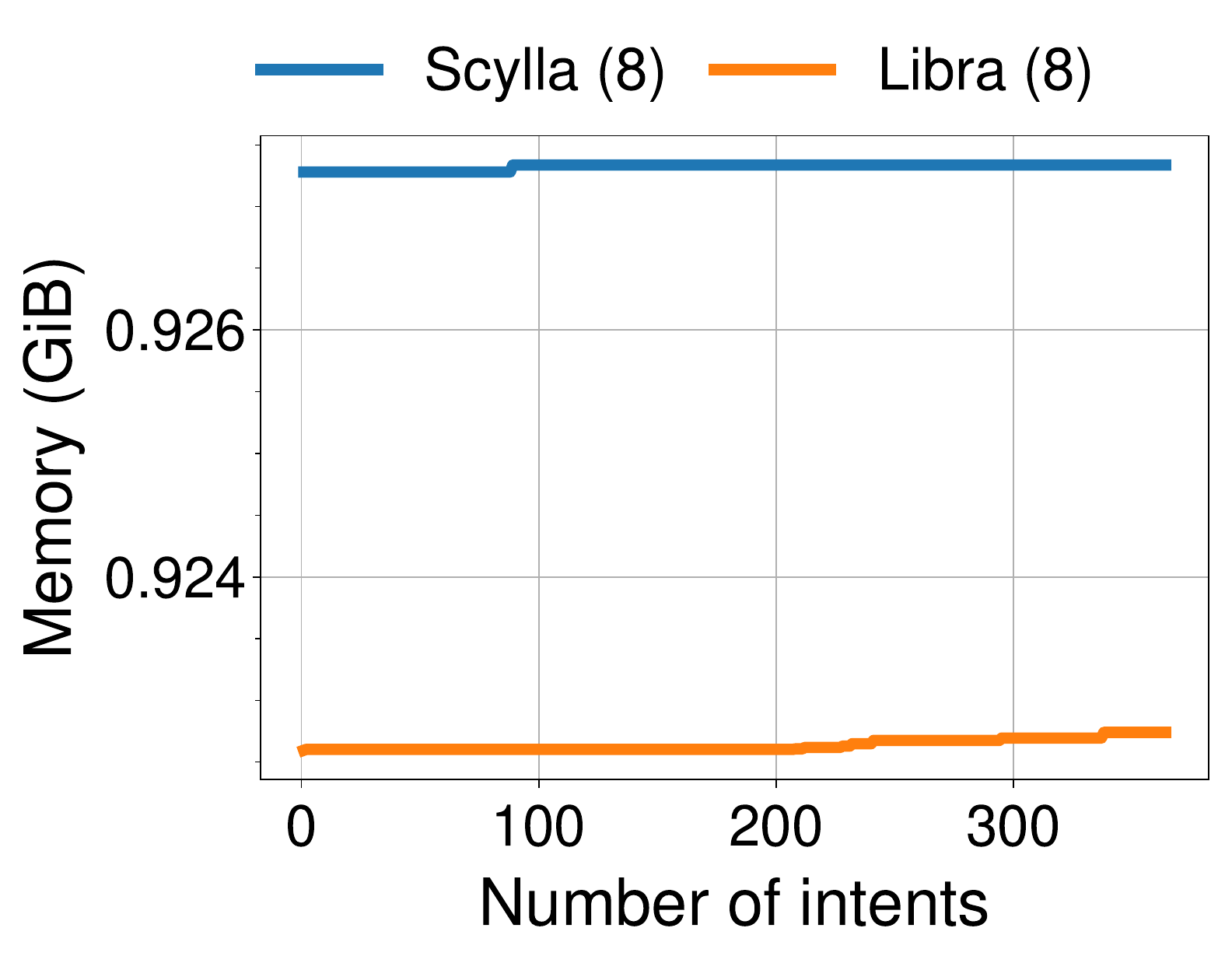}
        \caption{Memory (N1 underlay)}
    \end{subfigure}
    \hfill
    \begin{subfigure}[t]{0.195\linewidth}
        \includegraphics[width=\linewidth]{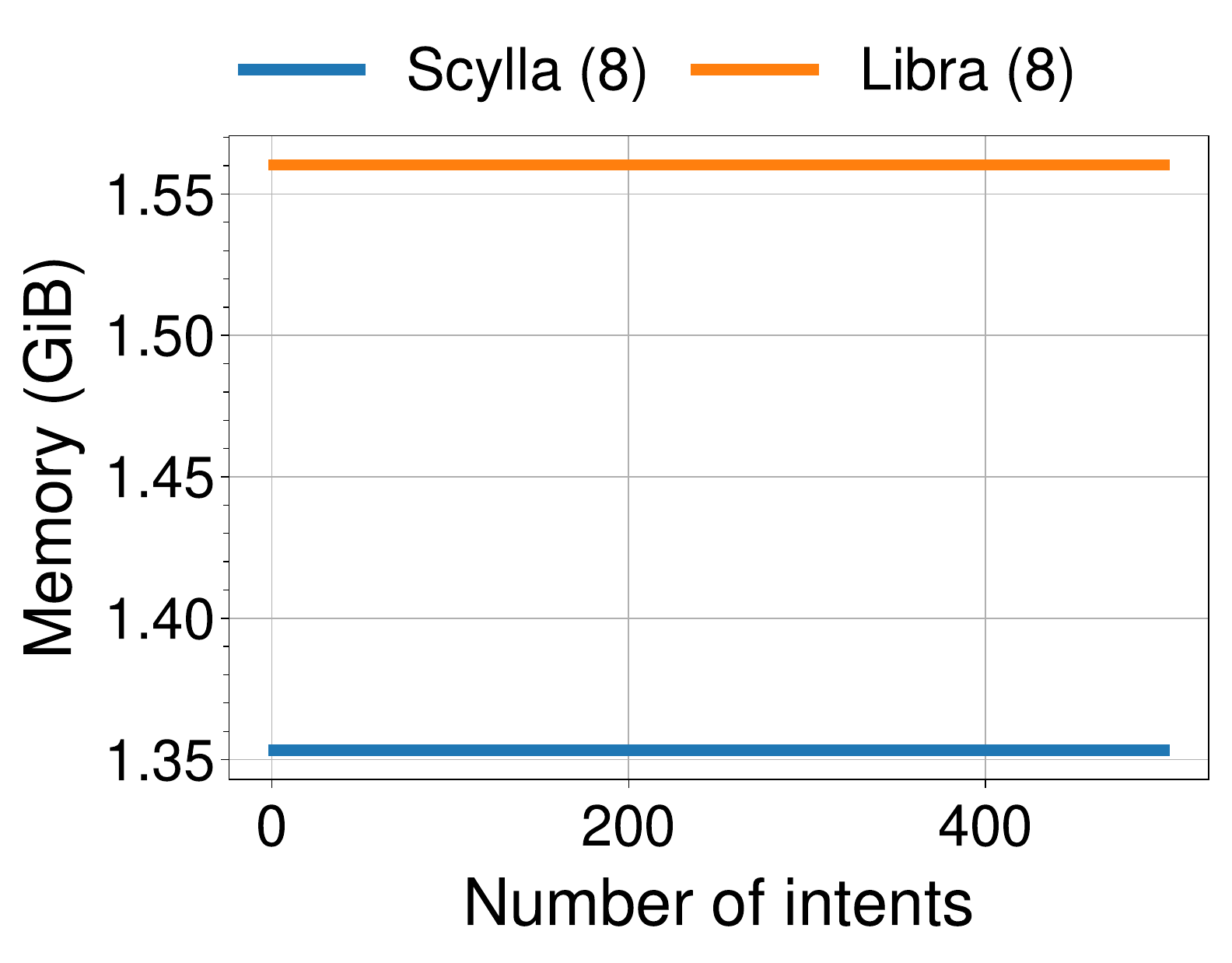}
        \caption{Memory (N2 underlay)}
    \end{subfigure}
    \hfill
    \begin{subfigure}[t]{0.195\linewidth}
        \includegraphics[width=\linewidth]{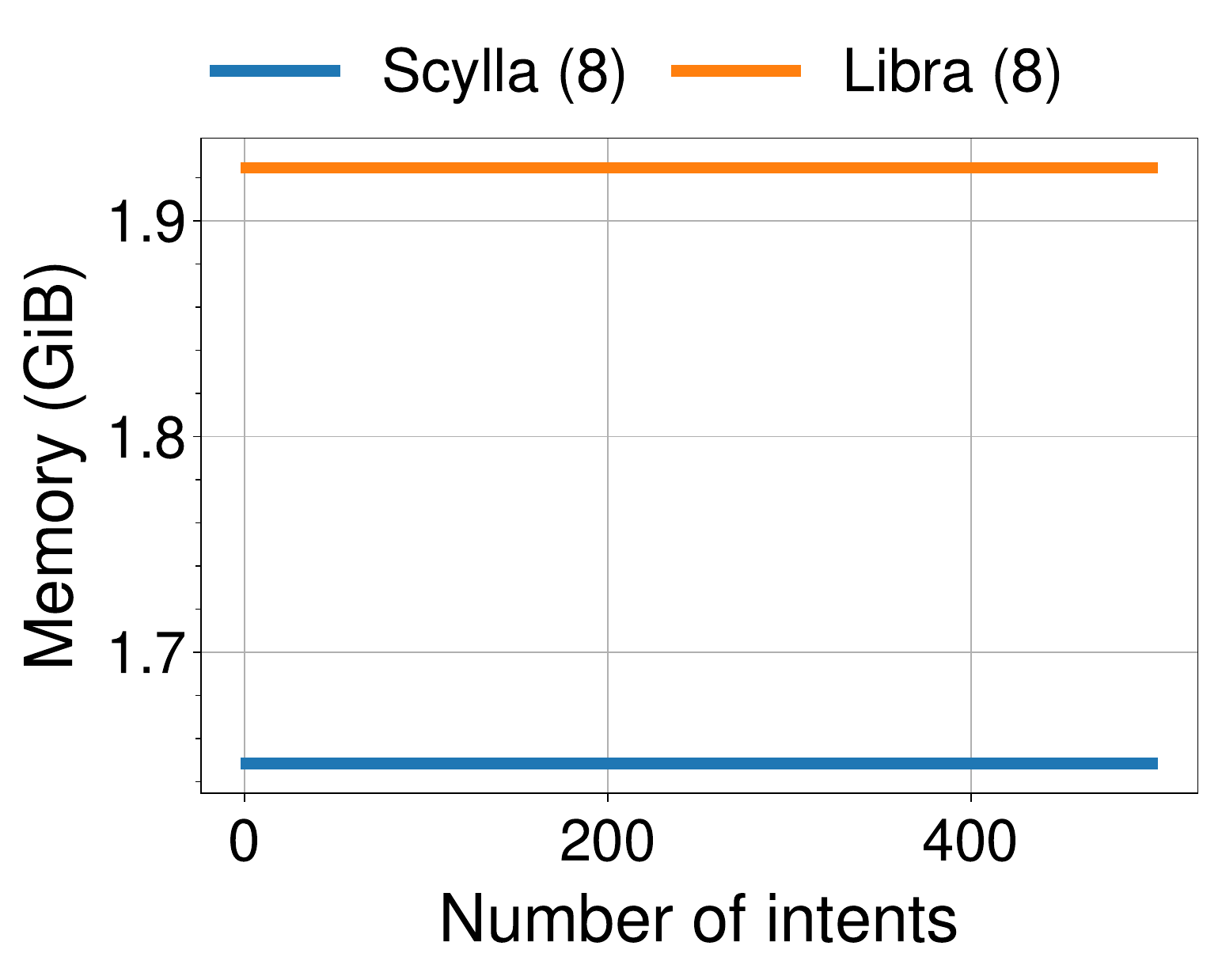}
        \caption{Memory (N3 underlay)}
    \end{subfigure}
    \hfill
    \begin{subfigure}[t]{0.195\linewidth}
        \includegraphics[width=\linewidth]{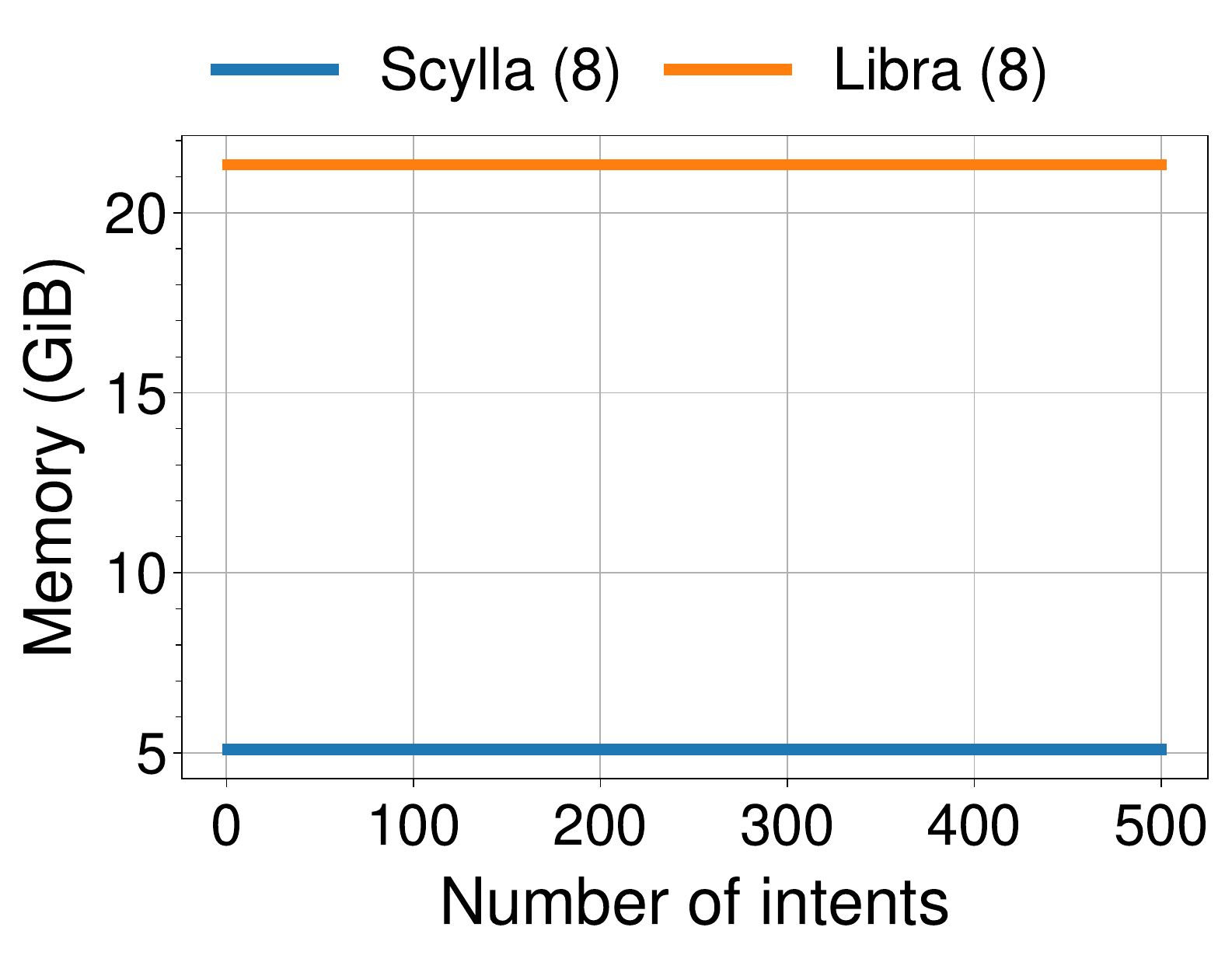}
        \caption{Memory (N4 underlay)}
    \end{subfigure}
    \hfill
    \begin{subfigure}[t]{0.195\linewidth}
        \includegraphics[width=\linewidth]{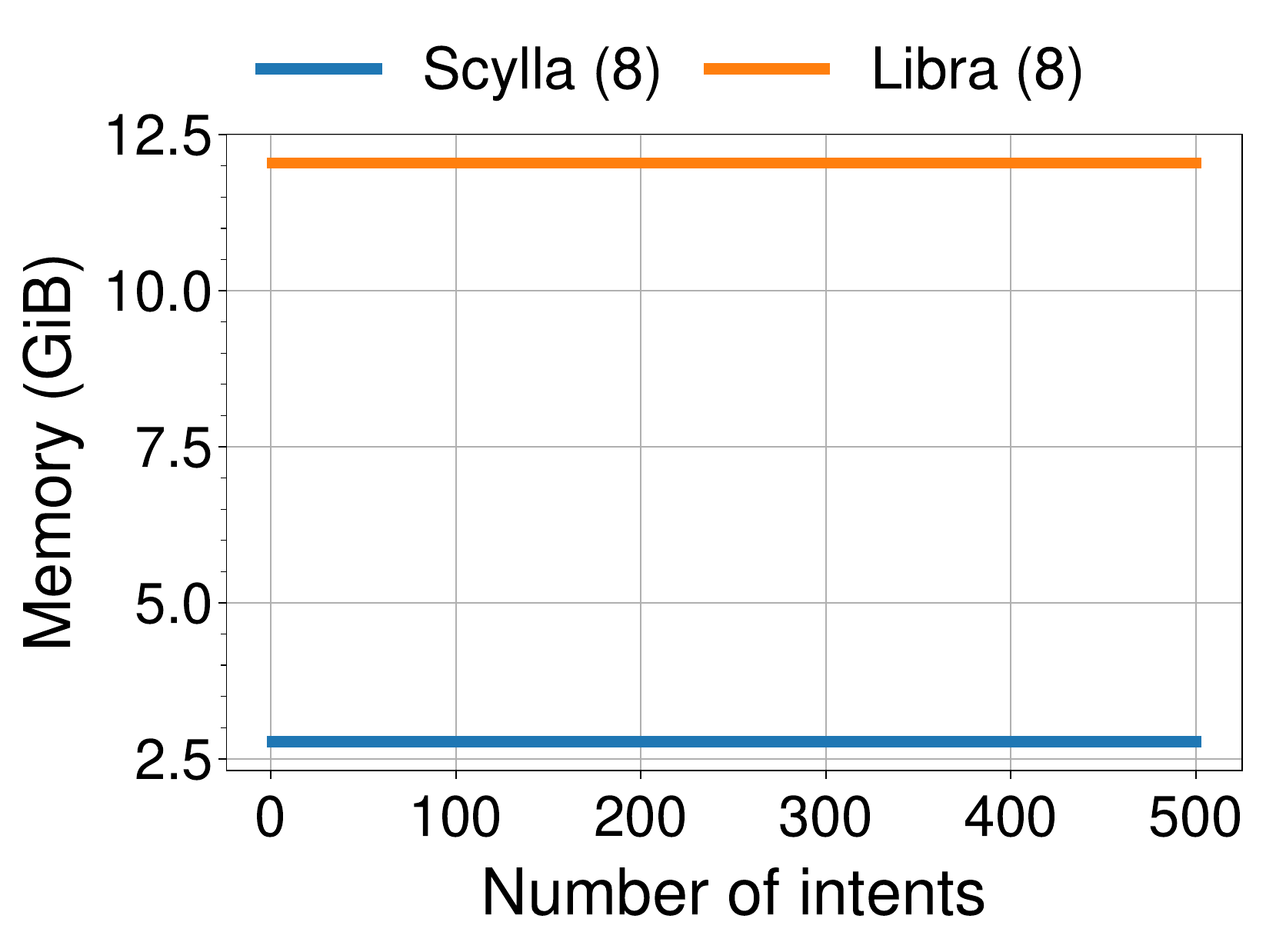}
        \caption{Memory (N5 underlay)}
    \end{subfigure}
    \caption{Comparison with Libra for maintaining 500 existing intents across network updates (time in log scale).}
    \label{app:fig:multi-intent.libra.update}
\end{figure*}

Figure~\ref{app:fig:multi-intent.libra.first-run} and
\ref{app:fig:multi-intent.libra.update} present the full results for
\S\ref{subsec:libra-compare}.

\subsection{Distributed loop detection}
\label{app:subsec:loop}

\begin{figure*}[ht]
    \centering
    \begin{subfigure}[t]{0.24\linewidth}
        \includegraphics[width=\linewidth]{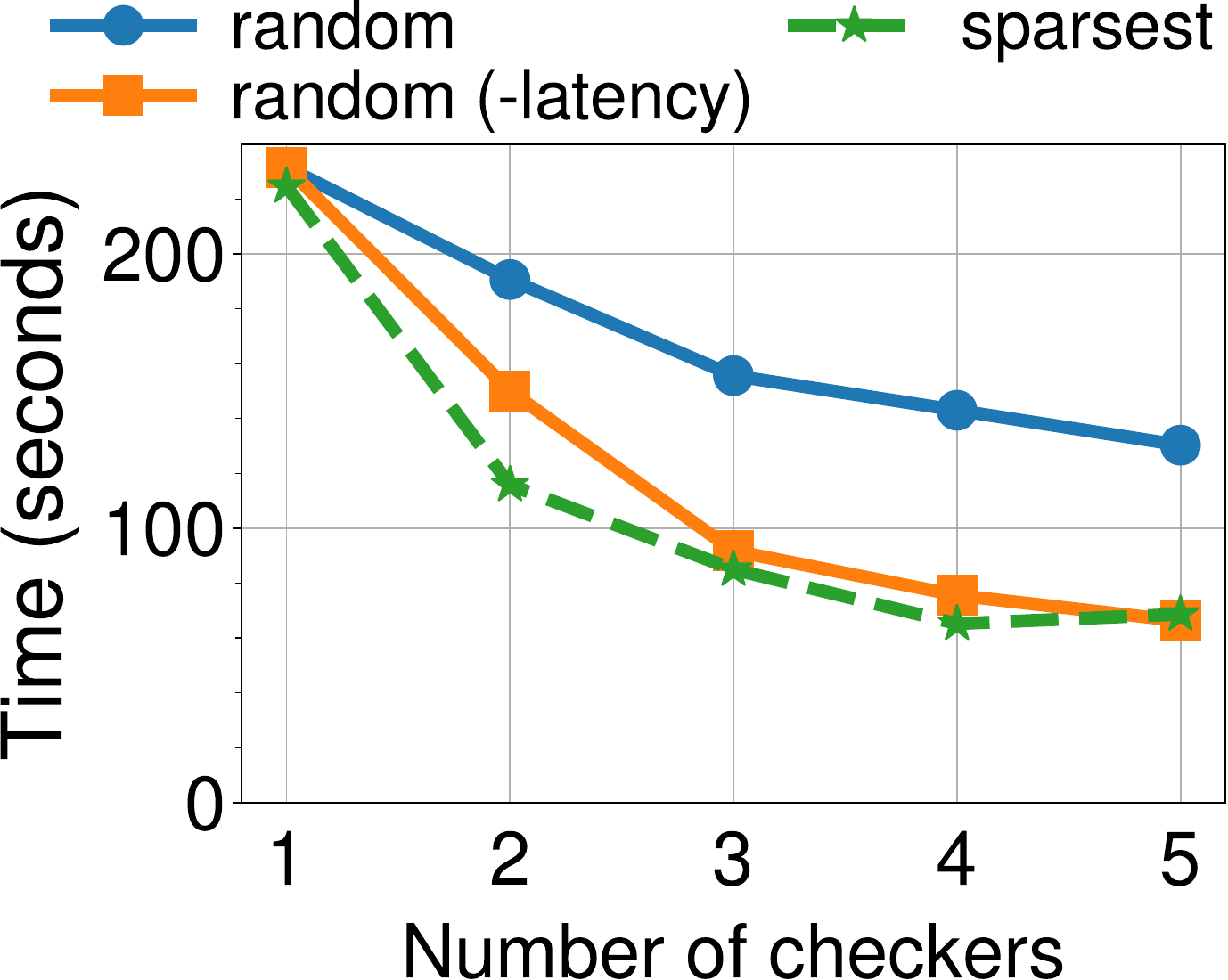}
        \caption{Total time (N4)}
    \end{subfigure}
    \hfill
    \begin{subfigure}[t]{0.24\linewidth}
        \includegraphics[width=\linewidth]{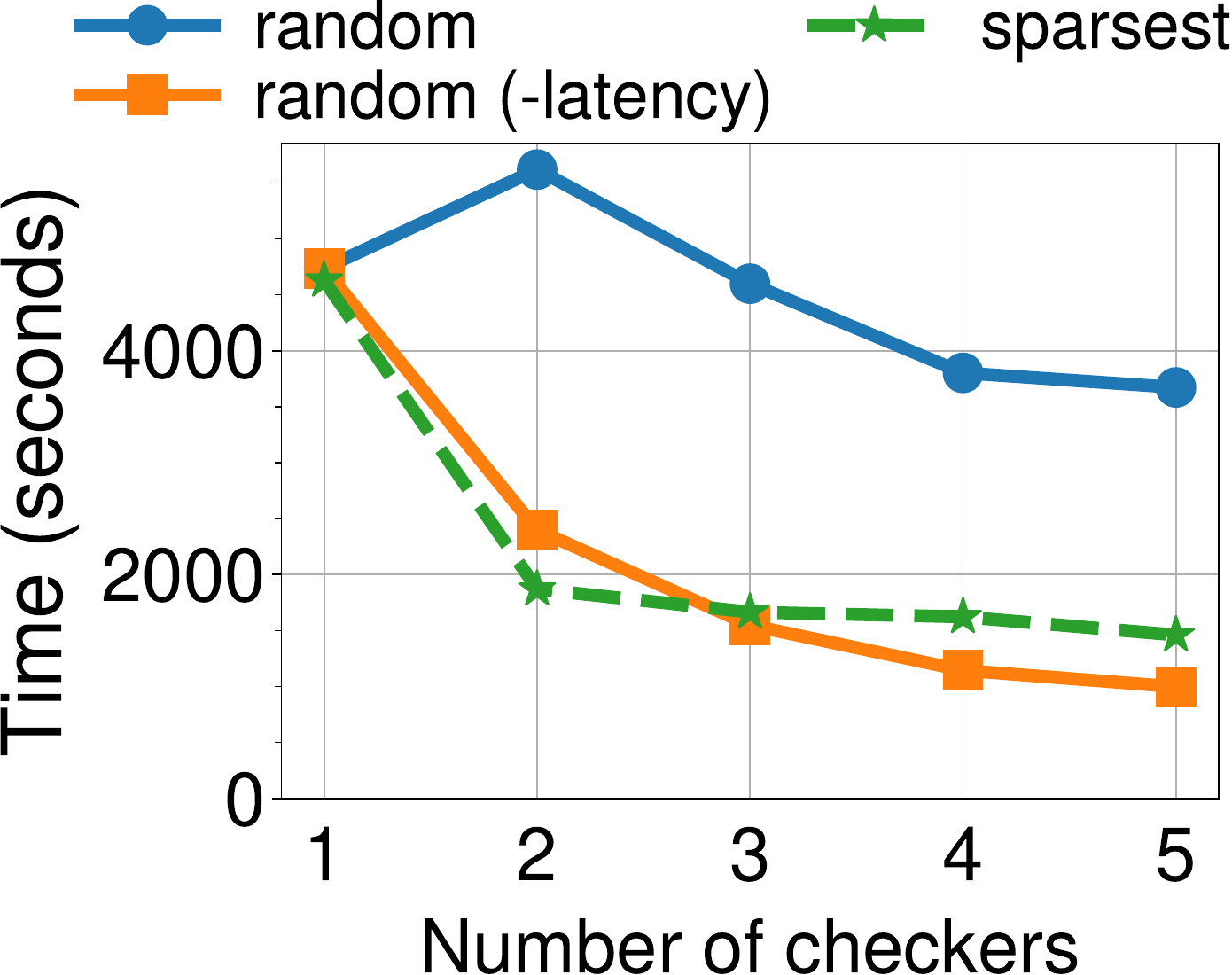}
        \caption{Total time (N5)}
    \end{subfigure}
    \hfill
    \begin{subfigure}[t]{0.24\linewidth}
        \includegraphics[width=\linewidth]{loop/loop.peak-memory.N4.pdf}
        \caption{Max memory/checker (N4)}
    \end{subfigure}
    \hfill
    \begin{subfigure}[t]{0.24\linewidth}
        \includegraphics[width=\linewidth]{loop/loop.peak-memory.N5.pdf}
        \caption{Max memory/checker (N5)}
    \end{subfigure}
    \caption{Performance of distributed loop detection including sparsest-cut partitioning.}
    \label{app:fig:loop.with-sparsest-cut}
\end{figure*}

Figure~\ref{app:fig:loop.with-sparsest-cut} includes the total run time for the sparsest cut method mentioned in \S\ref{subsec:loop-eval}. Since the sparsest cut method did not require communication between checkers, the total time remains the same with or without the messaging latency. This shows that for networks that are naturally segmented, \name can greatly improve the performance of verifying loop freedom by simply adding more checkers.

%%%%%%%%%%%%%%%%%%%%%%%%%%%%%%%%%%%%%%%%%%%%%%%%%%%%%%%%%%%%%%%%%%%%%%%%%%%%%%%%

\end{document}